\documentclass[]{article}

\newcommand{\sumal}{\sum_{\alpha=1}^{2}}
\usepackage{color}
\usepackage{float}
\usepackage{mathptmx}
\usepackage{helvet}
\usepackage{courier}
\usepackage{amsmath,amssymb}
\usepackage{amsfonts}
\usepackage{amsbsy}
\usepackage{bm}
\usepackage{type1cm}  
\usepackage{dcolumn}       
\usepackage{index}         
\usepackage{graphicx}        
\usepackage{multicol}        
\usepackage[bottom]{footmisc}
\usepackage{epstopdf}
\usepackage{subfigure}
\usepackage{mathtools}
\usepackage{color}
\mathtoolsset{centercolon}
\mathtoolsset{showonlyrefs=true}
\usepackage{cite}

\usepackage{authblk}
\usepackage{hyperref}

\begin{document}

\title{Effect of the dynamic pressure on the shock structure and sub-shocks formation in a mixture of polyatomic gases}

\author[1]{Tommaso Ruggeri\thanks{tommaso.ruggeri@unibo.it}}
\author[2]{Shigeru Taniguchi\thanks{taniguchi.shigeru@kct.ac.jp}}

\affil[1]{Department of Mathematics $\&$ Alma Mater Research Center on Applied Mathematics, University of Bologna, Via Saragozza, 8, Bologna, 40123, Italy}

\affil[2]{Department of Creative Engineering, National Institute of Technology, Kitakyushu College, 5-20-1 Shii, Kitakyushu, 802-0985, Japan}

\date{} 


\maketitle

\abstract{
We study the shock structure and the sub-shocks formation in a binary mixture of rarefied polyatomic gases, considering the dissipation only due to the dynamic pressure.
We classify the regions depending on the concentration and the Mach number for which there may exist the sub-shock in  the profile of shock structure in one or both constituents or not for prescribed values of the mass ratio of the constituents and the ratios of the specific heats. We compare the regions with the ones of the corresponding mixture of Eulerian gases and  
 we perform the numerical calculations of the shock structure for typical cases previously classified and confirm whether sub-shocks emerge. 
}

\bigskip

{\flushleft
{\bf keywords:} Shock structure, Mixture of gases, Rational extended thermodynamics, Polyatomic gases, Dynamic pressure, Sub-shock formation, Bulk viscosity
}

\bigskip

{\flushleft
{\bf MSC Classification:} 35L67, 76L05, 76N30, 35L02
}

\section{Introduction}
The qualitative and quantitative analysis of the shock structure is important for many applications and a challenging subject for physics to develop a valid theory to explain the rapid and steep change of the physical quantities near the shock front~\cite{VincentiKruger,Zeldovich}. 
In order to understand the structure of strong shock waves involving a highly-nonequilibrium state beyond the validity range of the \emph{local equilibrium} assumption, we need to establish an appropriate theory. 
Non-equilibrium thermodynamics based on the local equilibrium  assumption, including the Navier-Stokes-Fourier (NSF) theory, is not able to explain the shock structure with large Mach numbers. 

The kinetic approaches based on the Boltzmann equation, such as DSMC (Direct Simulation Monte Carlo), provide successful results on the shock structure in a rarefied monatomic gas~\cite{Bird}.  
However, due to internal modes, modeling the collision terms becomes difficult for polyatomic gases. 
Particularly difficult for a mixture of rarefied polyatomic gases object of the present study. On these subjects, see the recent works~\cite{PavicTorr,PavicGamba} and the review paper \cite{Pirner}.

Rational Extended Thermodynamics (RET)~\cite{RET,RuggeriSugiyama,BookNew} is a phenomenological theory that is valid beyond the validity range of the local equilibrium assumption. 
In addition to the usual equilibrium quantities, the RET theory adopts dissipative fluxes, such as the viscous stress and the heat flux, as independent variables. 
The closure of the system is achieved by using universal principles: \textit{objectivity}, \textit{entropy principle}, and \textit{thermodynamic stability}. 

In the case of rarefied polyatomic gas with large bulk viscosity, as the Mach number increases, the shock structure changes from the nearly symmetric profile (Type A) to an asymmetric profile (Type B) and further to the profile with thick and thin layers (Type C)~\cite{VincentiKruger}. 
RET can explain these changes of the shock wave structure in a unified way~\cite{ET14shock,ET6shock} in contrast to previous approaches such as the Bethe-Teller theory~\cite{BetheTeller} and the prediction by Navier-Stokes-Fourier theory~\cite{Gilbarg}. 
In particular, the shock structure in a rarefied polyatomic gas with large bulk viscosity (large relaxation time for the dynamic pressure) can be predicted by RET with only $6$ independent fields (RET$_6$)~\cite{6fields,6fields2}; the mass density $\rho$, the velocity $\mathbf{v}$, the temperature $T$, and the dynamic pressure $\Pi$. 
Within the limited resolution of RET$_6$, the thin layer is described as a sub-shock~\cite{ET6shock,NLET6shock} while RET$_{14}$ with $14$ independent variables can explain the fine structure with very steep but continuous change. 
After the successful explanation of the shock structure by RET, numerical analysis based on the kinetic theory was done. The agreement between the theoretical predictions by RET and kinetic theory strongly supports the usefulness of the analysis based on RET~\cite{Kosuge,2018kyoto}. 

The next step is to analyze the shock structure in a mixture of polyatomic gases. 
In a recent paper~\cite{GMixture}, by using the framework of RET, 
the field equations of a mixture of polyatomic gases are derived in the context of the multi-temperature model where each constituent has $6$ fields.

For a generic  quasi-linear hyperbolic system of partial differential equations
it has been shown that the profile of the shock structure may not be continuous only when the shock velocity meets a characteristic velocity~\cite{Ruggeri1993}. However, this necessary condition becomes also sufficient when the shock speed is greater than the maximum  characteristic velocity evaluated in the unperturbed field, as was proved in a theorem by  Boillat and Ruggeri~\cite{Breakdown}.
The present status of the sub-shock formation is summarized in a recent survey \cite{ShRu} and in the book \cite{BookNew}.

A sub-shock emerges only after the maximum characteristic velocity in the context of RET theory for a single fluid of rarefied gases according to the numerical calculations~\cite{Weiss,IJNLM2017}. 
On the other hand, sub-shock may appear before the maximum characteristic velocity, and multiple sub-shock may also arise in the context of mixture theories~\cite{FMR,Bisi1,Bisi2} and toy models~\cite{IJNLM2017,subshock2}.

In the case of the binary Eulerian mixture of rarefied polyatomic gases, the authors have given a complete classification of  the regions in the plane of concentration and Mach number where both scenarios of sub-shocks before and after the maximum characteristic velocity arises ~\cite{ShockBinaryEulerian_lincei,ShockBinaryEulerian_PhysFluids}.
For a dissipative mixture of gases, we started the equivalent analysis of Eulerian mixtures in a recent paper \cite{ShockBinaryET6_RdM} but examined only some particular values of the structural parameters and have not given a complete classification of the possible regions,  that is, the aim of the present paper.

The present paper is organized as follows: In Section \ref{sec:Model}, we summarize the basic equations for the present analysis. 
In Section \ref{sec:ShockStruct}, the shock structure problem is presented, and in Section \ref{sezione4} are classified the possible regions depending on the concentration and Mach number in which there is the possibility of sub-shock formation.
In Section \ref{sec:Numerical}, the numerical calculations of the shock structure are given and are confirmed whether the sub-shocks arise or not.
Section \ref{sec:Summary} shows the Conclusions. 

\section{Basic equations for a binary mixture of RET$_6$ gases} 
\label{sec:Model}

We consider a binary mixture of rarefied polytropic gases described by the following thermal and caloric equations of state: 
\begin{equation*}
p_{\alpha} = \frac{k_B}{m_{\alpha}} \rho_{\alpha} T_{\alpha}, \qquad 
\varepsilon_{\alpha} = \frac{k_B}{m_{\alpha}(\gamma_{\alpha} - 1)}T_{\alpha}, 
\end{equation*} 
where the quantities with suffix $\alpha$ $(= 1, 2)$ represent the corresponding quantities for the constituent $\alpha$ and $p_{\alpha}$, $k_B$, $m_\alpha$, $T_{\alpha}$, $\varepsilon_{\alpha}$, and $\gamma_{\alpha}$ are, respectively, the pressure, the Boltzmann constant, the mass of a molecule, the temperature, the specific internal energy, and the ratio of the specific heats.  
In the present analysis, we consider polytropic gases in which the specific heat is independent of the temperature, and therefore also $\gamma_\alpha$   is constant. 

In the present study, we analyze plane shock waves propagating along with $x$-direction and hereafter, we focus on the one-dimensional problem. Furthermore, it is assumed that the mixture is inert and no chemical reactions occur.

The system of a binary mixture of RET$_6$ gases without shear viscosity and heat conductivity was given in the paper \cite{GMixture} \footnote{In \cite{MRSimic} the production terms $\hat{e}_1$ and $\hat{\omega}_1$ differ from the present one by a factor $2$. The present choice is better to compare with previous results in the case of Eulerian gases.}: 
\begin{align}\label{finale1}
& \frac{\partial \rho_1 }{\partial t} + \frac{\partial \rho_1  v_1 }{\partial x} 
= 0,  \nonumber \\ 
&\frac{\partial \rho_1 v_1 }{\partial t}+ \frac{\partial}{\partial x} \left\{\rho_1  v^2_1 + p_1  + \Pi_1 \right\} = \hat{m}_1, \nonumber \\
&\frac{\partial}{\partial t} \left(\rho_1  v^2_1 + 2 \rho_1  \varepsilon_1 \right) +
\frac{\partial}{\partial x}\left\{\left[ \rho_1 v^2_1 + 2\rho_1  \varepsilon_1  + 2 p_1  + 2\Pi_1  \right] v_1  \right\} 
= 2(\hat{e}_1 +  \hat{m}_1 v), \nonumber
\\
&\frac{\partial }{\partial t}\left\{\rho_1  v_1^2 + 3(p_1 +\Pi_1 )\right\}
+ \frac{\partial }{\partial x}\left\{\left[\rho_1  v_1^2 + 5 (p_1 +\Pi_1 )\right] v_1 \right\} = - \frac{3 \Pi_1}{\tau_1}  +  2 (\hat{\omega}_1 +   \hat{m}_1 v), 
\nonumber \\
& \frac{\partial \rho_2 }{\partial t} + \frac{\partial \rho_2  v_2 }{\partial x} 
= 0,  \\
&\frac{\partial \rho_2 v_2 }{\partial t}+ \frac{\partial}{\partial x} \left\{\rho_2  v^2_2 + p_2  + \Pi_2 \right\} =  -\hat{m}_1, \nonumber \\
&\frac{\partial}{\partial t} \left(\rho_2  v^2_2 + 2 \rho_2  \varepsilon_2 \right) +
\frac{\partial}{\partial x}\left\{\left[ \rho_2 v^2_2 + 2\rho_2  \varepsilon_2  + 2 p_2  + 2\Pi_2  \right] v_2  \right\} 
=  -2(\hat{e}_1 +  \hat{m}_1 v), \nonumber \\
&\frac{\partial }{\partial t}\left\{\rho_2  v_2^2 + 3(p_2 +\Pi_2 )\right\}
+ \frac{\partial }{\partial x}\left\{\left[\rho_2  v_2^2 + 5 (p_2 +\Pi_2 )\right] v_2 \right\} = - \frac{3 \Pi_2}{\tau_2}   -2 (\hat{\omega}_1 +   \hat{m}_1 v), \nonumber
\end{align}
where $v$ is the velocity of center of mass and $\tau_{\alpha} > 0$ is the relaxation time for the dynamic pressure $\Pi_\alpha$ of the constituent $\alpha$. 
The production terms, $\hat{m}_{\alpha}$, $\hat{e}_{\alpha}$, and $\hat{\omega}_{\alpha}$ represent the interchange of the momentum, the energy, the momentum flux, respectively evaluated for $v=0$.

The expressions of $\hat{m}_1$, $\hat{e}_1$, and $\hat{\omega}_1$ are obtained in such that the entropy production   is positive and quadratic as usual in non-equilibrium thermodynamics \cite{GMixture} 
\begin{equation}\label{eq:production_interaction}
\begin{split}
\hat{m}_1 = \psi_{11} \hat{\Lambda}_1, \quad
\hat{e}_1 = 2(\theta_{11} \hat{\eta}_1 + \kappa_{11} \hat{\zeta}_1), \quad
\hat{\omega}_1 = 2(\kappa_{11} \hat{\eta}_1 + \phi_{11} \hat{\zeta}_1 ),
\end{split}
\end{equation}
with 
\begin{equation}\label{eq:main_field}
\begin{split}
&\hat{\Lambda}_1 = - \frac{u_1}{T_1}\left( 1- \frac{\Pi_1}{p_1}\right) + \frac{u_2}{T_2}\left( 1- \frac{\Pi_2}{p_2}\right), \\
&\hat{\eta}_1 = \frac{1}{2 T_1} \left( 1- \frac{\Pi_1}{p_1}\right) - \frac{1}{2 T_2} \left( 1- \frac{\Pi_2}{p_2}\right), \\
&\hat{\zeta}_1 = - \frac{1}{(5-3\gamma_1)T_1}\frac{\Pi_1}{p_1} + \frac{1}{(5-3\gamma_2)T_2}\frac{\Pi_2}{p_2}, 
\end{split}
\end{equation}
where $u_\alpha = v_\alpha - v$ is the diffusion velocity. 
The coefficient  $\psi_{11} >0$ and the matrix
\begin{align}\label{eq:stability_matrix}
\begin{pmatrix}
\theta_{11} & \kappa_{11}\\
\kappa_{11} & \phi_{11}
\end{pmatrix}, \qquad \text{is positive definite}.
\end{align}
In general, the phenomenological coefficients $\psi_{11}$, $\theta_{11}$, $\kappa_{11}$, and $\omega_{11}$ depend on the mass densities and temperatures and the functional forms of the coefficients may be determined by kinetic theoretical consideration  and/or experimental data. 

\subsection{Field equations for global quantities}

By taking sum of \eqref{finale1}, we have the following equivalent system  
\begin{align}\label{finale}
&\frac{\partial \rho}{\partial t} + \frac{\partial \rho v}{\partial x} = 0, \nonumber \\
&\frac{\partial \rho v}{\partial t}+ \frac{\partial }{\partial x} \left(\rho v^2 + p + \Pi - \sigma
\right) = 0, \nonumber \\
&\frac{\partial }{\partial t} \left(\frac{1}{2}\rho v^2 + \rho \varepsilon\right)
+\frac{\partial}{\partial x}\left\{\left(\frac{1}{2}\rho v^2 + \rho \varepsilon +  p + \Pi -\sigma
\right)v + q \right\} = 0, \nonumber \\
&\frac{\partial }{\partial t}\left\{\rho v^2 + 3(p+\Pi)\right\} + \frac{\partial }{\partial x}\left\{\left[\rho v^2 + 5 (p+\Pi) - 2 \sigma \right] v  + 2 Q \right\} = - 3\left(\frac{\Pi_{1}}{\tau_1}+\frac{\Pi_{2}}{\tau_2}\right),\nonumber \\
& \frac{\partial \rho_1 }{\partial t} + \frac{\partial \rho_1  v_1 }{\partial x} 
= 0,    \\
&\frac{\partial \rho_1 v_1 }{\partial t}+ \frac{\partial}{\partial x} \left\{\rho_1  v^2_1 + p_1  + \Pi_1 \right\} = \hat{m}_1, \nonumber \\
&\frac{\partial}{\partial t} \left(\rho_1  v^2_1 + 2 \rho_1  \varepsilon_1 \right) +
\frac{\partial}{\partial x}\left\{\left[ \rho v^2_1 + 2\rho_1  \varepsilon_1  + 2 p_1  + 2\Pi_1  \right] v_1  \right\} 
=  2(\hat{e}_1 +  \hat{m}_1 v), \nonumber \\
&\frac{\partial }{\partial t}\left\{\rho_1  (v_1 )^2 + 3(p_1 +\Pi_1 )\right\}
+ \frac{\partial }{\partial x}\left\{\left[\rho_1  v_1^2 + 5 (p_1 +\Pi_1 )\right] v_1 \right\} 
=  - \frac{3 \Pi_1}{\tau_1}   +2 (\hat{\omega}_1 +   \hat{m}_1 v), \nonumber
\end{align}
where the global quantities: mass density $\rho$,   pressure $p$,   specific internal energy $\varepsilon$,    shear stress $\sigma$, dynamic pressure $\Pi$,  heat flux $q$, and a new extra variable $Q$ are given by  
\begin{align*} 
&\rho = \sumal \rho_\alpha, \qquad p = \sumal p_\alpha, \nonumber\\
&\rho \varepsilon 
=  \rho \varepsilon_{\rm int} +\frac{1}{2} \sumal\rho_\alpha u_\alpha^2, \quad 
\left( \varepsilon_{\rm int} = \frac{1}{\rho} \sumal \rho_\alpha \varepsilon_\alpha \right),  \nonumber \\
&\sigma = { - \frac{2}{3} \sumal\rho_\alpha u_\alpha^2} =  - \frac{2}{3} \frac{\rho_{1} \rho}{\rho_{2}} u_1^2, \nonumber \\
&\Pi  =\sumal\left(\Pi_\alpha + \frac{1}{3}\rho_\alpha u_\alpha^2\right) = \Pi_1+\Pi_2 -\frac{\sigma}{2},  \\
&q = \sumal\left(\frac{1}{2}\rho_\alpha u_\alpha^2 + \rho_\alpha \varepsilon_\alpha + p_\alpha + \Pi_\alpha\right) u_\alpha, \nonumber \\
&Q = \sumal\left\{\frac{1}{2}\rho_\alpha u_\alpha^2 + \frac{5}{2}\left(p_\alpha + \Pi_\alpha\right)\right\} u_\alpha. \nonumber
\end{align*}
Here $\varepsilon_{\rm int}$  represents the intrinsic part of the global-specific internal energy. 
We note that the global shear stress and global heat flux appear in the system \eqref{finale} due to the diffusion velocity, even if we neglect these dissipative quantities in the field equations for each constituent. 

In order to discuss the behavior of  non-equilibrium  average temperature $\mathcal{T}$, we use the definition   proposed in~\cite{RuggeriSimic2008proceedings, GouinRuggeri,RuggeriSimic2009}. This average  temperature  is introduced such that the intrinsic internal energy $\varepsilon_\mathrm{int}$ of the multi-temperature mixture have the same  expression of  a single-temperature:
\begin{align*}
		\rho \varepsilon_{\rm int} 
		 = \rho_1 \varepsilon_1 (\rho_1, \mathcal{T}) + \rho_2 \varepsilon_2 (\rho_2, \mathcal{T})
 		= \rho_1 \varepsilon_1 (\rho_1 , T_1) + \rho_2 \varepsilon_2 (\rho_2, T_2) . 
\end{align*}

\subsection{Equilibrium subsytem}

The production terms $\hat{m}_{\alpha}$, $\hat{e}_{\alpha}$, and $\hat{\omega}_{\alpha}$ vanish in equilibrium and from \eqref{eq:production_interaction} and \eqref{eq:main_field}, we obtain $v_1 = v_2 = v$, $T_1 = T_2 = T$, and $\Pi_1 = \Pi_2 = 0$, where $T$ is the average temperature in equilibrium.
In a generic equilibrium state, the total (equilibrium) pressure and the total internal intrinsic energy are the same forms as in the case of a single-component gas 
\begin{equation*} 
        p^{(\rm eq)} 
        = \frac{k_{B}}{m} \rho  T, \qquad
	\varepsilon^{(\rm eq)}_{\rm int} 
	= \frac{k_{B}}{m(\gamma-1)}  T 
\end{equation*}
provided that we introduce following \cite{SimicMT} an average mass $m \equiv m(c)$ and an average ratio of the specific heats $\gamma \equiv \gamma(c)$ 
\begin{equation}\label{equ:avgTemp2}
	\begin{split}
		\frac{1}{m} = \frac{c}{  m_{1}} + \frac{1-c}{  m_{2}},\quad
		\frac{1}{\gamma-1} = \frac{m }{m_{1}} \frac{c}{\gamma_1 - 1} +  \frac{m }{m_{2}} \frac{1 - c}{\gamma_2 - 1}, 
	\end{split}	
\end{equation}
where $c=\rho_1/\rho \in ]0,1[$  denoting the concentration. 

The associate equilibrium subsystem of \eqref{finale} according to the definition given in \cite{Boillat-1997} is given by
\begin{align}\label{subsystem:finale}
\begin{split}
&\frac{\partial \rho}{\partial t} + \frac{\partial \rho v}{\partial x} = 0,  \\
&\frac{\partial \rho v}{\partial t}+ \frac{\partial }{\partial x} \left(\rho v^2 + p^{(\rm eq)} \right) = 0,\\
&\frac{\partial }{\partial t} \left(\frac{1}{2}\rho v^2 + \rho \varepsilon^{(\rm eq)}_{\rm int} \right)
+\frac{\partial}{\partial x}\left\{\left(\frac{1}{2}\rho v^2 + \rho \varepsilon^{(\rm eq)}_{\rm int} +  p^{(\rm eq)} \right)v \right\} = 0,   \\
& \frac{\partial \rho_1 }{\partial t} + \frac{\partial \rho_1  v }{\partial x} = 0.  
\end{split}
\end{align}

\section{Shock Structure}
\label{sec:ShockStruct}

The shock structure is a traveling wave solution with constant velocity $s$ in which all components of the field depend on a single variable $\varphi = x - s t$ connecting two constant equilibrium state $\mathbf{u}_{0}$ and $\mathbf{u}_\mathrm{I}$ at $\varphi = \pm \infty$ respectively.

We can choose without loss of generality the unperturbed velocity $v_0=0$.
Taking into account the conservation laws, we have the following boundary conditions
\begin{equation*}
\mathbf{u}_{0} =
\left[%
\begin{array}{c}
\rho_{0} \\
v_{0}  \\
T_{0} \\
\Pi_{0}\\
\left(\rho_1\right)_{0} \\
\left(v_1\right)_{0} \\
\left(T_1\right)_{0} \\
\left(\Pi_1\right)_{0}\\
\end{array}%
\right], \quad
\mathbf{u}_\mathrm{I} =
\left[%
\begin{array}{c}
\rho_\mathrm{I} \\
v_\mathrm{I} \\
T_\mathrm{I} \\
\Pi_\mathrm{I}\\
\left(\rho_1\right)_\mathrm{I} \\
\left(v_1\right)_\mathrm{I} \\
\left(T_1\right)_\mathrm{I} \\
\left(\Pi_1\right)_\mathrm{I}\\
\end{array}%
\right]. 
\end{equation*}
In the unperturbed equilibrium state $\mathbf{u}_{0}$, we have:
\begin{equation}\label{eq:RH0}
\left(\rho_1\right)_0 = c_{0} \rho_0, \quad \left(T_1\right)_0 =  T_0, \quad 
v_0  = \left(v_1\right)_0 = 0, \quad
\Pi_0 = \left(\Pi_1\right)_0 = 0, 
\end{equation}
and the perturbed equilibrium state $\mathbf{u}_\mathrm{I}$ is the solution of the Rankine-Hugoniot equations of the equilibrium sub-system \eqref{subsystem:finale}:
\begin{align}\label{eq:RH1}
&\rho_\mathrm{I} = \frac{(\gamma_0 + 1) M_{0}^{2}}{2 + (\gamma_0 - 1) M_{0}^{2}}\rho_0, \quad
c_\mathrm{I}=c_0, \quad \left(\rho_1\right)_\mathrm{I} = c_{0} \rho_\mathrm{I} \nonumber \\
&T_\mathrm{I} ={T_1}_\mathrm{I} = \frac{\{2 + (\gamma_0 - 1) M_0^2\}\{1 + \gamma_0(2 M_0^2 - 1)\}}{(\gamma_0 + 1)^2 M_0^2} T_0,  
 \\
& v_\mathrm{I}  = {(v_1)}_\mathrm{I} =\frac{2 (M_{0}^{2} - 1)}{(\gamma_0 + 1) M_{0}} a_0, \quad \Pi_\mathrm{I} = \left(\Pi_1\right)_\mathrm{I} = 0,  
\nonumber 
\end{align}
where $\gamma_0$ is the function $\gamma$ given in \eqref{equ:avgTemp2} evaluated in $c_0$:  $\gamma_0 = \gamma(c_0)$. 
The unperturbed Mach number $M_0$ is defined by
\begin{equation*}
M_{0} =\frac{s-v_0}{a_{0}}, 
\end{equation*}
with $a_0$ being the sound velocity in the unperturbed state 
\begin{equation*}
a_{0} = \sqrt{ \gamma_0 \frac{k_{B}}{m_{0}} T_{0} },
\end{equation*}
and $m_{0} = m(c_{0})$ being the equilibrium average mass
of the mixture.

\section{Regions classified by the possibility of sub-shock formation}\label{sezione4}
Let us analyze whether the continuous shock-structure solution exists or not. 
Without any loss of generality, hereafter, we assume $m_{1} \leq m_{2}$. 
Following the paper \cite{ShockBinaryET6_RdM}, we will analyze the shock wave propagating in the positive direction with respect to the fluid flow  introducing  the characteristic velocities in an equilibrium state for constituents $1$ and $2$
\begin{equation}\label{l1l2}
    \lambda_1 = v + \sqrt{\frac{5}{3}  \frac{k_{\mathrm{B}}}{m_{1}} T}, \quad 
    \lambda_2 = v + \sqrt{\frac{5}{3}  \frac{k_{\mathrm{B}}}{m_{2}} T},
\end{equation}
and the characteristic velocities of the equilibrium subsystem
\begin{equation}\label{ll}
    \bar{\lambda} = v + \sqrt{\gamma  \frac{k_{\mathrm{B}}}{m} T}, 
\end{equation}
with $m$ and $\gamma$ given in \eqref{equ:avgTemp2}.

Let's introduce the dimensionless characteristic velocities at the unperturbed and perturbed states
\begin{equation*}
    M_{10} \equiv \frac{\lambda_{10}}{a_0}, \quad 
    M_{20} \equiv \frac{\lambda_{20}}{a_0}, \quad 
    M_\mathrm{1I} \equiv \frac{\lambda_\mathrm{1I}}{a_0}, \quad 
    M_\mathrm{2I} \equiv \frac{\lambda_\mathrm{2I}}{a_0}, \quad 
\end{equation*}
where, taking into account \eqref{l1l2}, \eqref{ll} and \eqref{eq:RH0}, \eqref{eq:RH1} the explicit expressions of these quantities are given by 
\begin{align}\label{MMM}
  \begin{split}
          &M_{10} = \sqrt{\frac{5 }{3  \gamma_0 \left\{c_{0} + (1 - c_{0}) \mu\right\}}}, \quad
    M_{20} =\sqrt{\frac{5 \mu}{3 \gamma_0 \left\{c_{0} + (1 - c_{0}) \mu \right\}}},   \\
    &M_\mathrm{1I} = \frac{2 (M_{0}^{2} - 1)}{(\gamma_0 + 1) M_{0}} + \sqrt{ \frac{5 \{ 2 + (\gamma_0-1) M_0^2 \}\{ 1 + \gamma_0 (2 M_0^2 - 1) \}}{3 \gamma_0 (\gamma_0+1)^2 M_0^2}\frac{m_0}{m_{1}}}, \\
    &M_\mathrm{2I} = \frac{2 (M_{0}^{2} - 1)}{(\gamma_0 + 1) M_{0}} + \sqrt{ \frac{5 \{ 2 + (\gamma_0-1) M_0^2 \}\{ 1 + \gamma_0 (2 M_0^2 - 1) \}}{3 \gamma_0 (\gamma_0+1)^2 M_0^2}\frac{m_0}{m_{2}}},
  \end{split} 
\end{align}
with $\mu$ being the ratio of the masses of the constituents defined by
\begin{equation*}
\mu = \frac{m_{1}}{m_{2}}, \quad 0 < \mu \leq 1. 
\end{equation*}

The analogous expressions for a mixture of Eulerian gases are given in  \cite{ShockBinaryET6_RdM} and indicated with an apex ${(E)}$ in the Figures \ref{fig:d1d2_5and7_subshock_regions_mu01_ET6andEulerian}-\ref{fig:subshock_regions_mu09_ET6andEulerian}.

In the previous paper~\cite{ShockBinaryET6_RdM}, we have identified sufficient and necessary conditions for the sub-shock formation: 

According to the theorem given in~\cite{Breakdown},
\emph{the necessary and sufficient condition for the sub-shock formation for the shock structure of the constituent $1$ (light species) is given by }
\begin{equation*}
    s > \lambda_{10} \quad \Leftrightarrow \quad M_{0} > M_{10}. 
\end{equation*}

While, \emph{the necessary condition but in general not sufficient for the sub-shock formation for the shock structure of  constituent $2$ (heavy species) is given by }
 \begin{align}\label{state}
 		  \textbf{Case A) }  & \, \text{If} \quad  \mu \leq {3 \gamma_1}/{5},   \text{and } \nonumber \\
 		&   \text{when}\,\,\, 0 < c_0 \leq  c_0^* , \,\, \text{for } \, \, s > \lambda_{20} \quad \Leftrightarrow \quad \text{for } \, \, M_{0} > M_{20} , \nonumber  \\
 		&   \text{when}\,\,\,  c_0^* < c_0 < 1, \,\, \text{for } \, \, s > s^*, \,\, \text{solution of }   \,\, \lambda_{\rm 2I}(s^*)=s^*, \nonumber\\
   &   \qquad \Leftrightarrow \quad \text{for } \, \,M_{0} > M^*_{20},  \,\, \text{solution of }   \,\, M_{\rm 2I}(M^*_{20}) = M^*_{20},  
   \\ &  \text{where}  \,\, {c}_0^*(\mu) \,\,  \text{is the solution of} \,\,  \lambda_{20} = \bar{\lambda}_{0} \,\, \Leftrightarrow M_{20}(c_0^*,\mu) = 1. \nonumber\\
 		 \textbf{Case B)}  &  \, \text{If} \quad  \mu >{3\gamma_1}/{5}, \,\, \text{for } \, \, s > \lambda_{20} \, \Leftrightarrow \, \text{for } \, \, M_{0} > M_{20}, \, \text{for any} \,\, c_0 \in \, ]0,1[ \, . \nonumber
\end{align}

\subsection{Regions in Case A)}
\begin{figure}[h]
	\centering
	\includegraphics[width=0.75\linewidth]{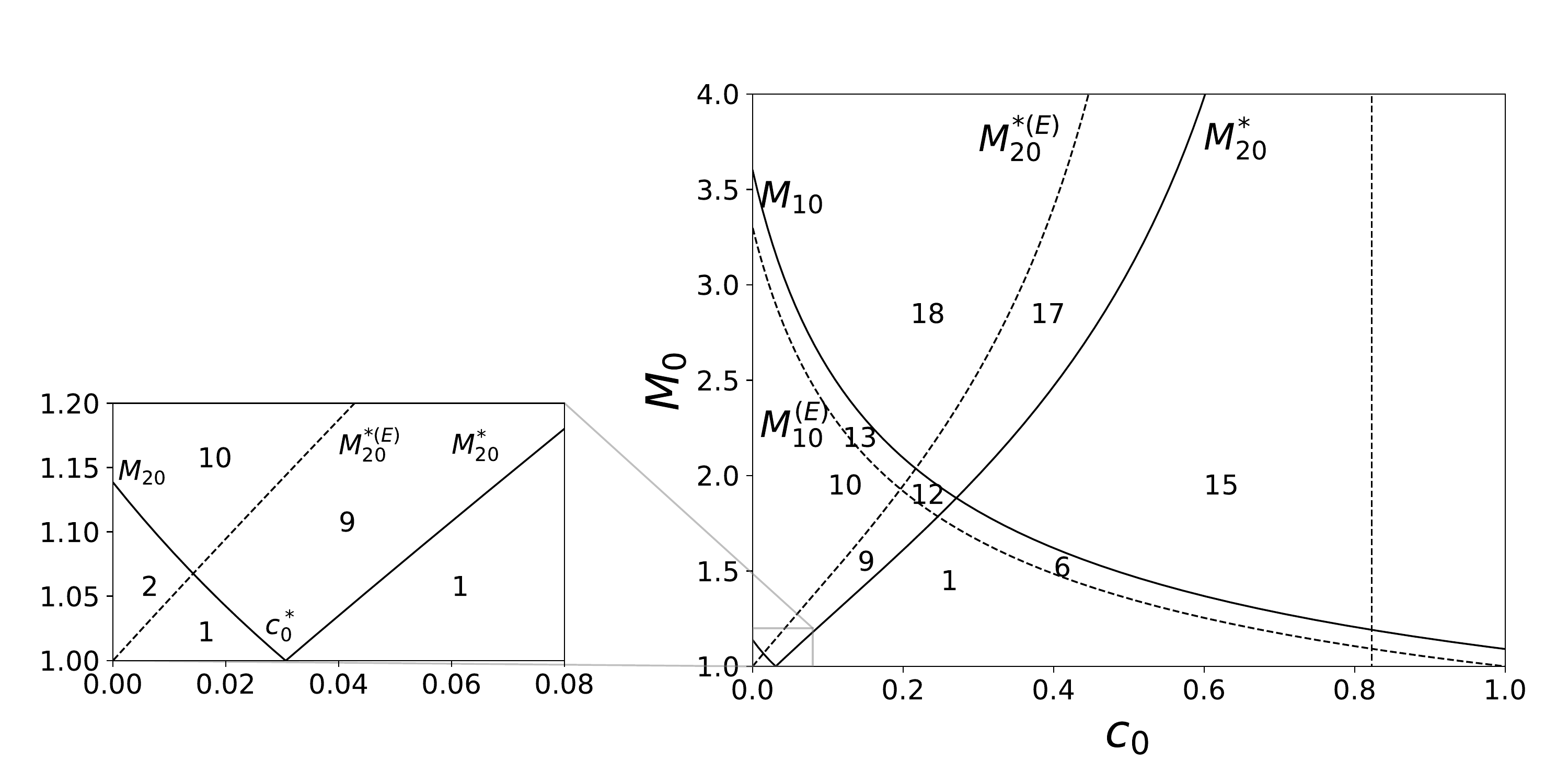}
	\caption{Regions in the plane $(c_0,M_0)$ of possible sub-shocks in Case A$_1$ (right);  
 $\gamma_1=7/5$, $\gamma_2=9/7$, and $\mu = 0.1$.  
 The vertical dotted line represents the asymptote of $M^*_{20}$. The magnification of the regions around $(c_0, M_0) = (0, 1)$ is also shown (left).
 }
	\label{fig:d1d2_5and7_subshock_regions_mu01_ET6andEulerian}
\end{figure}

\begin{figure}[h]
	\centering
	\includegraphics[width=0.49\linewidth]{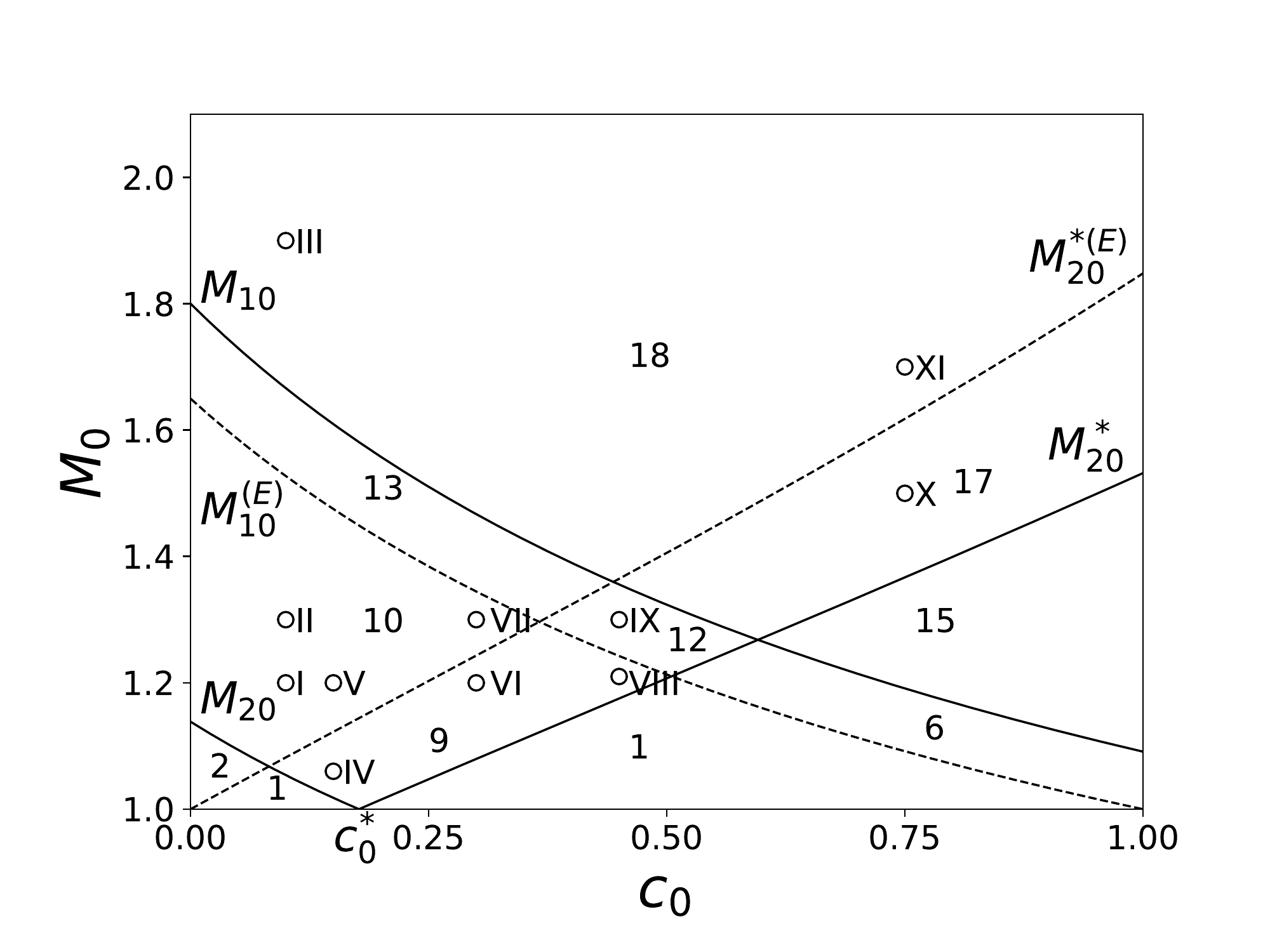}
	\caption{Regions in the plane $(c_0,M_0$) of possible sub-shocks in Case A$_2$;  
 $\gamma_1=7/5$, $\gamma_2=9/7$, and $\mu = 0.4$. 
	The circles represent the parameters adopted for numerical calculations shown in Sec. \ref{sec:Numerical}. }
	\label{fig:subshock_regions_mu04_ET6andEulerian2}
\end{figure}

\begin{figure}[h]
	\begin{center}
		\includegraphics[width=0.5\linewidth]{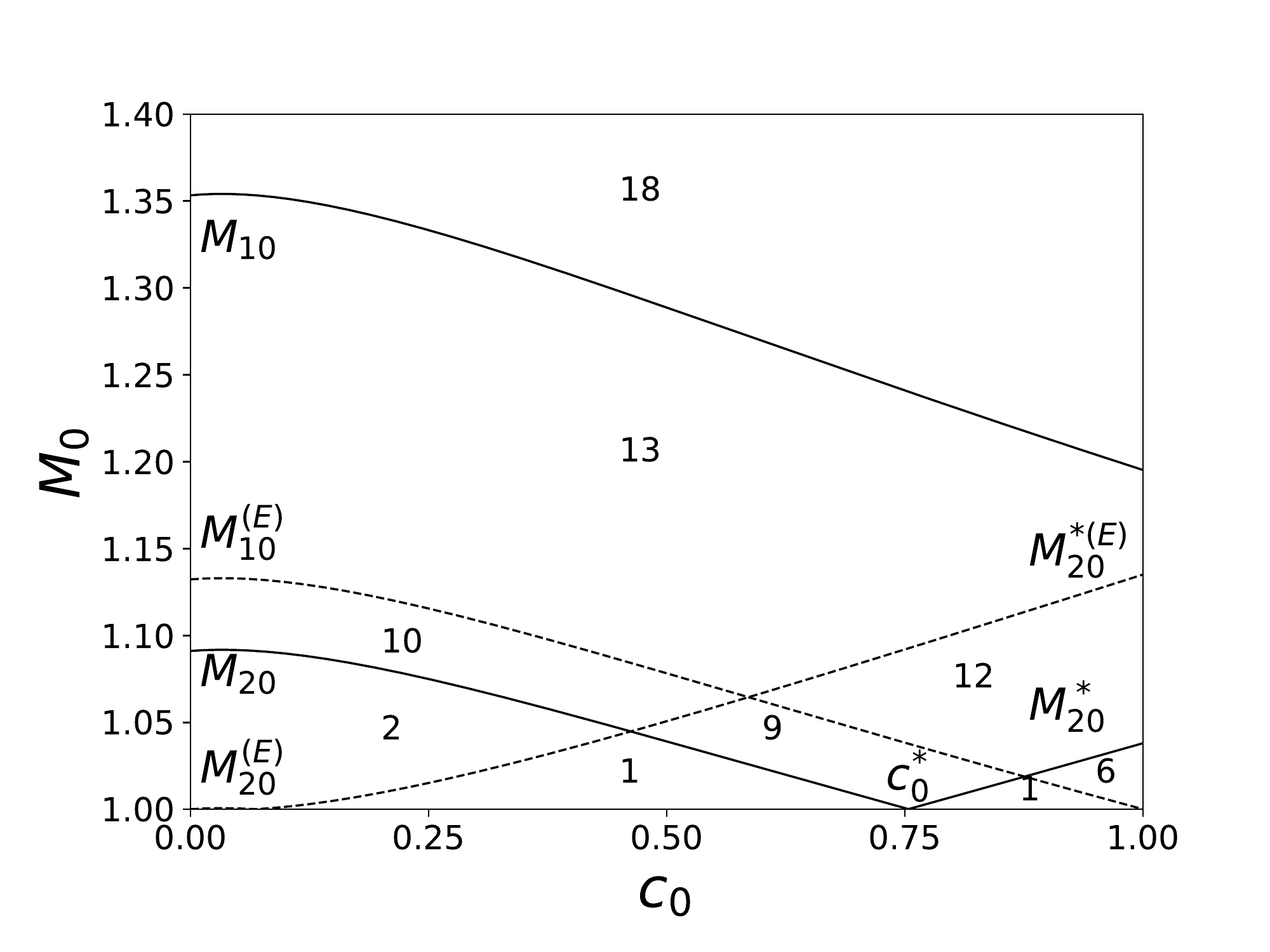} %
	\end{center}
	\caption{Regions classified by the possibility of the sub-shock formation in the $M_0$ -- $c_0$ plane in Case A$_3$; 
 $\gamma_1 = 7/6$, $\gamma_2 = 7/5$, and $\mu = 0.65$.}
	\label{fig:subshock_regions_mu065_ET6andEulerian}
\end{figure}

\begin{figure}[h]
	\begin{center}
		\includegraphics[width=0.5\linewidth]{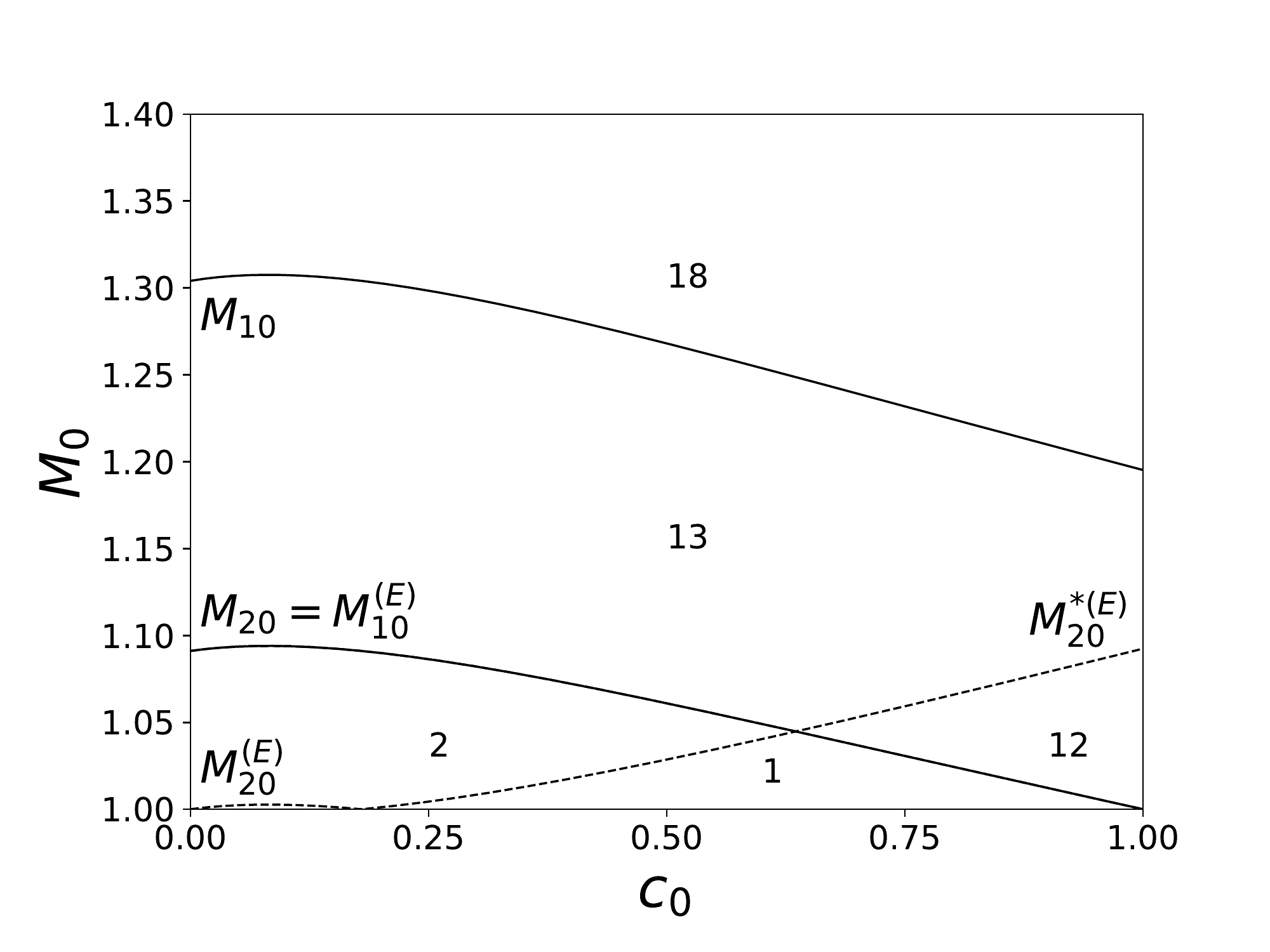} %
	\end{center}
	\caption{The regions classified by the possibility of the sub-shock formation in the $M_0$ -- $c_0$ plane in Case A$_4$; $\gamma_1 = 7/6$, $\gamma_2 = 7/5$, and $\mu = 0.7$. In this case, $c_0^* = 1$. } 
	\label{fig:subshock_regions_mu07_ET6andEulerian}
\end{figure}

Case A) has four possible types of regions depending on $\mu$:
\begin{itemize}
    \item A$_1$) in which $M^*_{20}$ intersects $M_{10}$ and has an asymptote like in the example  reported in Fig.\ref{fig:d1d2_5and7_subshock_regions_mu01_ET6andEulerian}. Taking into account   \eqref{MMM} this situation is possible if 
    \begin{equation*}
        0 < \mu \leq \frac{3}{10}(\gamma_1 - 1).
    \end{equation*}
      \item A$_2$) in which $M^*_{20}$ intersects $M_{10}$ and doesn't have an asymptote like in the example  reported in Fig.\ref{fig:subshock_regions_mu04_ET6andEulerian2}. Taking into account   \eqref{MMM} this situation is possible if 
    \begin{equation*}
       \frac{3}{10}(\gamma_1 - 1) < \mu < \frac{15-33\gamma_1}{15\gamma_1-65}.
    \end{equation*}
    \item A$_3$) in which $M^*_{20}$ does not intersect $M_{10}$  like in the example  reported in Fig.\ref{fig:subshock_regions_mu065_ET6andEulerian}. Taking into account   \eqref{MMM} this situation is possible if 
    \begin{equation*}
      \frac{15-33\gamma_1}{15\gamma_1-65} \leq \mu < \frac{3}{5}\gamma_1.
    \end{equation*}
     \item A$_4$) in which 
      $M^*_{20}$ does not intersect $M_{10}$ and the curves of $M_{20}$ and $M_{10}^{(E)}$ coincide with each other and $c_0^* = 1$
       like in the example  reported in Fig.\ref{fig:subshock_regions_mu07_ET6andEulerian}. Taking into account   \eqref{MMM} this situation is possible if 
    \begin{equation*}
       \mu = \frac{3}{5}\gamma_1.
    \end{equation*}
\end{itemize}

\subsection{Regions in Case B)}
\begin{figure}[h]
	\begin{center}
		\includegraphics[width=0.5\linewidth]{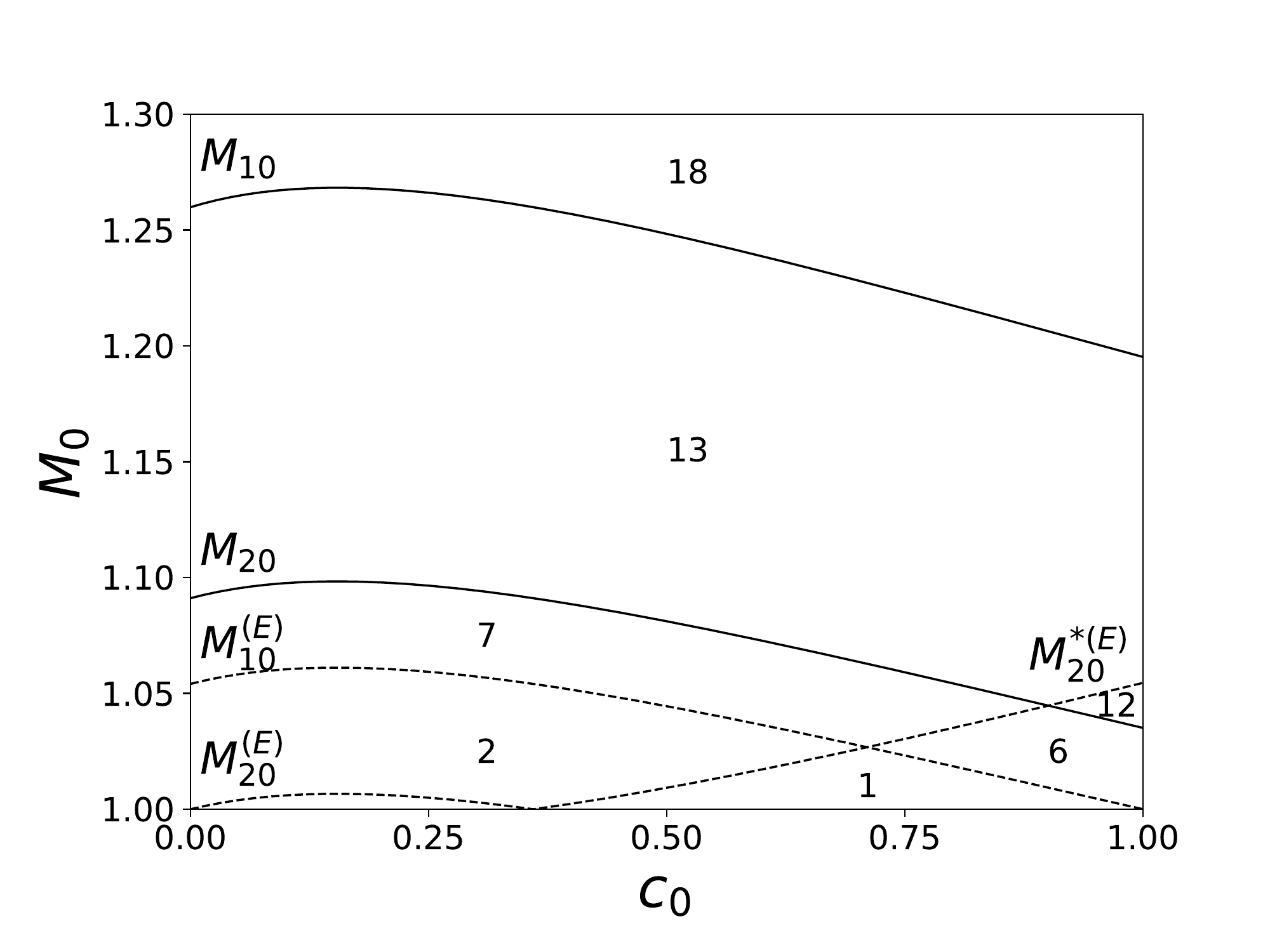} %
	\end{center}
	\caption{The regions classified by the possibility of the sub-shock formation in the $M_0$ -- $c_0$ plane in Case B$_1$;  $\gamma_1 = 7/6$, $\gamma_2 = 7/5$, and $\mu = 0.75$.}
	\label{fig:subshock_regions_mu075_ET6andEulerian}
\end{figure}

\begin{figure}[h]
	\begin{center}
		\includegraphics[width=0.5\linewidth]{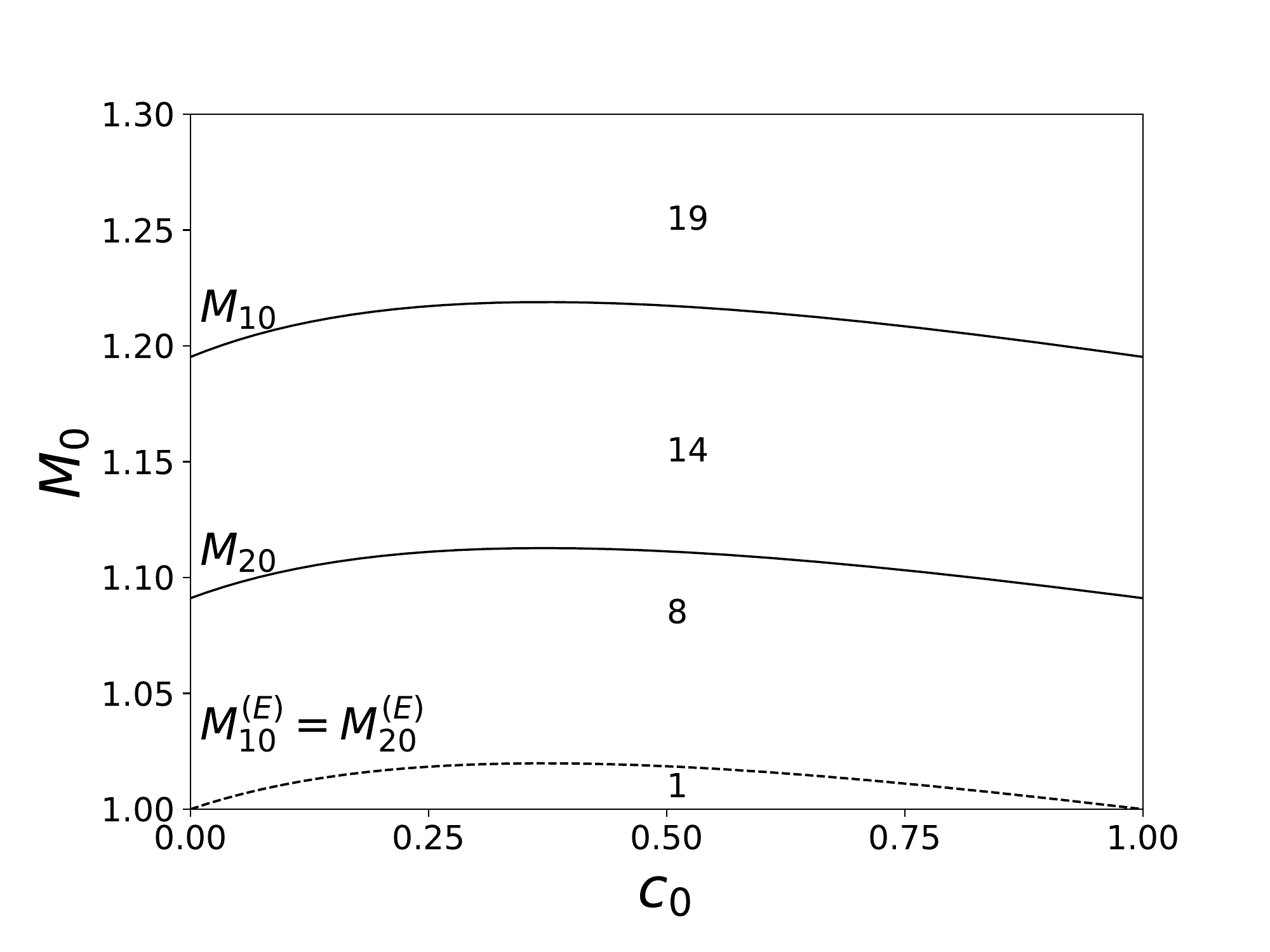} %
	\end{center}
	\caption{Regions classified by the possibility of the sub-shock formation in the $M_0$ -- $c_0$ plane in Case B$_2$;  $\gamma_1 = 7/6$, $\gamma_2 = 7/5$, and $\mu = 5/6$.}
	\label{fig:subshock_regions_mu0833_ET6andEulerian}
\end{figure}

\begin{figure}[htbp]
	\begin{center}
		\includegraphics[width=0.5\linewidth]{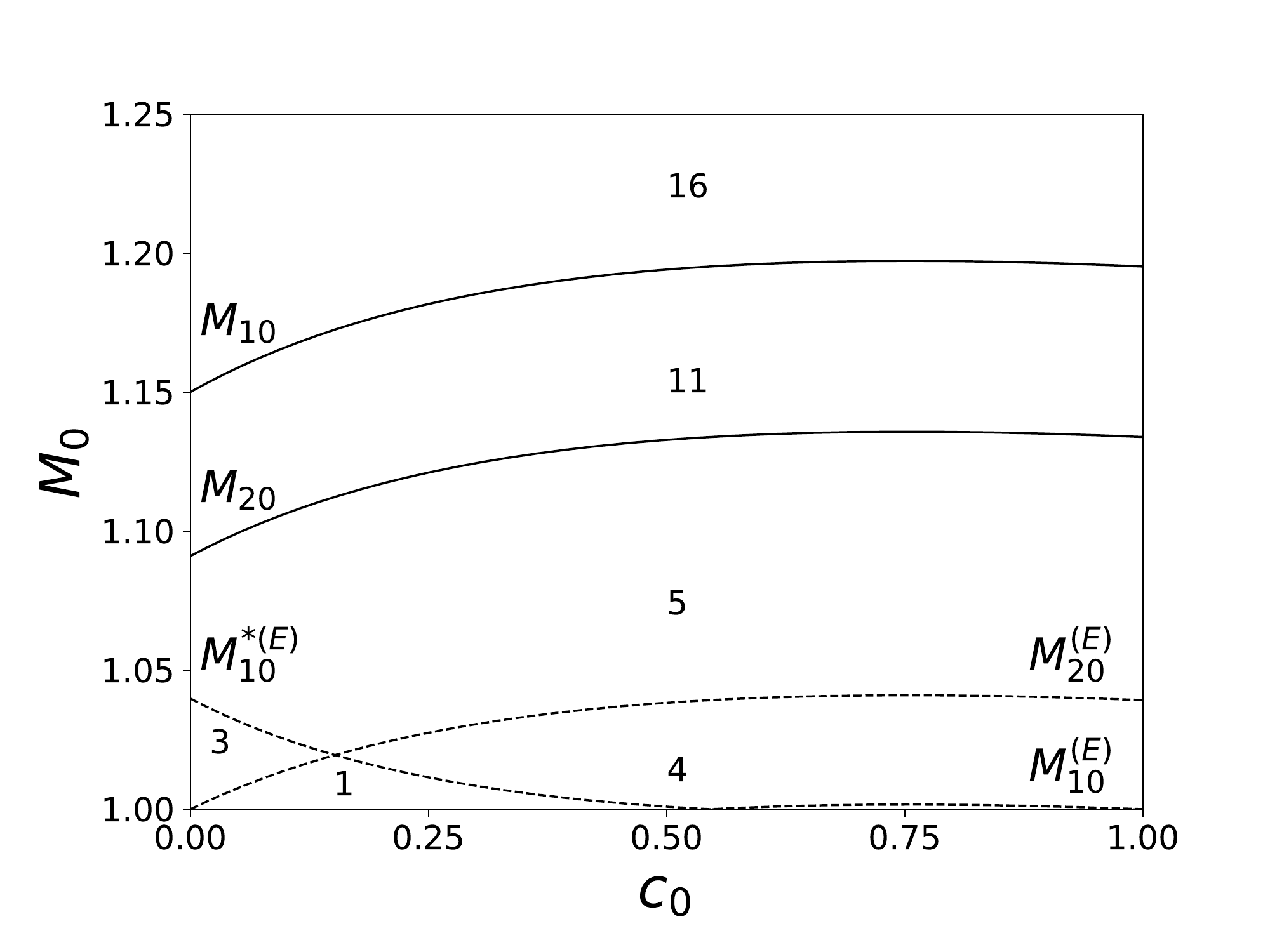} %
	\end{center}
	\caption{The regions classified by the possibility of the sub-shock formation in the $M_0$ -- $c_0$ plane in Case B$_3$;  $\gamma_1 = 7/6$, $\gamma_2 = 7/5$, and $\mu = 0.9$.}
	\label{fig:subshock_regions_mu09_ET6andEulerian}
\end{figure}

Case B) has three possible topologies  of regions depending on $\mu$:
\begin{itemize}
	\item B$_1$) in which $M_{20}$ is always greater than one in the whole range of $0 < c_0 < 1$, and therefore $M_{20}^*$ does not appear  like in the example  reported in Fig.\ref{fig:subshock_regions_mu075_ET6andEulerian}. Taking into account   \eqref{MMM}, this situation is possible if 
	\begin{equation*}
	\frac{3}{5}\gamma_1< \mu < \frac{\gamma_1 }{\gamma_2}.
	\end{equation*}
	\item B$_2$) as in case B$_1$) with $M^{(E)}_{10}=M^{(E)}_{20}$ like in the exapmle reported in Fig. \ref{fig:subshock_regions_mu0833_ET6andEulerian}. 
 Taking into account   \eqref{MMM} and the results in~\cite{ShockBinaryEulerian_PhysFluids}, this situation is possible if 
	\begin{equation*}
	\mu = \frac{\gamma_1 }{\gamma_2}.
	\end{equation*}
	\item B$_3$) as in case B$_1$) but $M^{(E)}_{10}<M^{(E)}_{20}$ like in the exapmle reported in Fig. \ref{fig:subshock_regions_mu09_ET6andEulerian}. 
 Taking into account  
 \eqref{MMM} and the results in~\cite{ShockBinaryEulerian_PhysFluids},
	this situation is possible if 
	\begin{equation*}
	\mu > \frac{\gamma_1 }{\gamma_2}.
	\end{equation*}
\end{itemize}

\begin{table}[htbp]
	\caption{Summary of the possibility of the sub-shock formation for the classified regions in the plane $(c_0, M_0$). 
		The symbols $S_1$ and $S_2$ represent the sub-shock for the shock structure of the constituent $1$ and  $2$, respectively. 
		The corresponding conclusions for a binary mixture of Eulerian gases are also shown in the columns $S_1^{(E)}$ and $S_2^{(E)}$. }
	\label{tbl:classification}
	\centering
	\begin{tabular}{c | c c | c c}
		Region & $S_1$  & $S_2$ & $S_1^{(E)}$  & $S_2^{(E)}$ \\\hline
		1       & No  & No & No  & No \\  
		2      & No  & No  & No  & Maybe \\ 
		3     & No & No & No & Yes \\ 
		4     & No & No & Maybe & No \\ 
		5     & No & No & Maybe & Yes \\ 
		6      & No & No & Yes  & No \\  
		7     & No & No & Yes & Maybe \\ 
		8     & No & No & Yes & Yes \\ 
		9      & No  & Maybe & No  & No  \\ 
		10      & No  & Maybe  & No  & Maybe  \\ 
		11     & No & Maybe & Maybe & Yes \\ 
		12      & No & Maybe & Yes  & No \\ 
		13      & No & Maybe  & Yes  & Maybe  \\ 
		14     & No & Maybe & Yes & Yes \\ 
		15     & Yes & No & Yes & No \\  
		16     & Yes & Maybe & Maybe & Yes \\ 
		17     & Yes & Maybe & Yes & No  \\ 
		18      & Yes & Maybe  & Yes & Maybe  \\ 
		19     & Yes & Maybe & Yes & Yes \\ 
	\end{tabular}
\end{table}

In Figures \ref{fig:d1d2_5and7_subshock_regions_mu01_ET6andEulerian}-\ref{fig:subshock_regions_mu09_ET6andEulerian}, we enumerate the different regions, and the conclusions whether we can have or no sub-shock according to \eqref{state} are summarized in Table \ref{tbl:classification}. The word ``Maybe" indicates that the condition is necessary but not sufficient for the existence of a sub-shock. In fact, the singularity is maybe apparent due to an indeterminate form $0/0$. For more details, see \cite{ShockBinaryEulerian_PhysFluids,ShockBinaryET6_RdM}.

\section{Numerical simulation of shock structure}
\label{sec:Numerical}
In the present section, we solve the system of field equations numerically and clarify the possibility of the sub-shock formation. 

For convenience, we introduce the following dimensionless variables to scale the state variables and the independent variable $\varphi$ with the values evaluated in the unperturbed state:
\begin{align*}
&\hat{\rho} = \frac{\rho}{\rho_{0}}, \quad
\hat{v} = \frac{v}{a_{0}}, \quad
\hat{\mathcal{T}} = \frac{\mathcal{T}}{T_{0}}, \quad
\hat{\Pi} = \frac{\Pi}{\rho_0\frac{k_B}{m_0}T_{0}}, \nonumber \\
&\hat{\varphi} = \frac{\varphi}{t_c \, a_0}, \quad \hat{x} = \frac{x}{t_c \, a_0}, \quad \hat{t} = \frac{t}{t_c}, \quad \hat{\tau} = \frac{\tau}{t_c}, \\
&\hat{\psi} = \frac{t_c}{\rho_0 T_0}\psi_{11}, \quad
\hat{\theta} = \frac{t_c}{\rho_0 \frac{k_B}{m_0} T_0^2}\theta_{11}, \quad
\hat{\kappa} = \frac{t_c}{\rho_0 \frac{k_B}{m_0} T_0^2}\kappa_{11}, \quad
\hat{\phi} = \frac{t_c}{\rho_0 \frac{k_B}{m_0} T_0^2}\phi_{11}, \quad \nonumber
\end{align*}
where $t_c$ is an arbitrary characteristic time for numerical computations. 

The stability conditions, namely, $\psi_{11} >0$ and that the matrix of \eqref{eq:stability_matrix} is positive definite, can be written in terms of the dimensionless phenomenological coefficients as follows: $\hat{\psi} > 0, \, \hat{\theta}  > 0,\, \hat{\theta} \, \hat{\phi} - \hat{\kappa}^2 > 0$.

As done previously in the literature~\cite{Brini_Osaka, Brini_Wascom,MentrelliRuggeri,IJNLM2017,subshock2,ShockBinaryEulerian_PhysFluids}, we perform numerical calculations on the Riemann problem consisting of two equilibrium states $\mathbf{u}_0$ and $\mathbf{u}_{\rm I}$ satisfying \eqref{eq:RH0} and \eqref{eq:RH1} and obtain the profiles of the physical quantities after a long time. 
Following an idea of Liu~\cite{Liu_conjecture}, a conjecture about the large-time behavior of the Riemann problem and the Riemann problem with structure \cite{Liu_struct1,Liu_struct2} for a system of balance laws was proposed by Ruggeri and Coworkers~\cite{Brini_Osaka,BriniRuggeri,MentrelliRuggeri}.
According to this conjecture, if the Riemann initial data correspond to a shock family $\mathcal{S}$ of the equilibrium sub-system, for a large time, the solution of the Riemann problem of the full system converges to the corresponding shock structure.

We have developed numerical code adopting the Uniformly accurate Central Scheme of order 2 (UCS2)~\cite{UCS2} to solve the Riemann problem numerically. 
As we do not know to refer in this paper to a particular real mixture of gases also because it seems difficult to find experimental data for shock profile in a mixture of polyatomic gases with large bulk viscosity, we only analyze the qualitative behavior of shock and sub-shocks, and we chose for our numerical experiments the following parameters: 
$\hat{\psi} = \hat{\theta} = \hat{\kappa} = 0.2$, $\hat{\phi} = 0.3$, $\hat{\tau}_1 = 1$, and $\hat{\tau}_2 = 2$.
 
\begin{figure}[h]
	\begin{center}
		\includegraphics[width=0.3\linewidth]{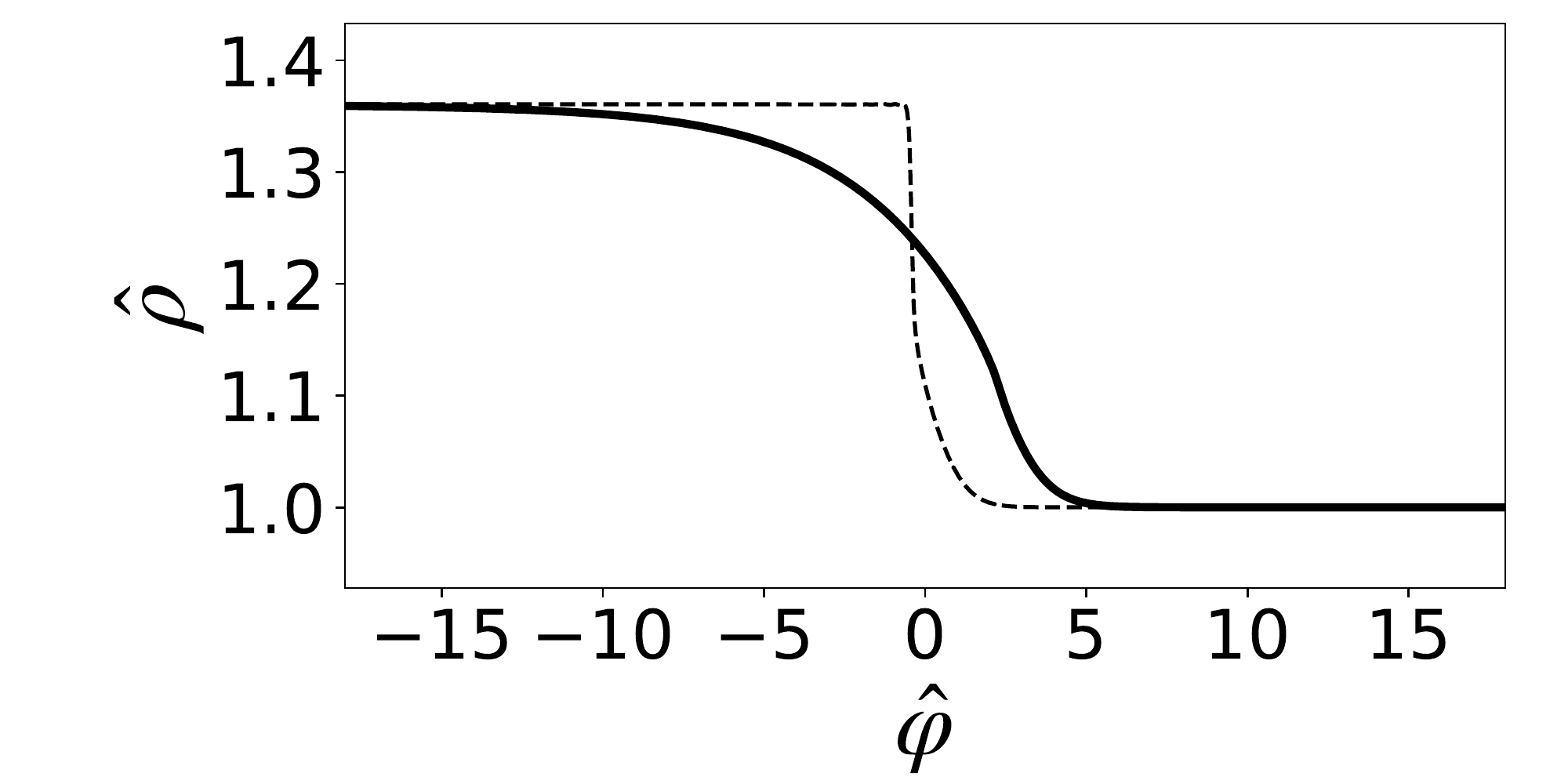}\,%
		\includegraphics[width=0.3\linewidth]{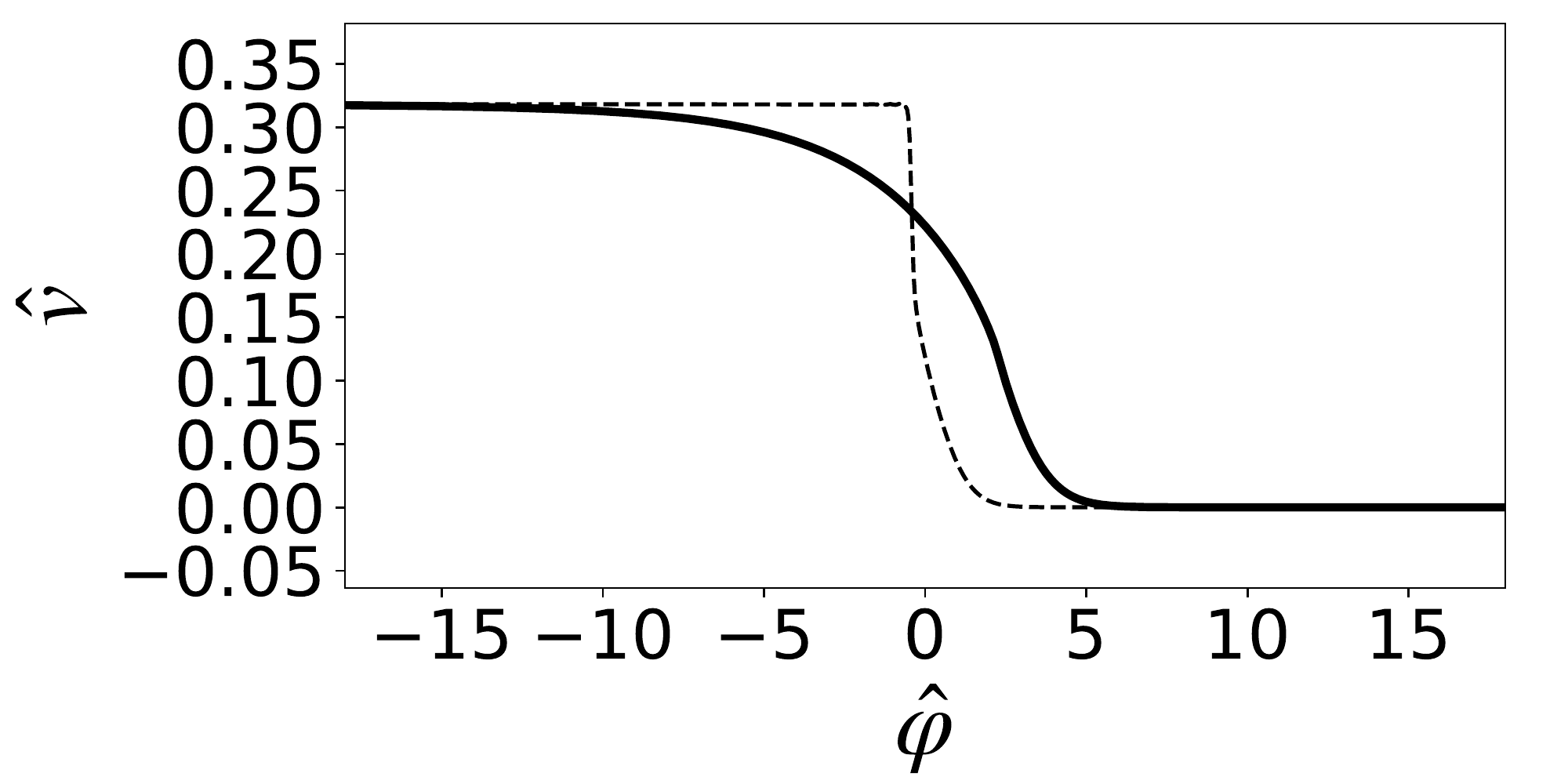}\,%
		\includegraphics[width=0.3\linewidth]{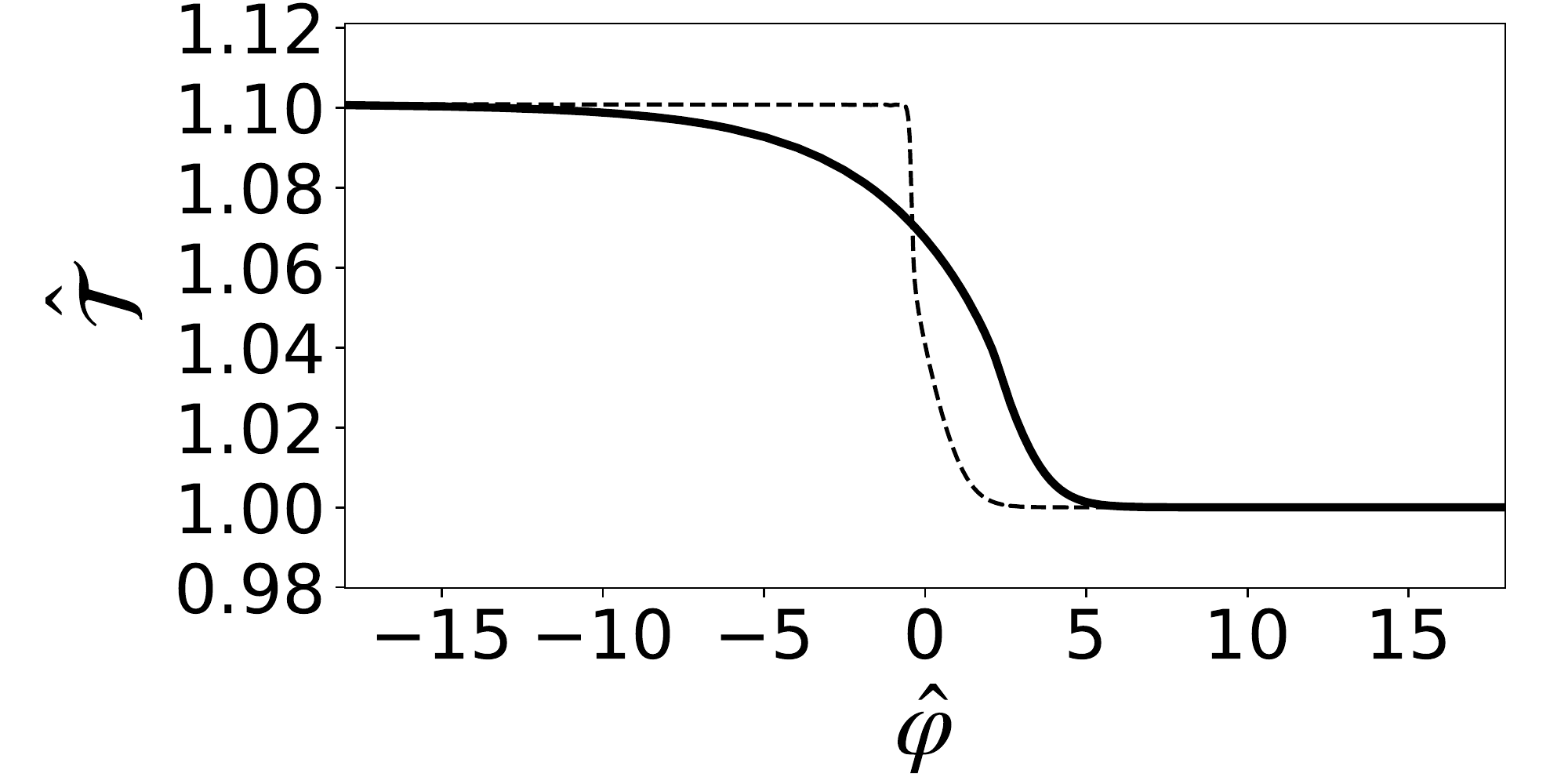}\\
        \includegraphics[width=0.3\linewidth]{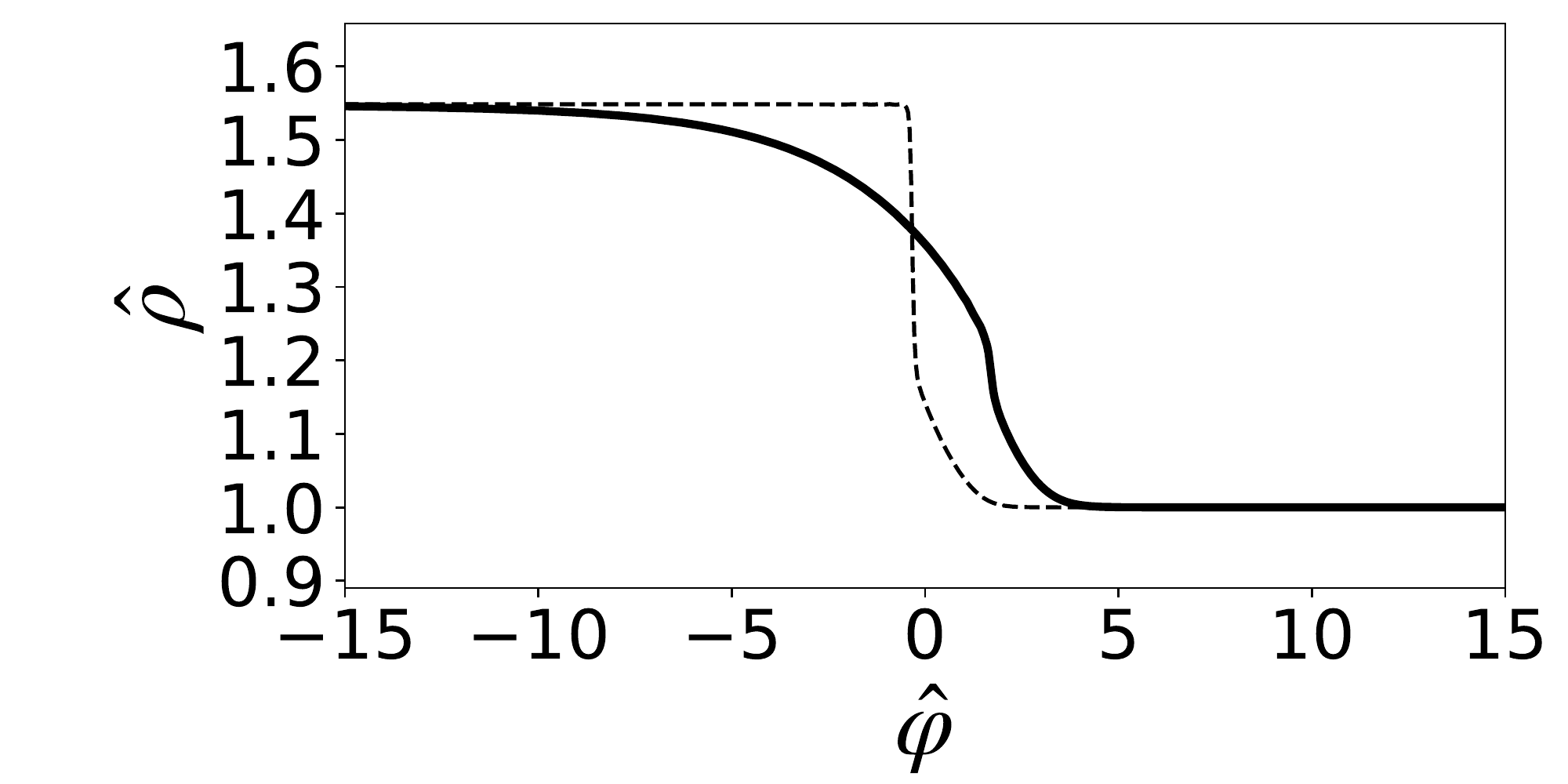}\,%
 		\includegraphics[width=0.3\linewidth]{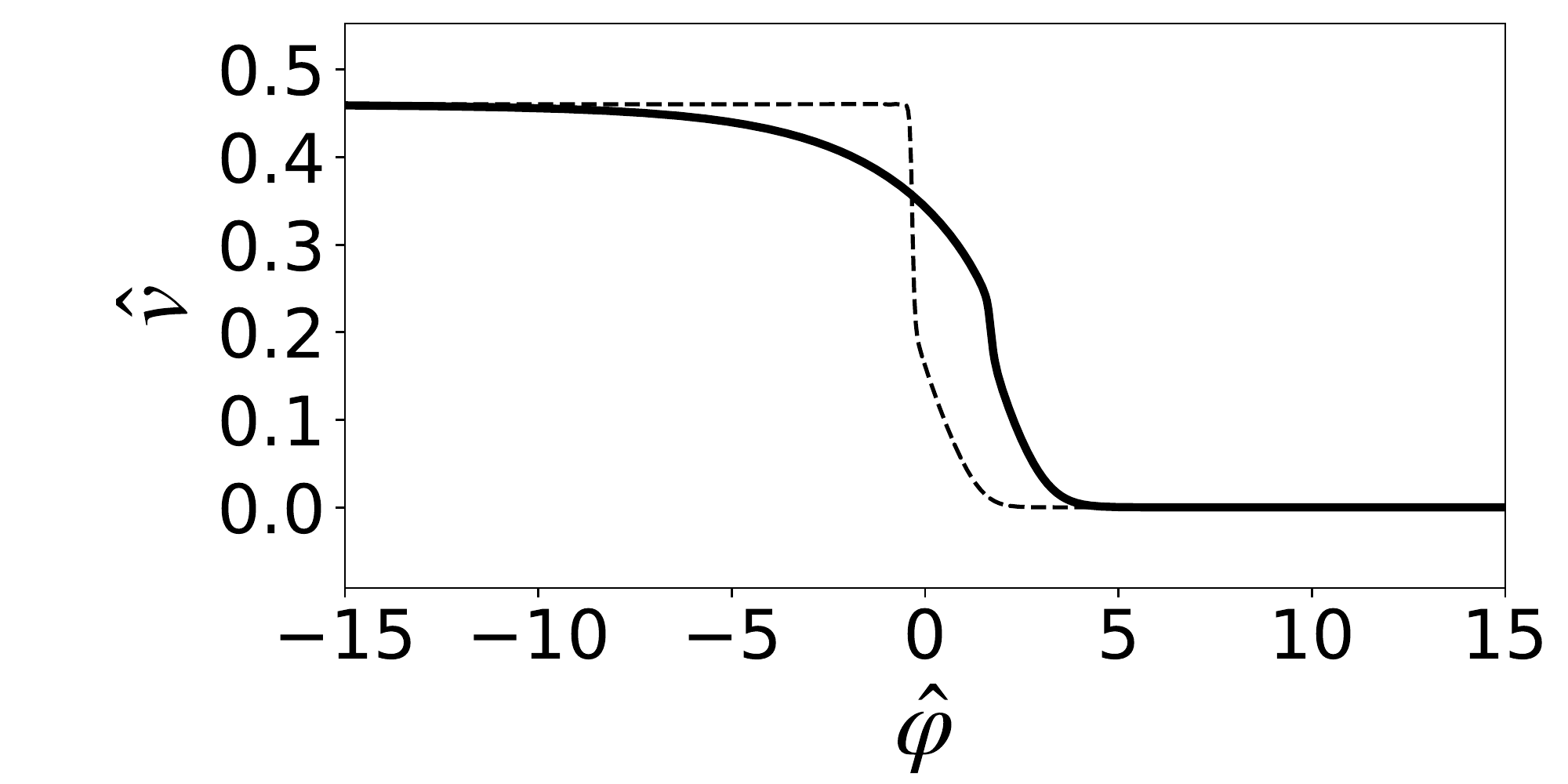}\,%
 		\includegraphics[width=0.3\linewidth]{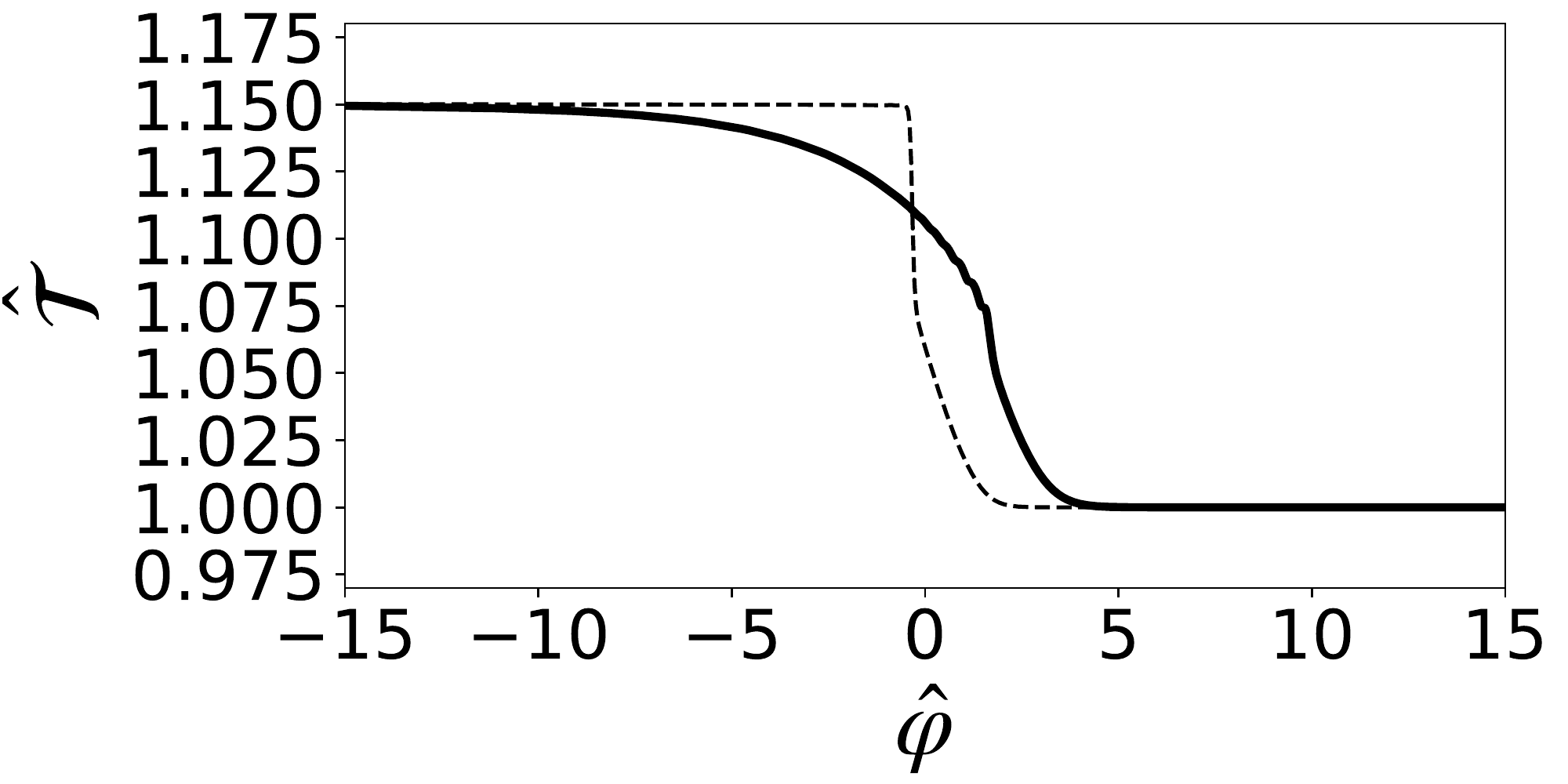}\\%
   		\includegraphics[width=0.3\linewidth]{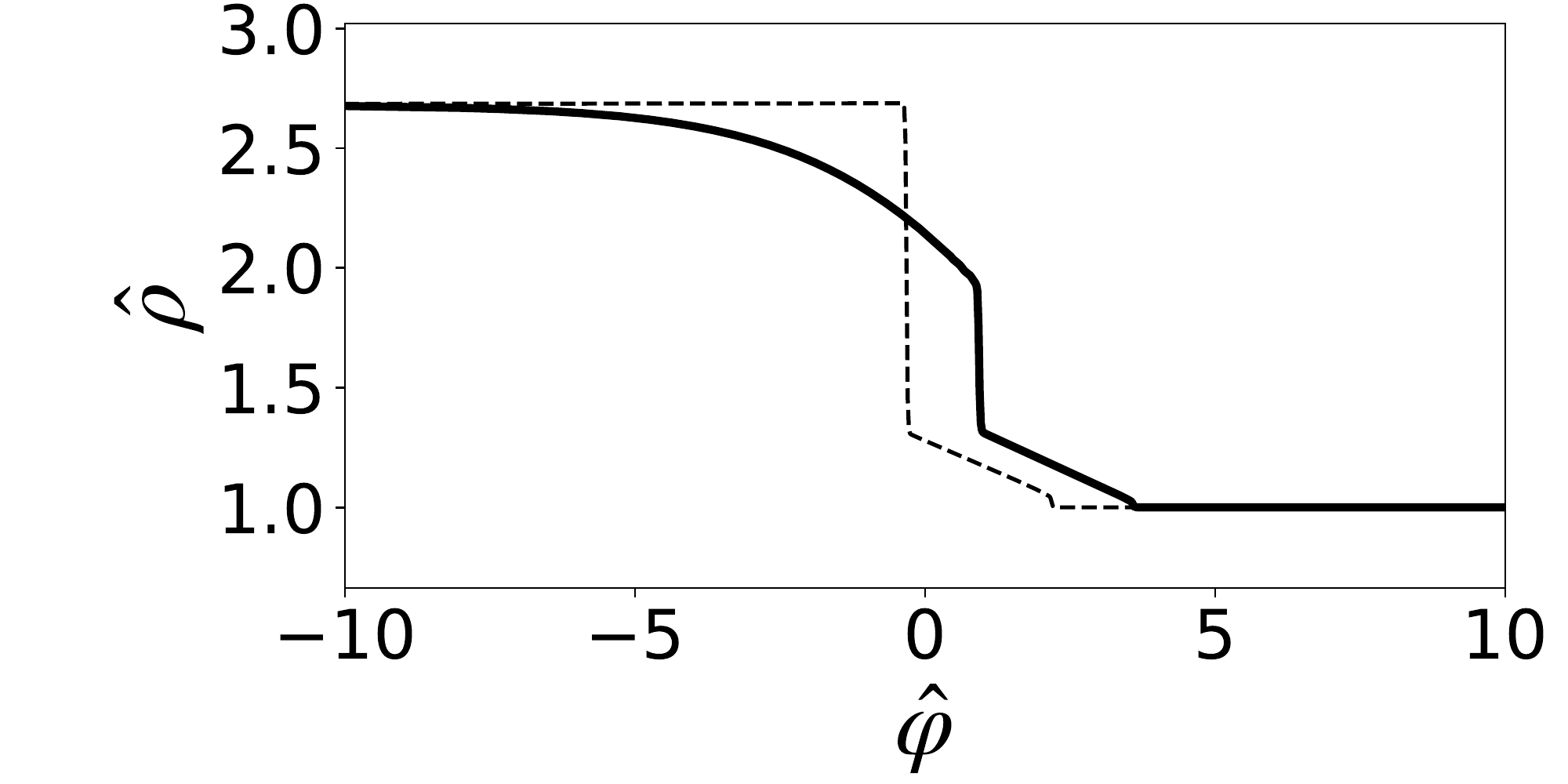}\,%
 		\includegraphics[width=0.3\linewidth]{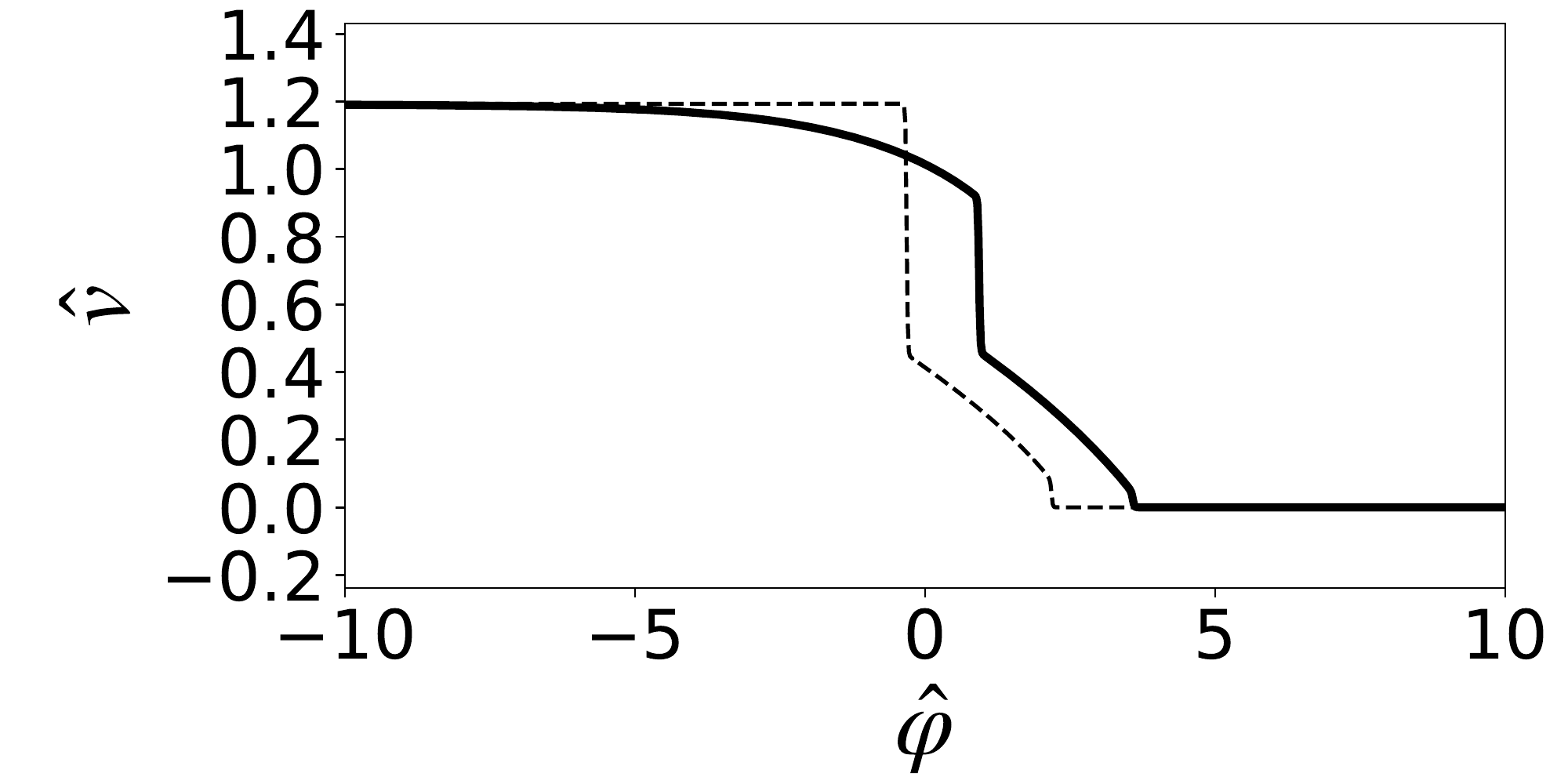}\,%
 		\includegraphics[width=0.3\linewidth]{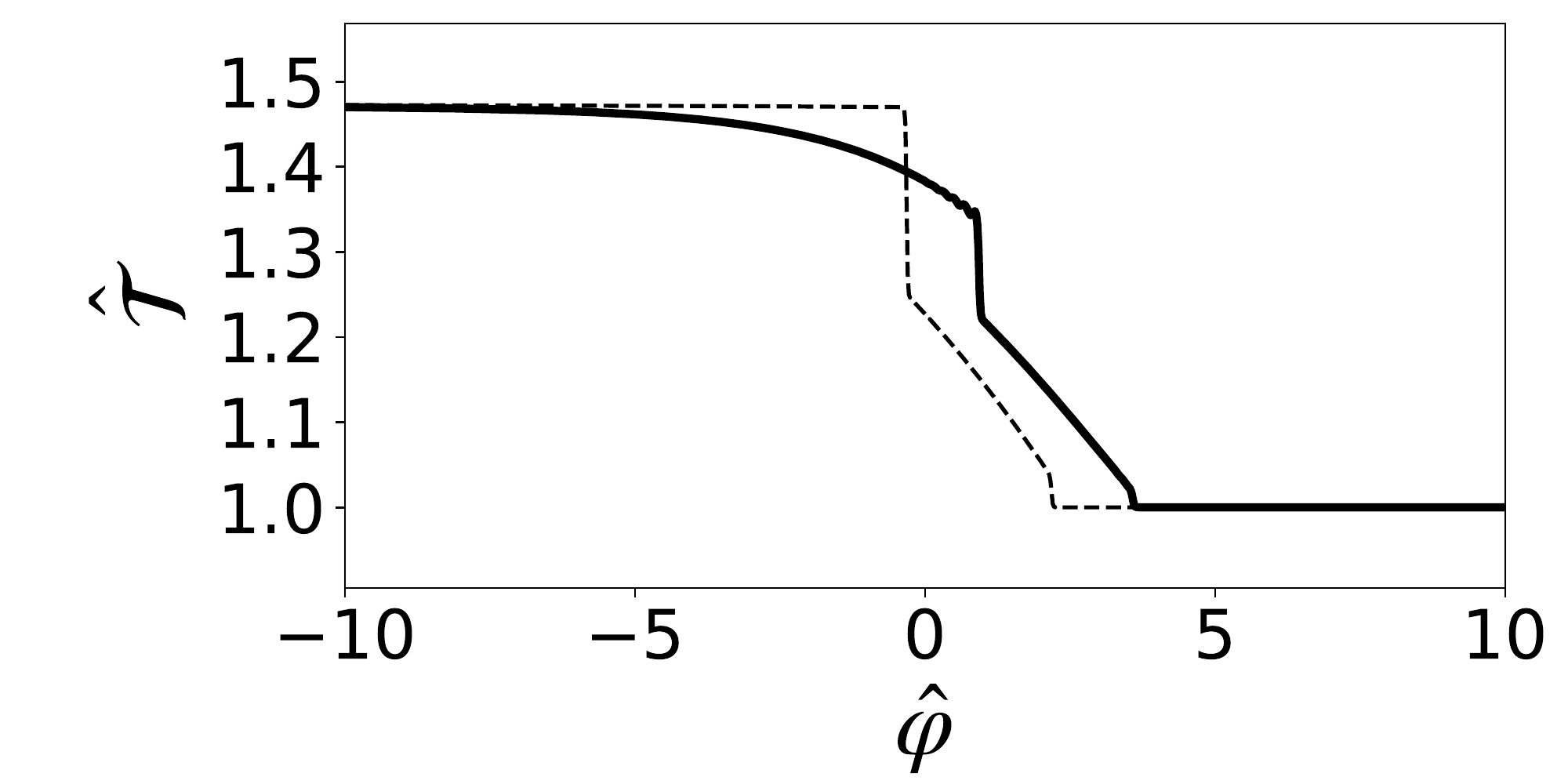}%
	\end{center}
	\caption{Shock structure of the dimensionless global mass density, mixture velocity, and average temperature (Solid curves) with  $\gamma_1=7/5$, $\gamma_2=9/7$, $\mu = 0.4$, $c_0 = 0.1$ for several Mach numbers; $M_0 = 1.2$ indicated by circle No. I in Region $10$ in Figure \ref{fig:subshock_regions_mu04_ET6andEulerian2} (top row), $M_0 = 1.3$ indicated by circle No. II in Region $10$ (middle row), and $M_0 = 1.9$ indicated by circle No. III in Region $18$ (bottom row). 
  The corresponding predictions by the theory of a mixture of Eulerian gases are also shown (dashed curves). }
	\label{fig:c01_M0-1_2}
\end{figure}

We show the shock structure for several Mach numbers in the case of $\gamma_1 = 7/5$, $\gamma_2 = 9/7$, $\mu = 0.4$, and $c_0 = 0.1$ in Figure \ref{fig:c01_M0-1_2}. 
These parameters correspond to circles Nos. I -- III in Figure \ref{fig:subshock_regions_mu04_ET6andEulerian2}. 
It is confirmed that the shock-structure solution predicted by the RET$_6$ theory is continuous for $M_0=1.2$ while a sub-shock for the shock structure of the constituent $2$ emerges for $M_0=1.3$. 
If we increase the Mach number more, we observe another sub-shock for the shock structure of the constituent $1$ as shown in Figure \ref{fig:c01_M0-1_2}    for $M_0=1.9$. 

For comparison, we also depicted the theoretical predictions by the Eulerian theory with the same parameters. 
As expected, the shock structure predicted by the RET$_6$ theory is broader than the ones predicted by the Eulerian theory.
Moreover, although the necessary condition for the sub-shock formation for constituent $2$ is satisfied in both theories for $M_0=1.2$, the sub-shock is observed only in the Eulerian theory. 
These results indicate the dissipative effect of the dynamic pressure on the shock structure and the sub-shock formation. 

\begin{figure}[htbp]
	\begin{center}
  		\includegraphics[width=0.3\linewidth]{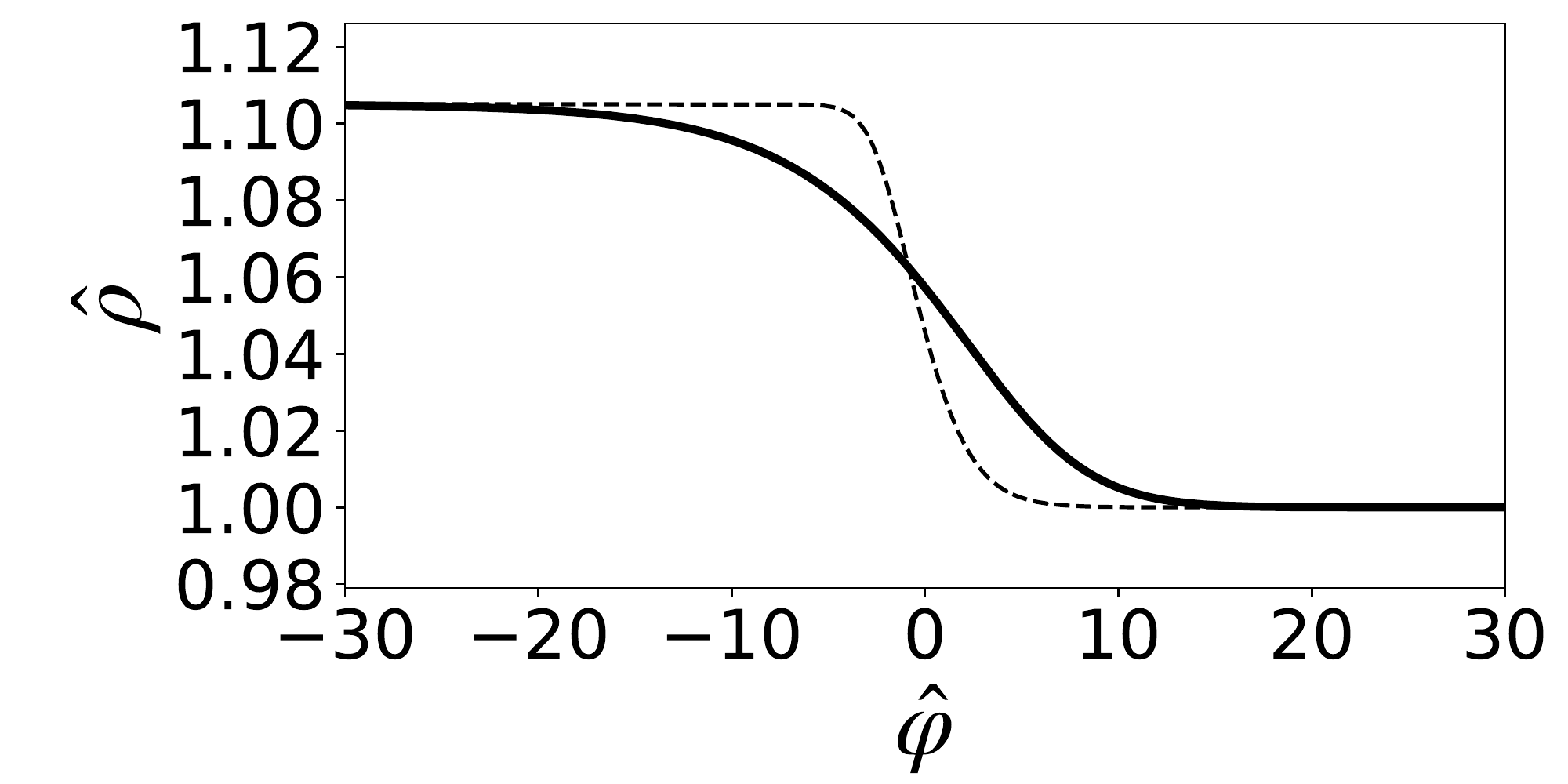}\,%
		\includegraphics[width=0.3\linewidth]{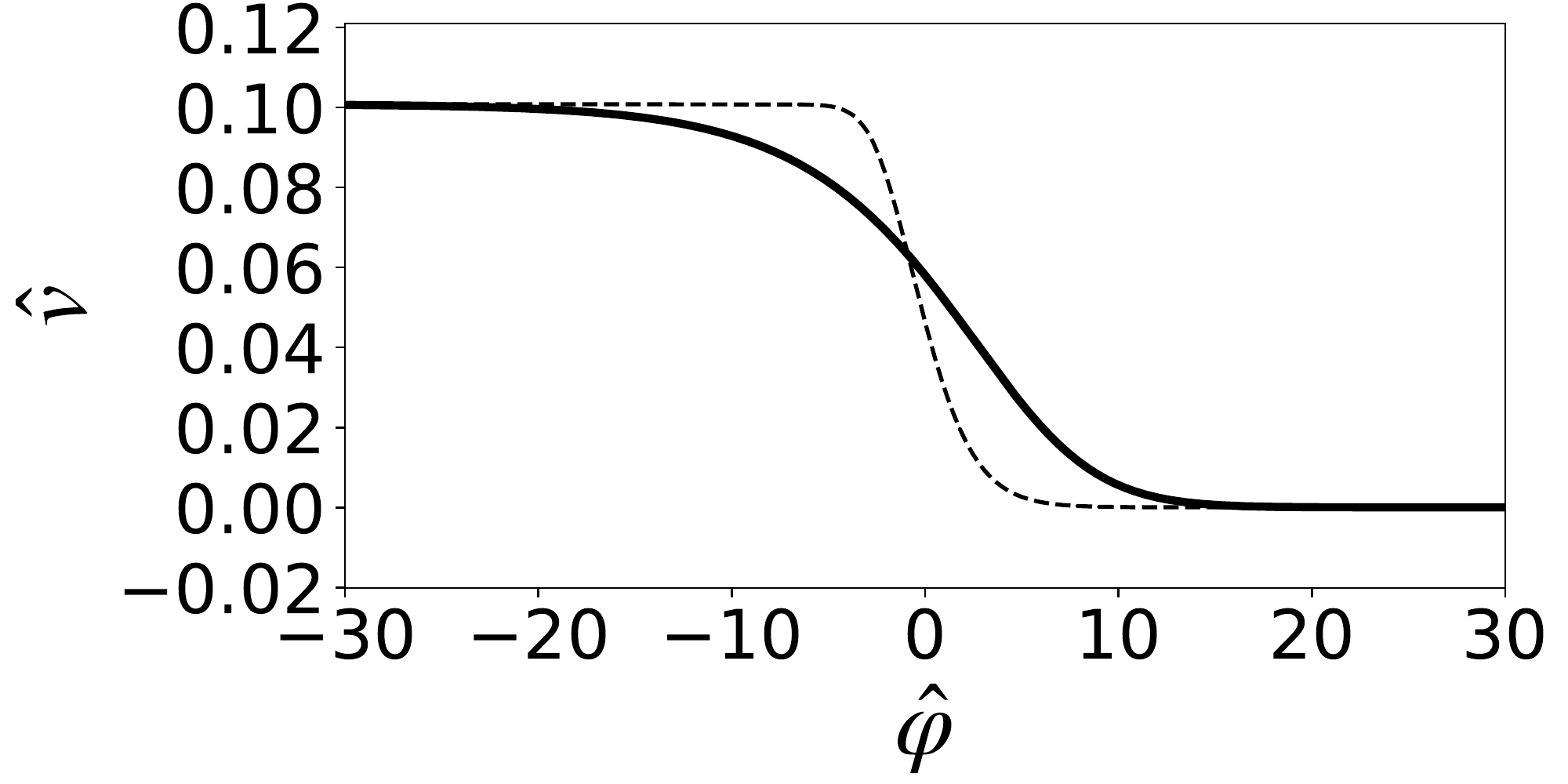}\,%
		\includegraphics[width=0.3\linewidth]{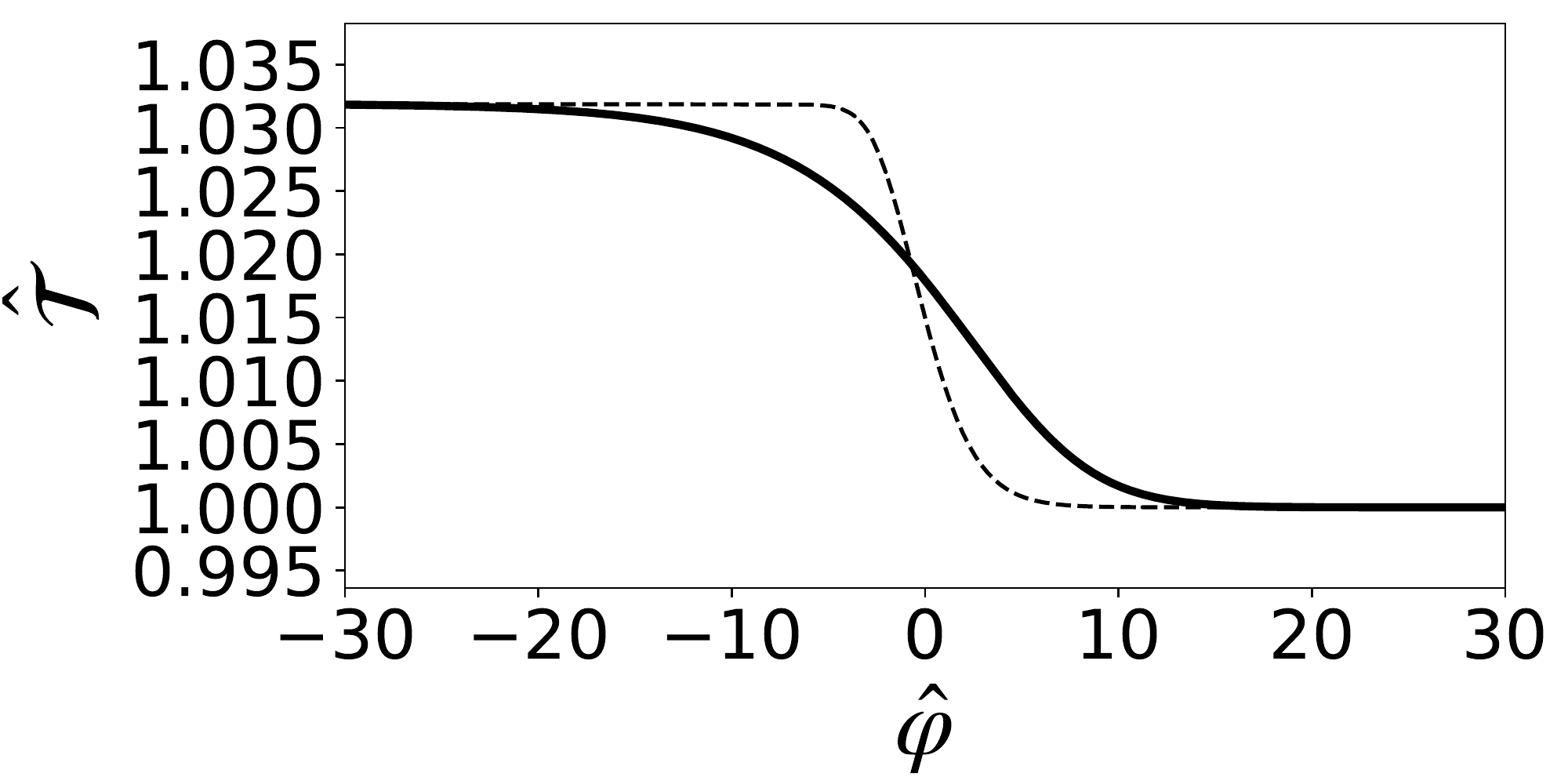}\\%
  		\includegraphics[width=0.3\linewidth]{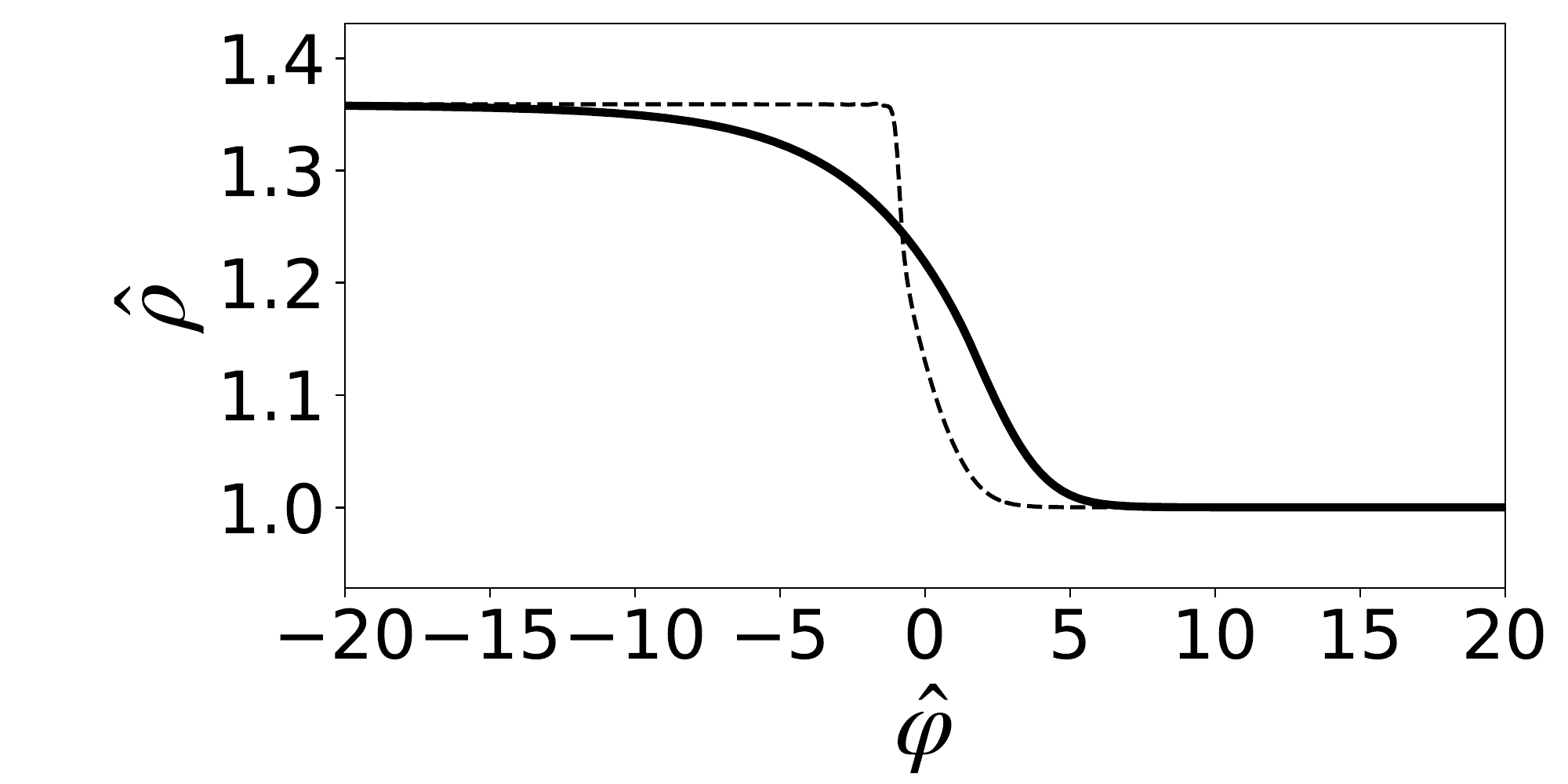}\,%
		\includegraphics[width=0.3\linewidth]{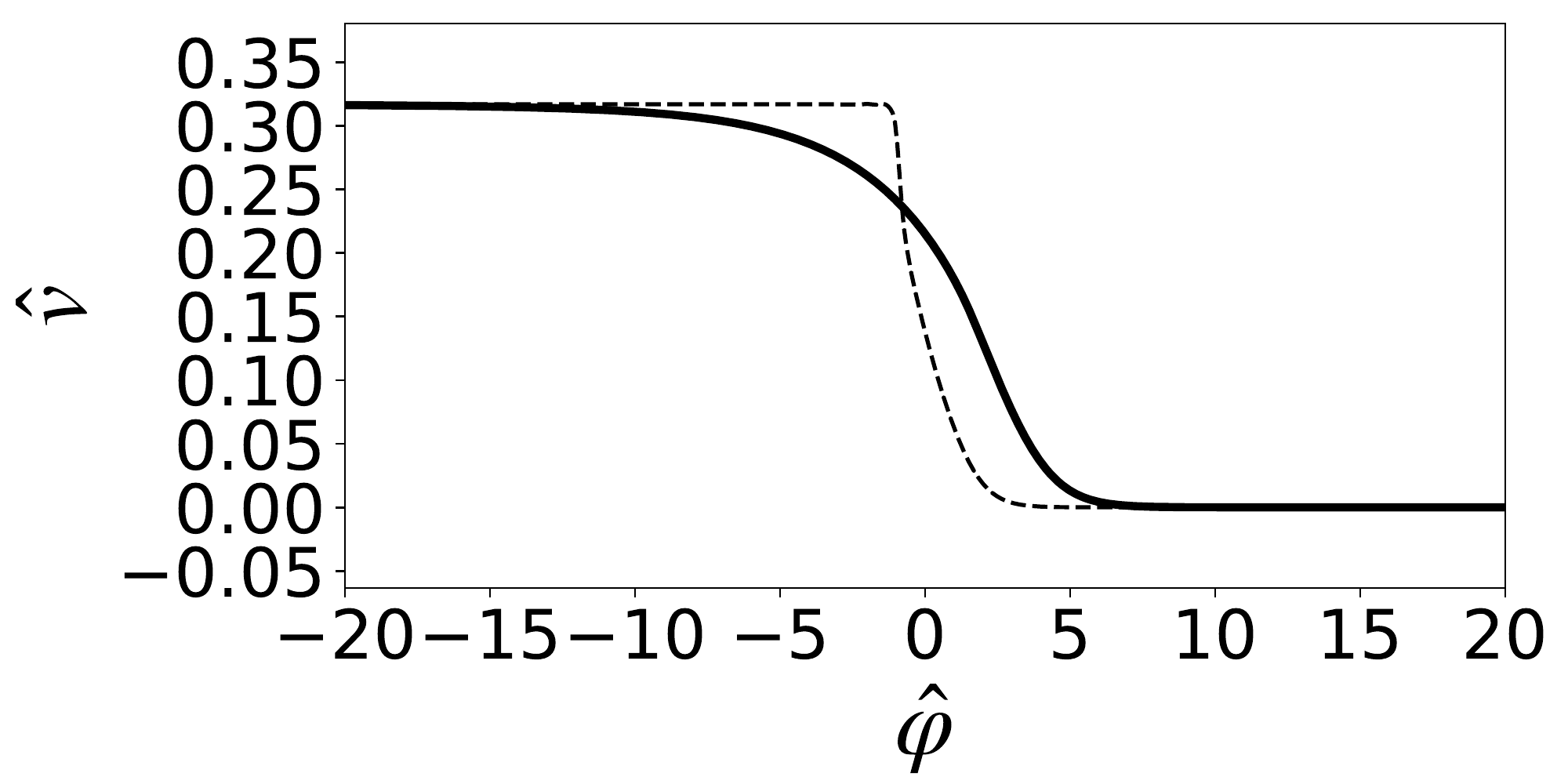}\,%
		\includegraphics[width=0.3\linewidth]{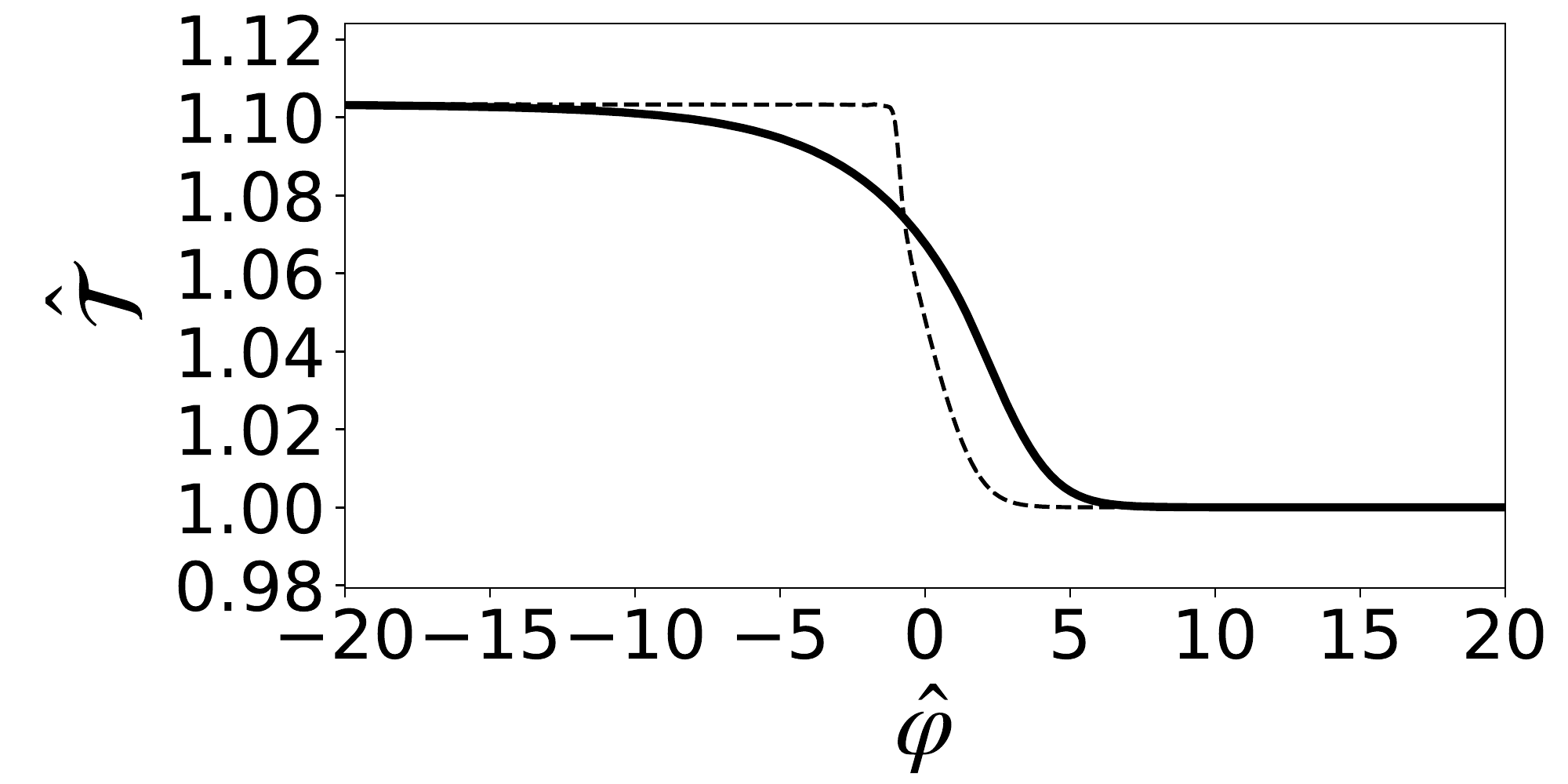}%
	\end{center}
	\caption{Shock structure predicted by the RET$_6$ theory (Solid curves) and the Eulerian theory (dashed curves) with $\gamma_1=7/5$, $\gamma_2=9/7$, $\mu = 0.4$, $c_0 = 0.15$ for  $M_0 = 1.06$ indicated by circle No. IV in Region $9$ (top row) and for $M_0=1.2$ indicated by circle No. V in Region $10$ (bottom row).  
}
	\label{fig:c015_M0-1_06}
\end{figure}

Next, let us analyze the shock structure with $\gamma_1 = 7/5$, $\gamma_2 = 9/7$, $\mu = 0.4$, and $c_0 = 0.15$ for $M_0 = 1.06$ and for $M_0 = 1.2$ shown in Figures \ref{fig:c015_M0-1_06}. 
The parameters correspond to circles Nos. IV and V in Figure \ref{fig:subshock_regions_mu04_ET6andEulerian2}. 
The result for $M_0 = 1.06$ is interesting because the necessary condition $M_0>M_{20}$ is satisfied only in the RET$_6$ theory, in spite of the fact that the RET$_6$ theory incorporates the dissipative effect. 
Figure \ref{fig:c015_M0-1_06} shows that both theories predict continuous solutions without sub-shock formation for $M_0=1.06$. 
If we increase the Mach number, the sub-shock formation is observed in a mixture of the Eulerian gases immediately after $M_{20}^{*(E)}$ while the RET$_6$ theory predicts a continuous shock-structure solution for $M_0 = 1.2$. 

\begin{figure}[htbp]
	\begin{center}
  		\includegraphics[width=0.3\linewidth]{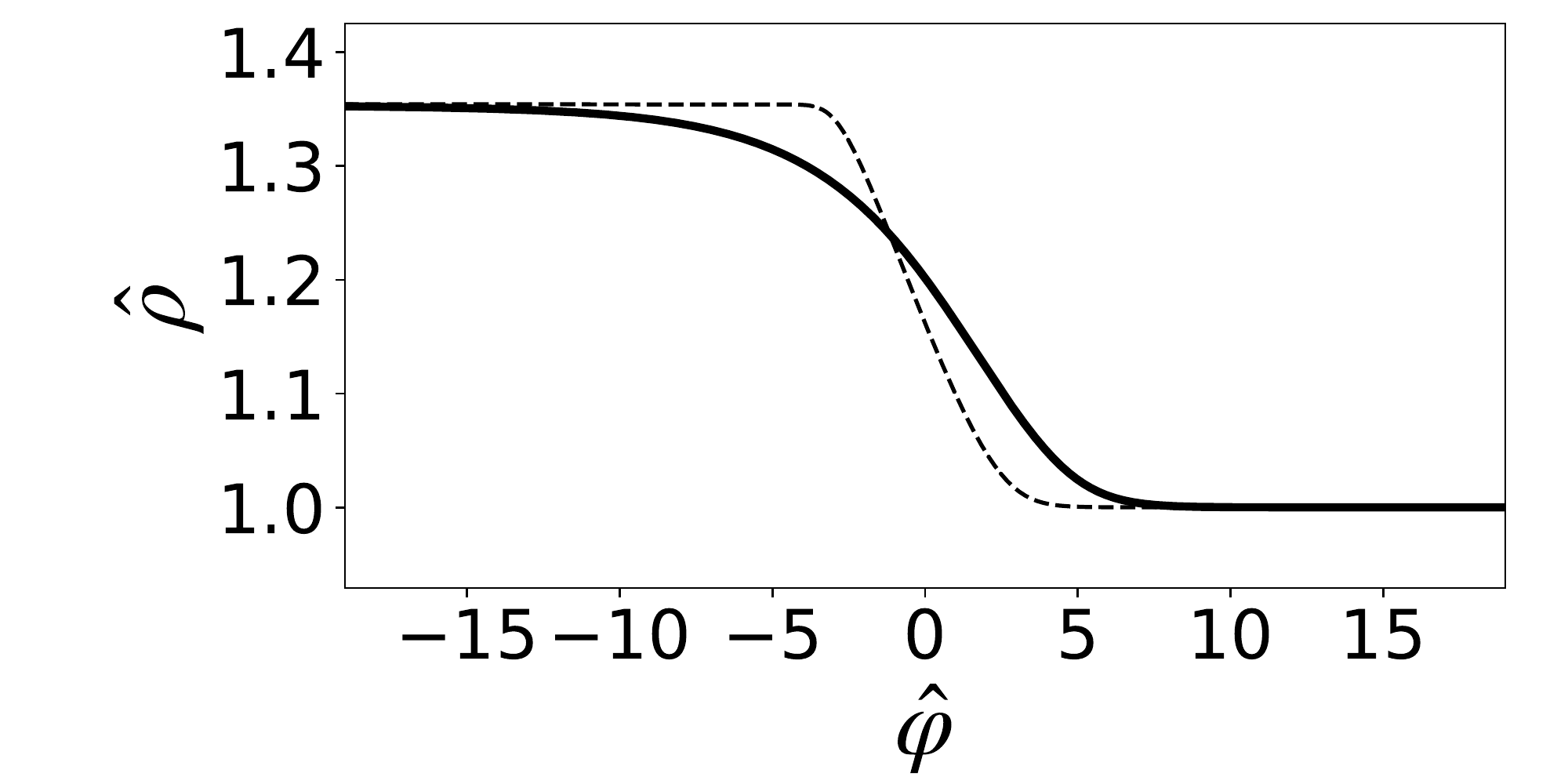}\,%
		\includegraphics[width=0.3\linewidth]{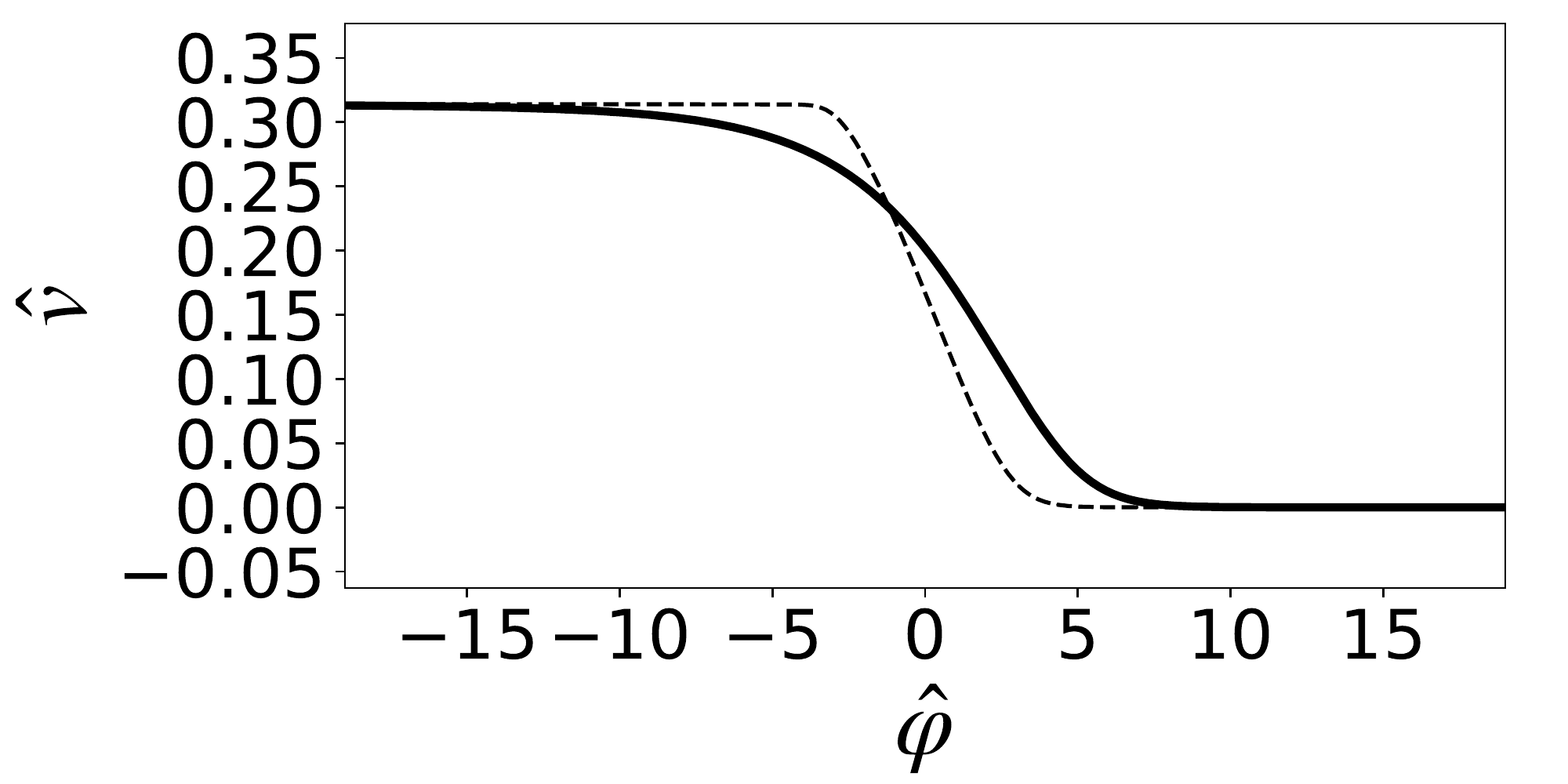}\,%
		\includegraphics[width=0.3\linewidth]{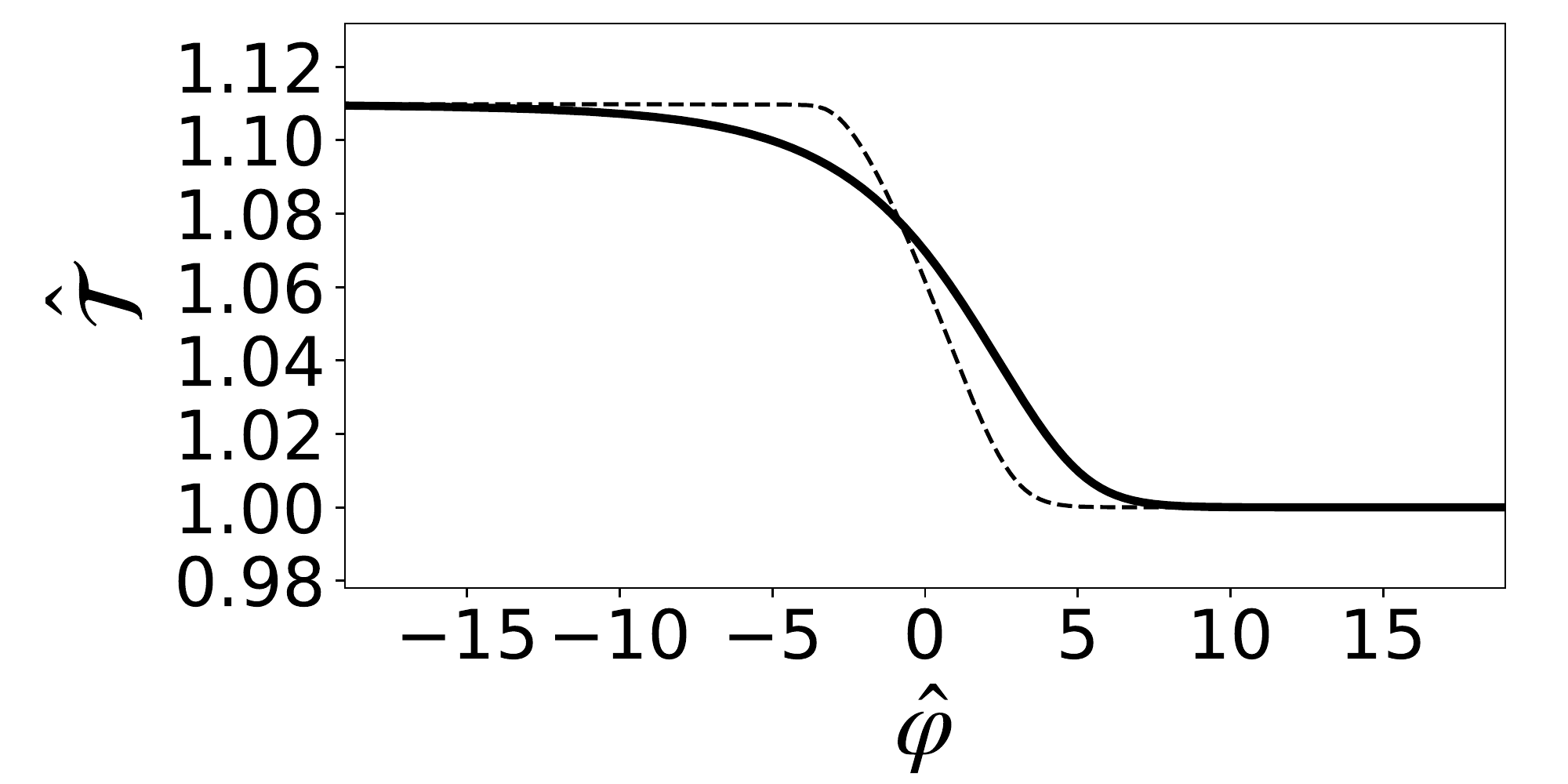}\\%
  		\includegraphics[width=0.3\linewidth]{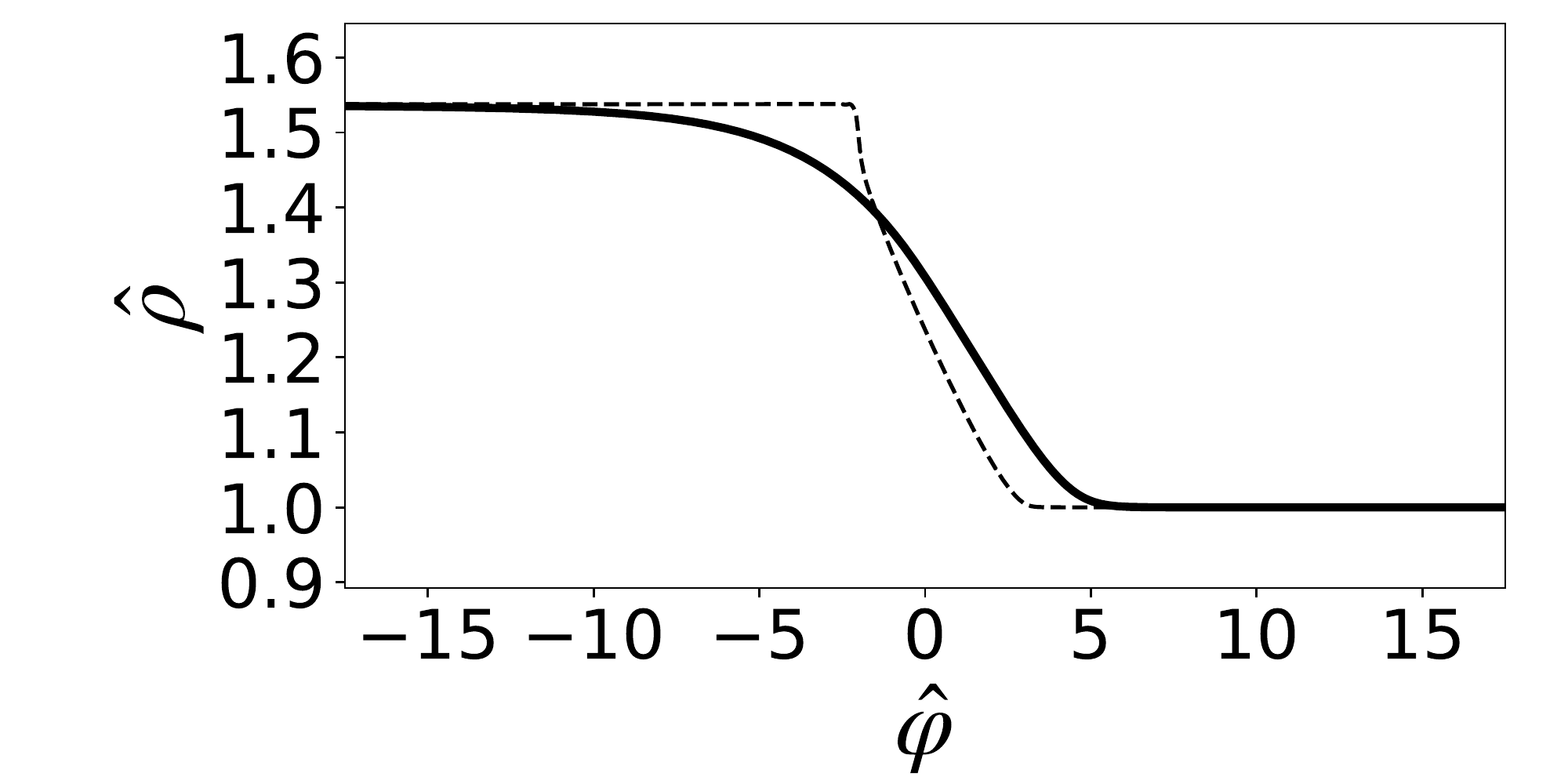}\,%
		\includegraphics[width=0.3\linewidth]{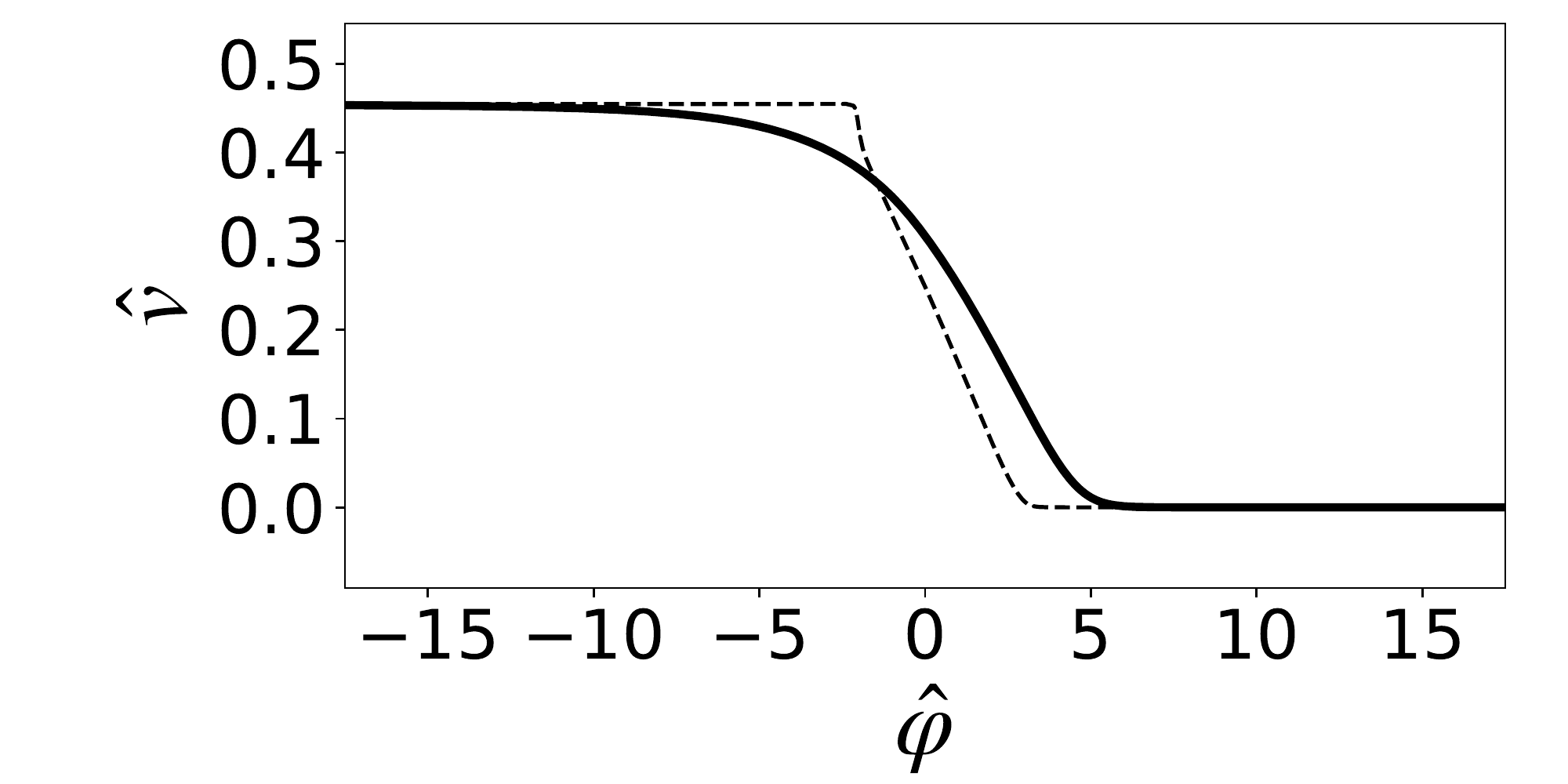}\,%
		\includegraphics[width=0.3\linewidth]{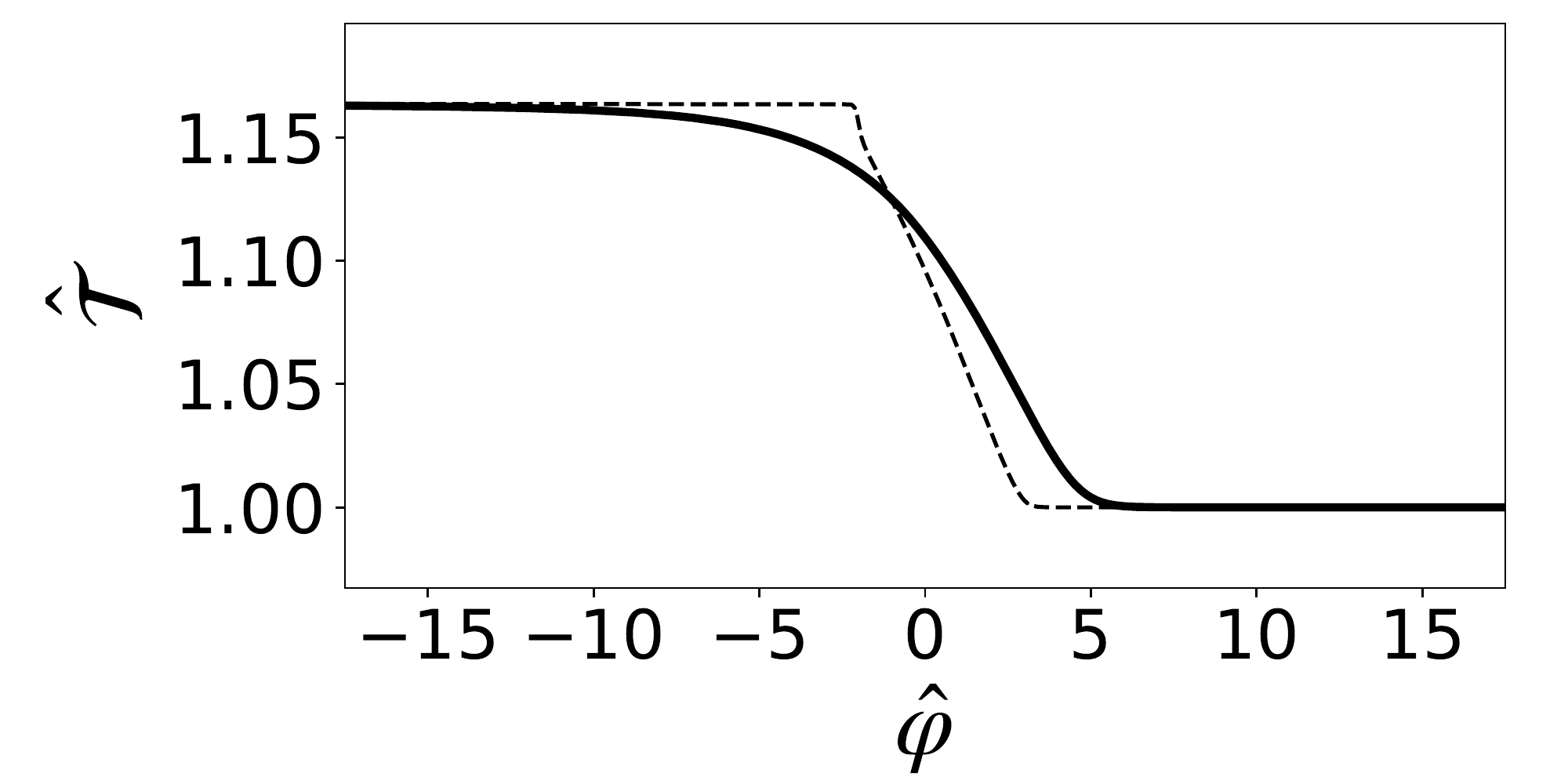}%
	\end{center}
	\caption{Shock structure predicted by the RET$_6$ theory (Solid curves) and the Eulerian theory (dashed curves) with $\gamma_1=7/5$, $\gamma_2=9/7$, $\mu = 0.4$, $c_0 = 0.3$ for $M_0 = 1.2$ indicated by circle No. VI in Region $9$ in Figure \ref{fig:subshock_regions_mu04_ET6andEulerian2} (top row), and for $M_0=1.3$ indicated by circle No. VII in Region $10$ (bottom row).
    }
	\label{fig:c03_M0-1_2}
\end{figure}

Similar situation is observed in the the shock structure with $\gamma_1 = 7/5$, $\gamma_2 = 9/7$, $\mu = 0.4$, and $c_0 = 0.3$ for $M_0 = 1.2$ and for $M_0 = 1.3$ shown in Figure \ref{fig:c03_M0-1_2}. 
In these cases, parameters are indicated by circles No. VI and VII in Figure \ref{fig:subshock_regions_mu04_ET6andEulerian2}. 
Although the necessary condition $M_0>M_{20}^*$ is satisfied in the RET$_6$ theory, the theoretical predictions by the RET$_6$ theory are continuous for both $M_0=1.2$ and $M_0=1.3$. 
The sub-shock is observed only in the Eulerian theory for $M_0=1.3$. 

\begin{figure}[htbp]
	\begin{center}
  		\includegraphics[width=0.3\linewidth]{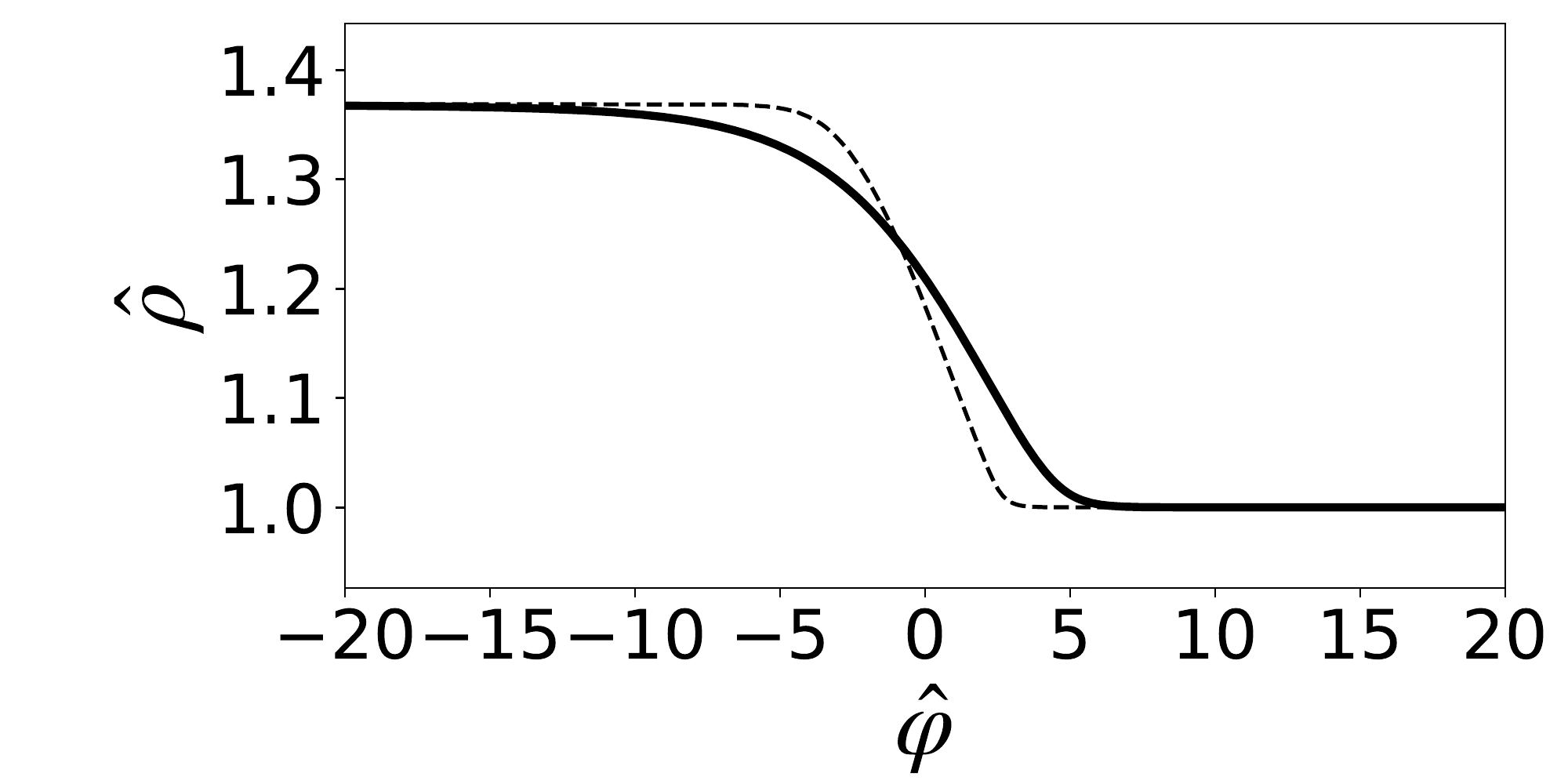}\,%
		\includegraphics[width=0.3\linewidth]{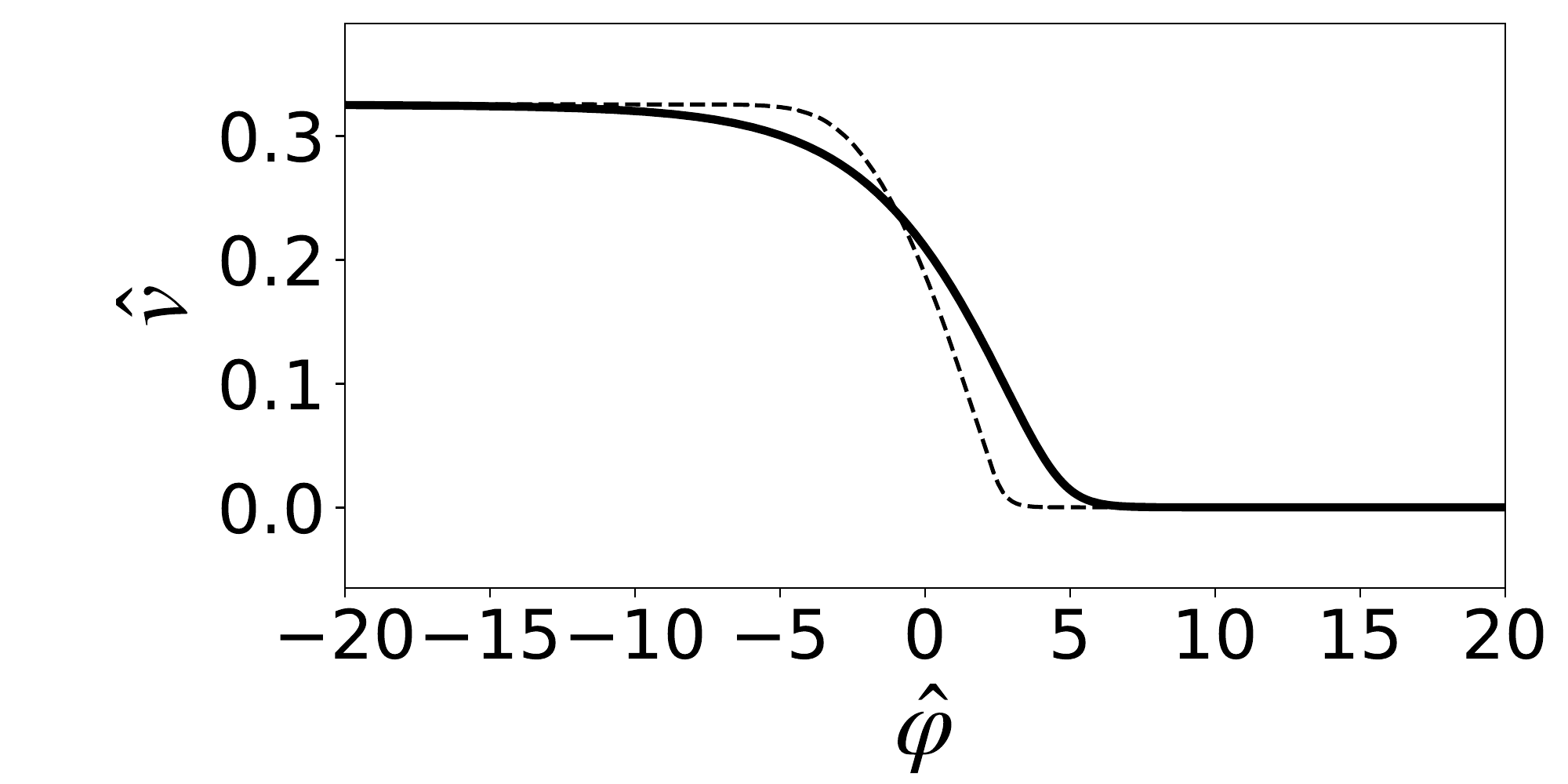}\,%
		\includegraphics[width=0.3\linewidth]{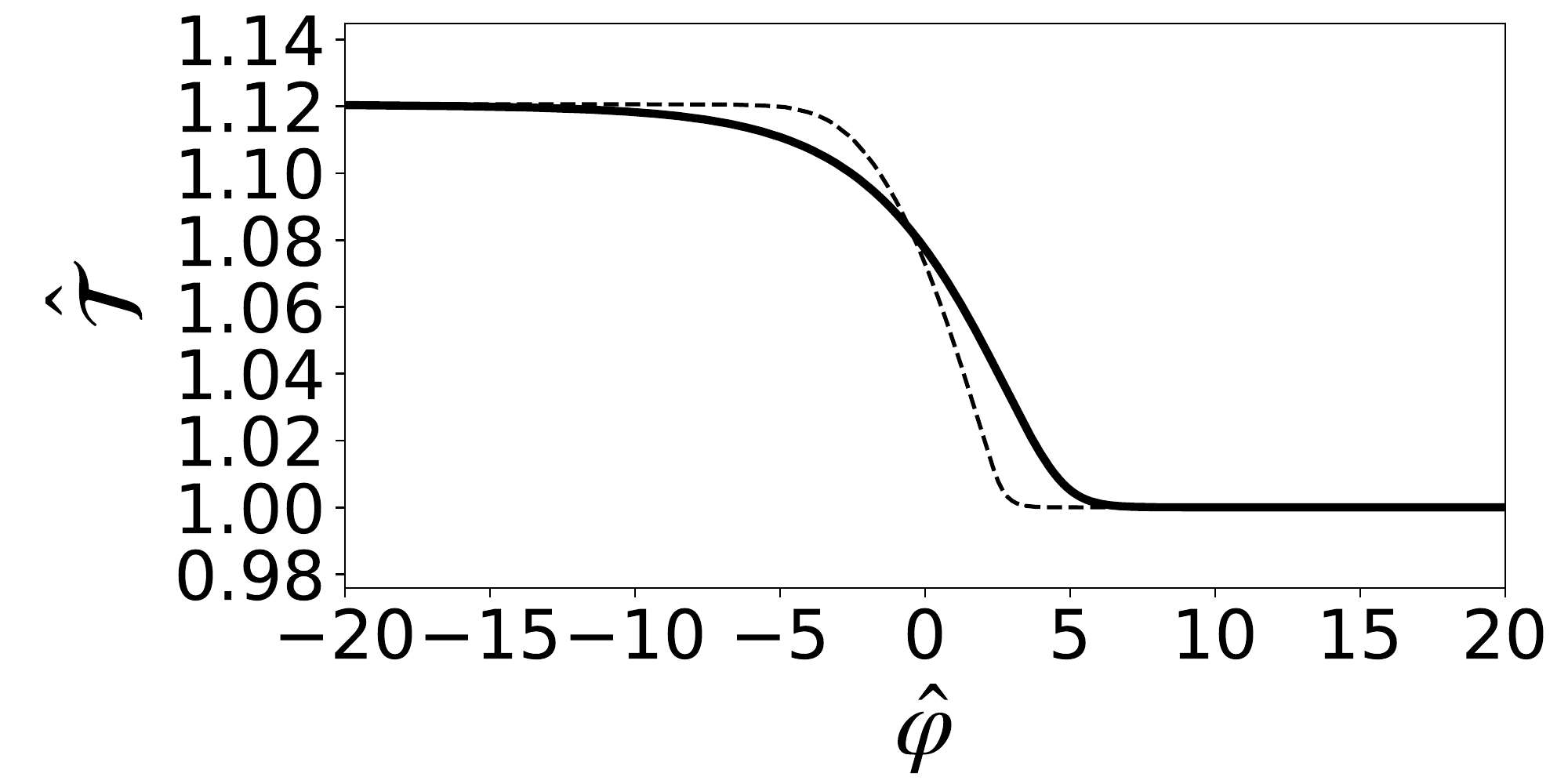}\\%
  		\includegraphics[width=0.3\linewidth]{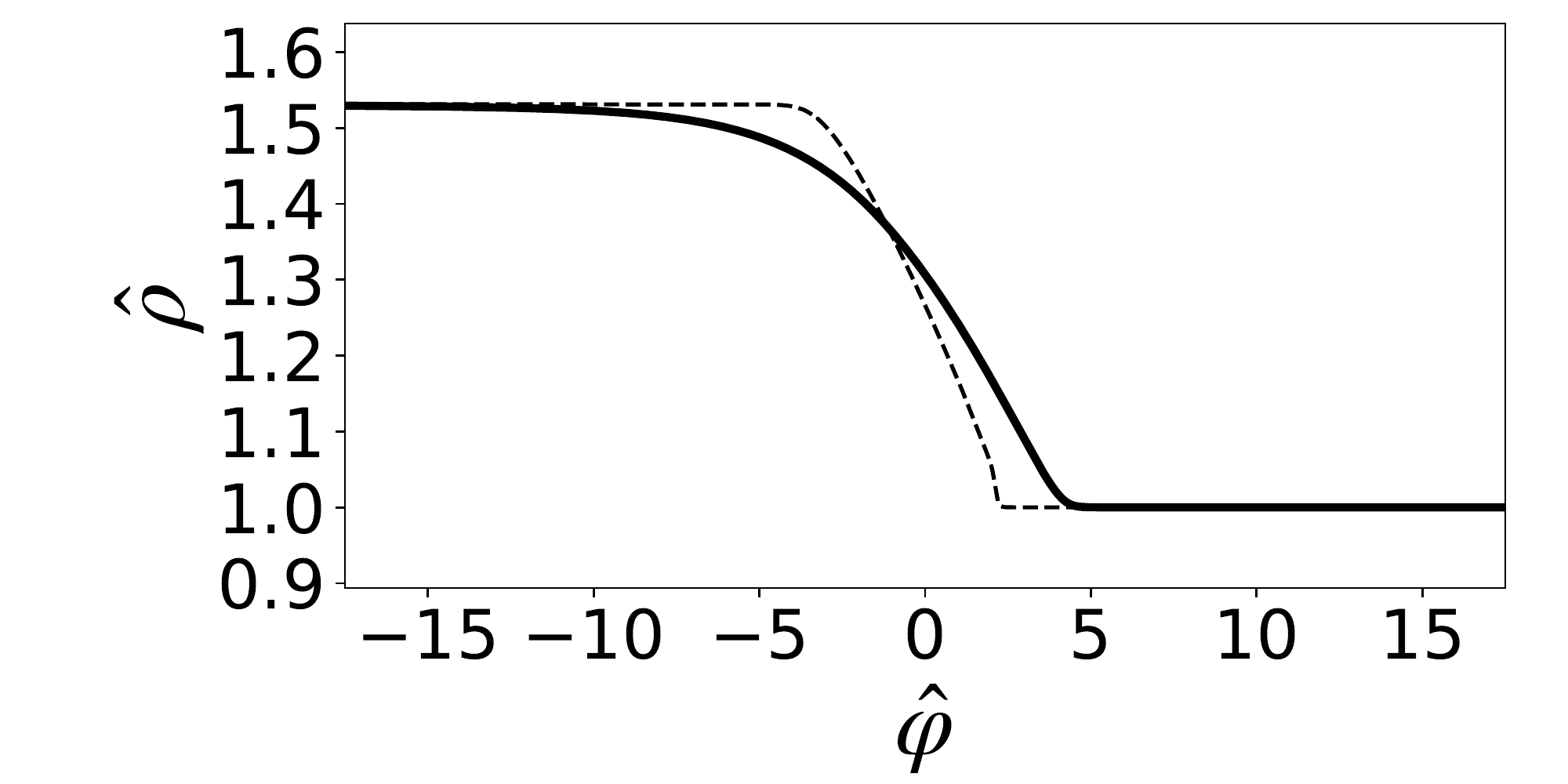}\,%
		\includegraphics[width=0.3\linewidth]{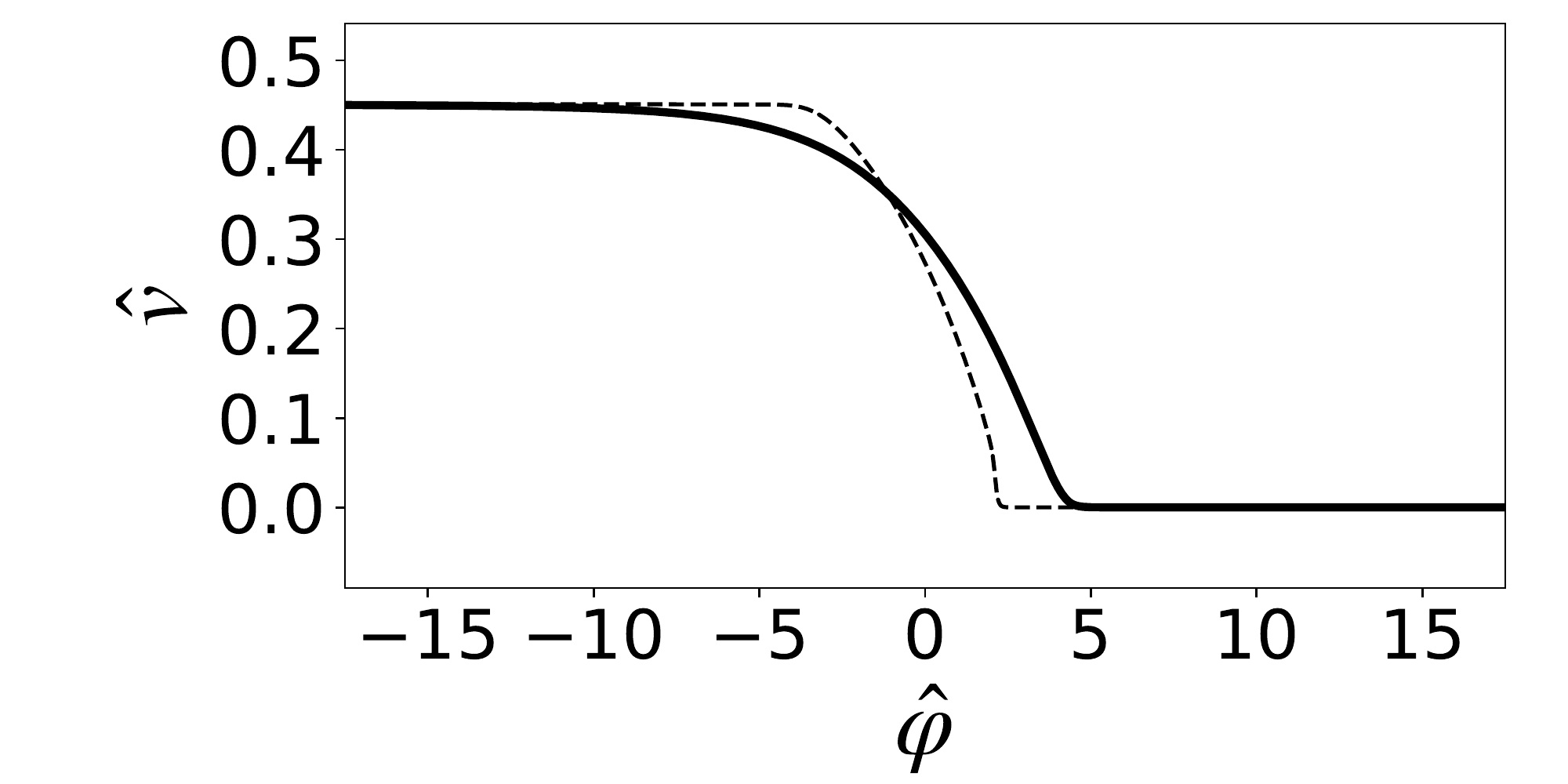}\,%
		\includegraphics[width=0.3\linewidth]{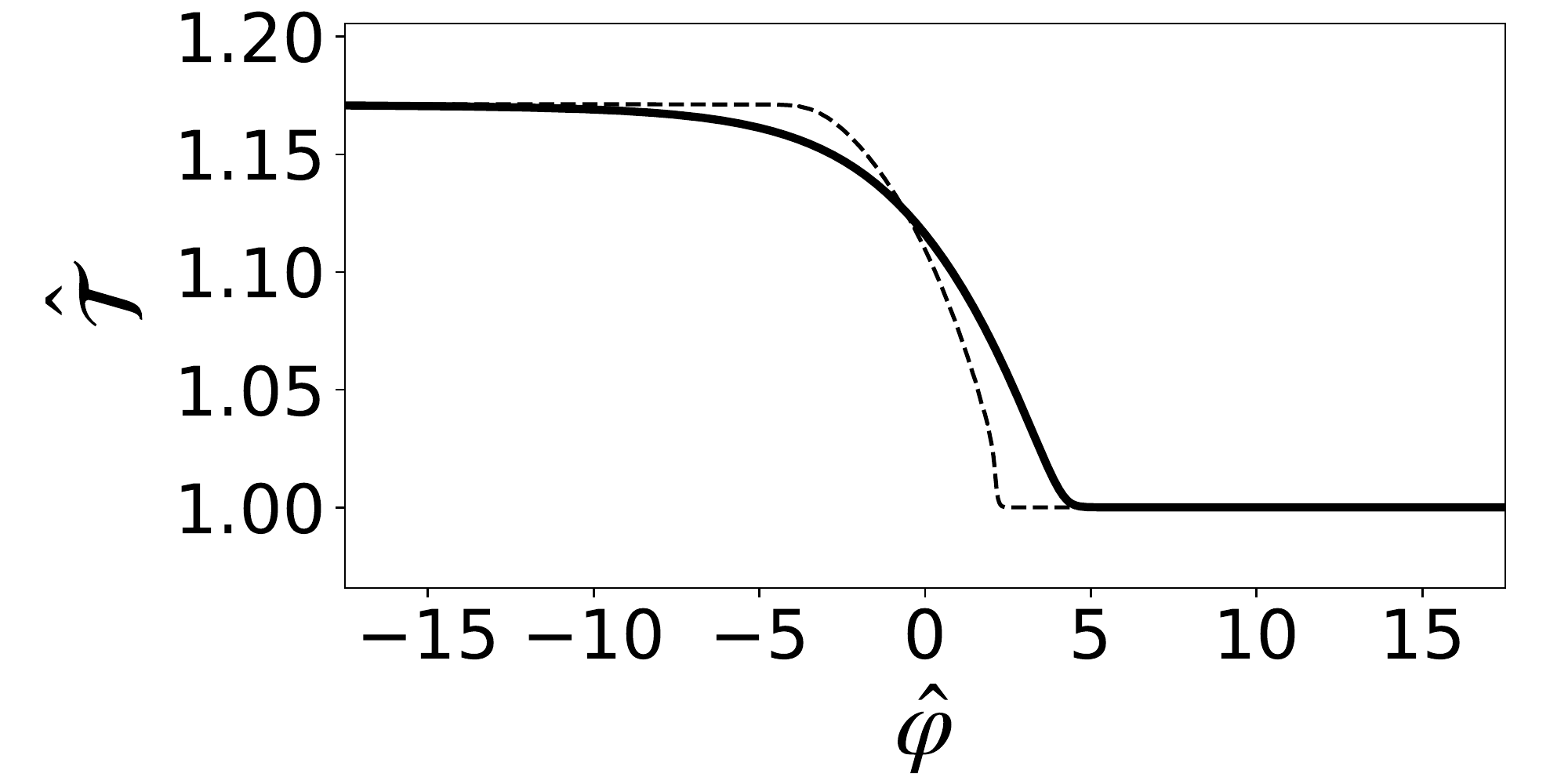}%
        \end{center}
	\caption{Shock structure predicted by the RET$_6$ theory (Solid curves) and the Eulerian theory (dashed curves) with $\gamma_1=7/5$, $\gamma_2=9/7$, $\mu = 0.4$, $c_0 = 0.45$ for $M_0 = 1.21$ indicated by circle No. VIII in Region $9$ in Figure \ref{fig:subshock_regions_mu04_ET6andEulerian2} (top row), and for $M_0 = 1.3$ indicated by circle No. IX in Region $12$ (bottom row). }
	\label{fig:c045_M0-1_21}
\end{figure}

Another example is the shock structure with $\gamma_1 = 7/5$, $\gamma_2 = 9/7$, $\mu = 0.4$, and $c_0 = 0.45$ for $M_0 = 1.21$ and for $M_0 = 1.3$ shown in Figure \ref{fig:c045_M0-1_21}. 
These parameters correspond to circles No. VIII and IX in Figure \ref{fig:subshock_regions_mu04_ET6andEulerian2}. 
Again, the RET$_6$ theory predicts continuous solutions for both both $M_0=1.2$ and $M_0=1.3$ while the Eulerian theory predicts a sub-shock $M_0=1.3$ in agreement with the theorem. 

\begin{figure}[htbp]
	\begin{center}
  		\includegraphics[width=0.3\linewidth]{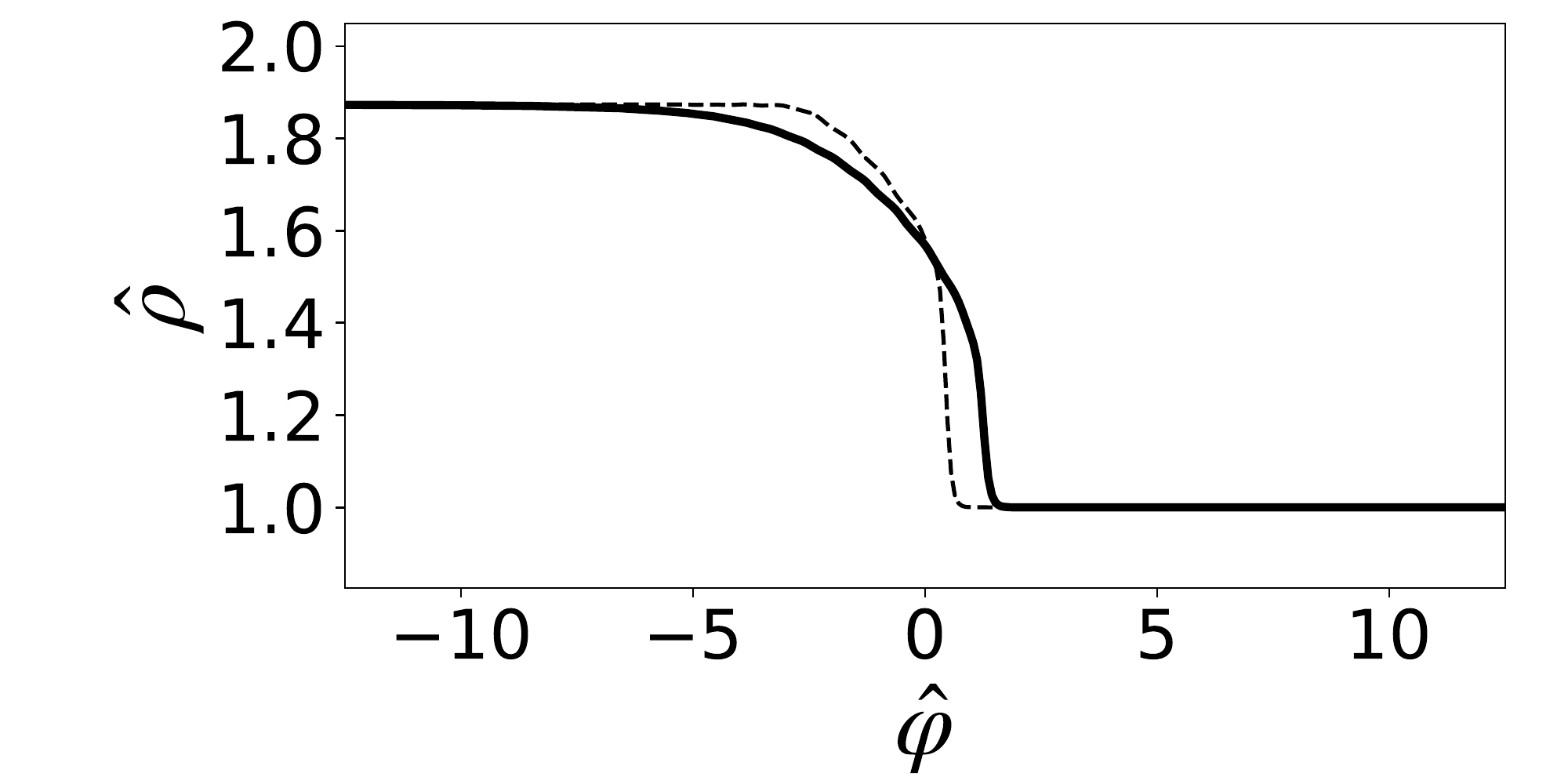}\,%
		\includegraphics[width=0.3\linewidth]{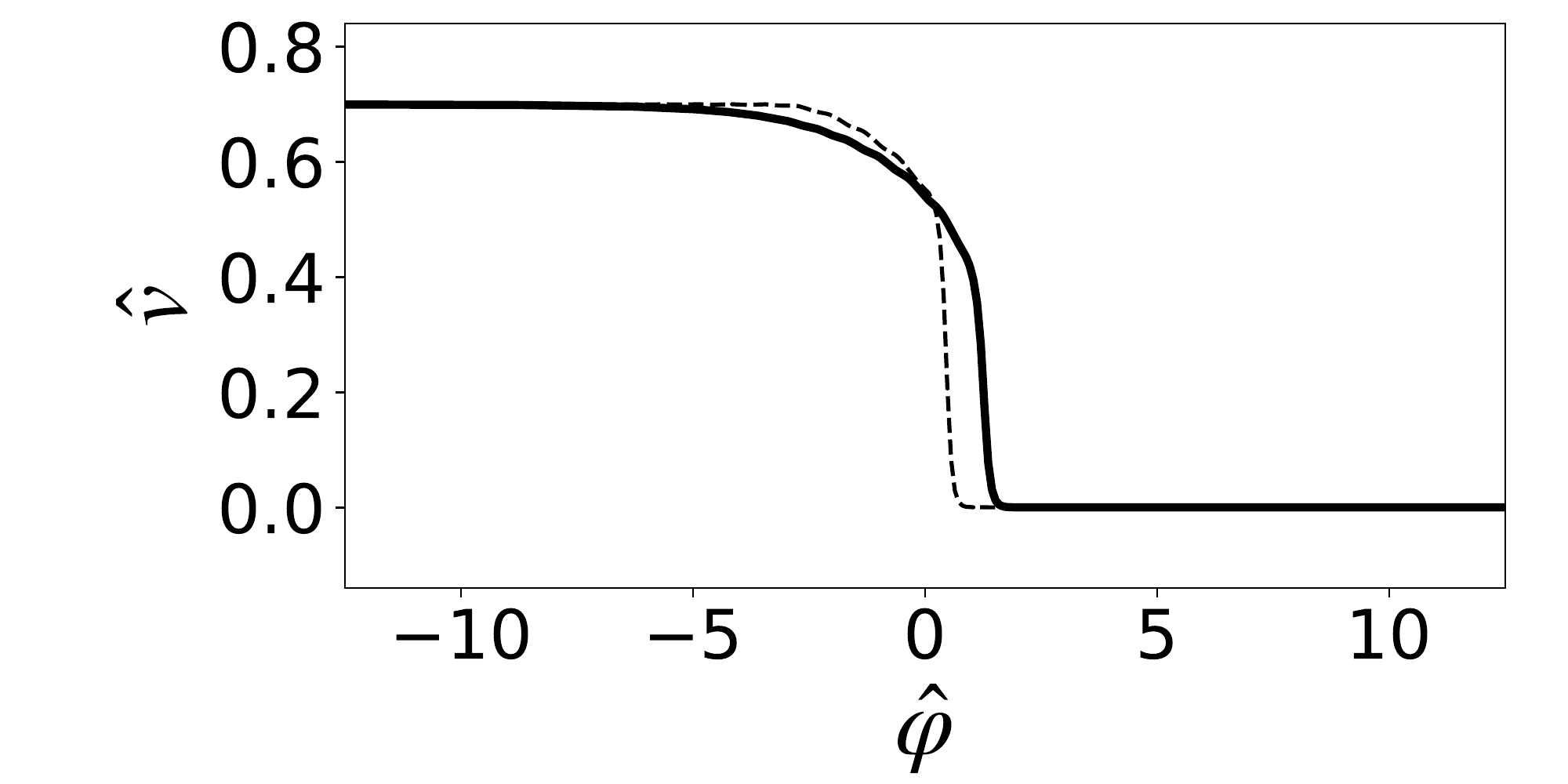}\,%
		\includegraphics[width=0.3\linewidth]{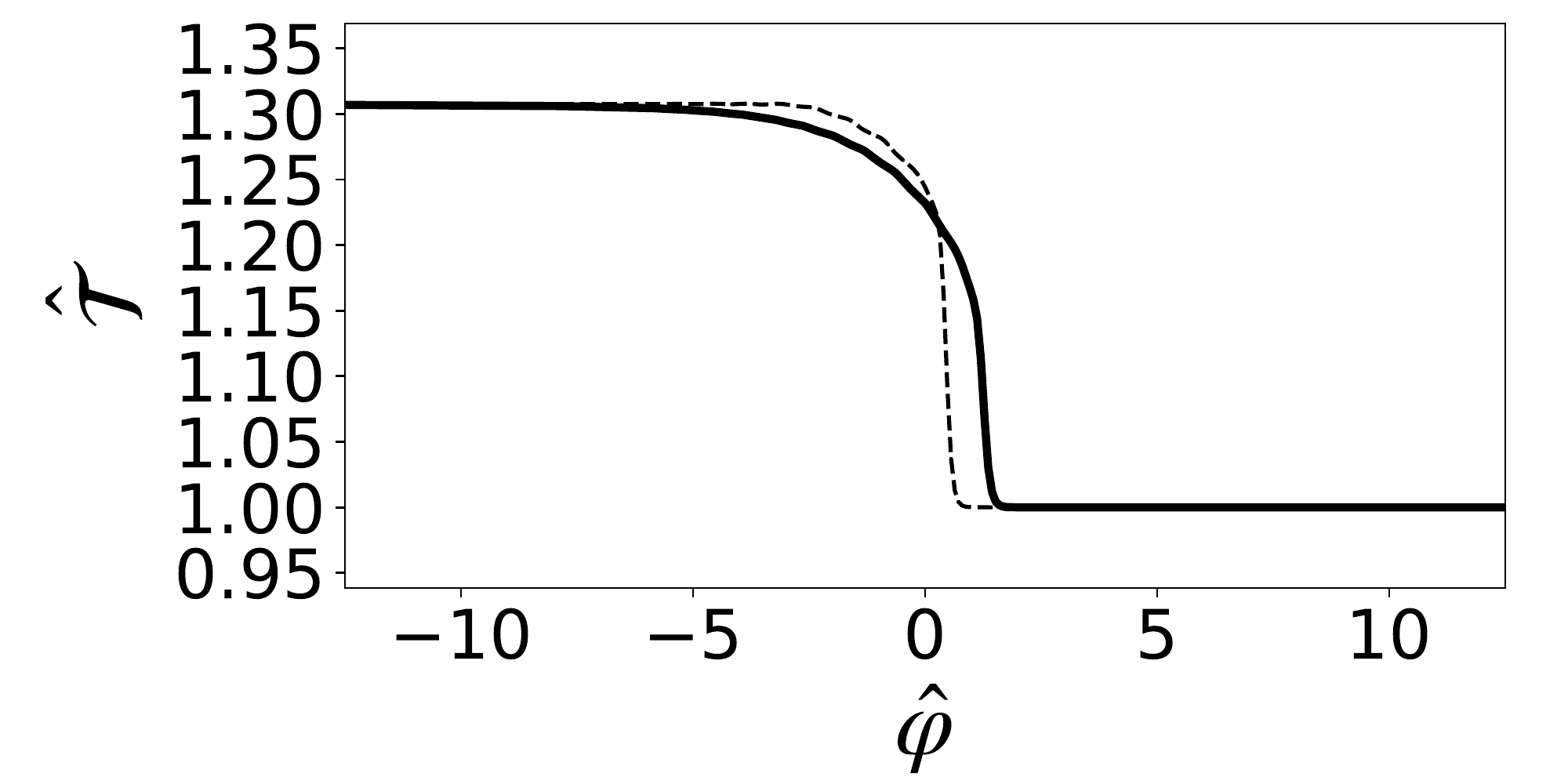}\\%
  		\includegraphics[width=0.3\linewidth]{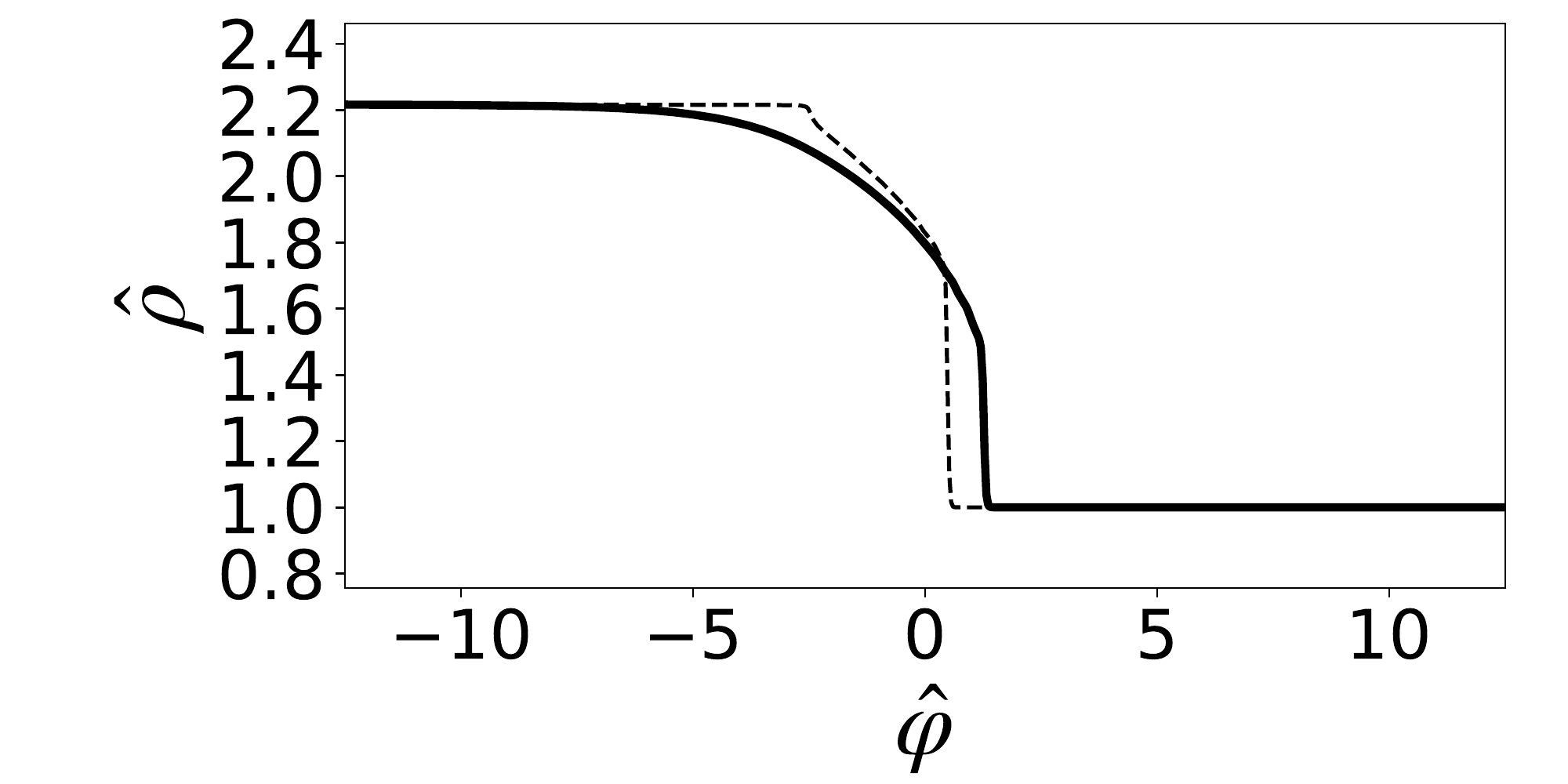}\,%
		\includegraphics[width=0.3\linewidth]{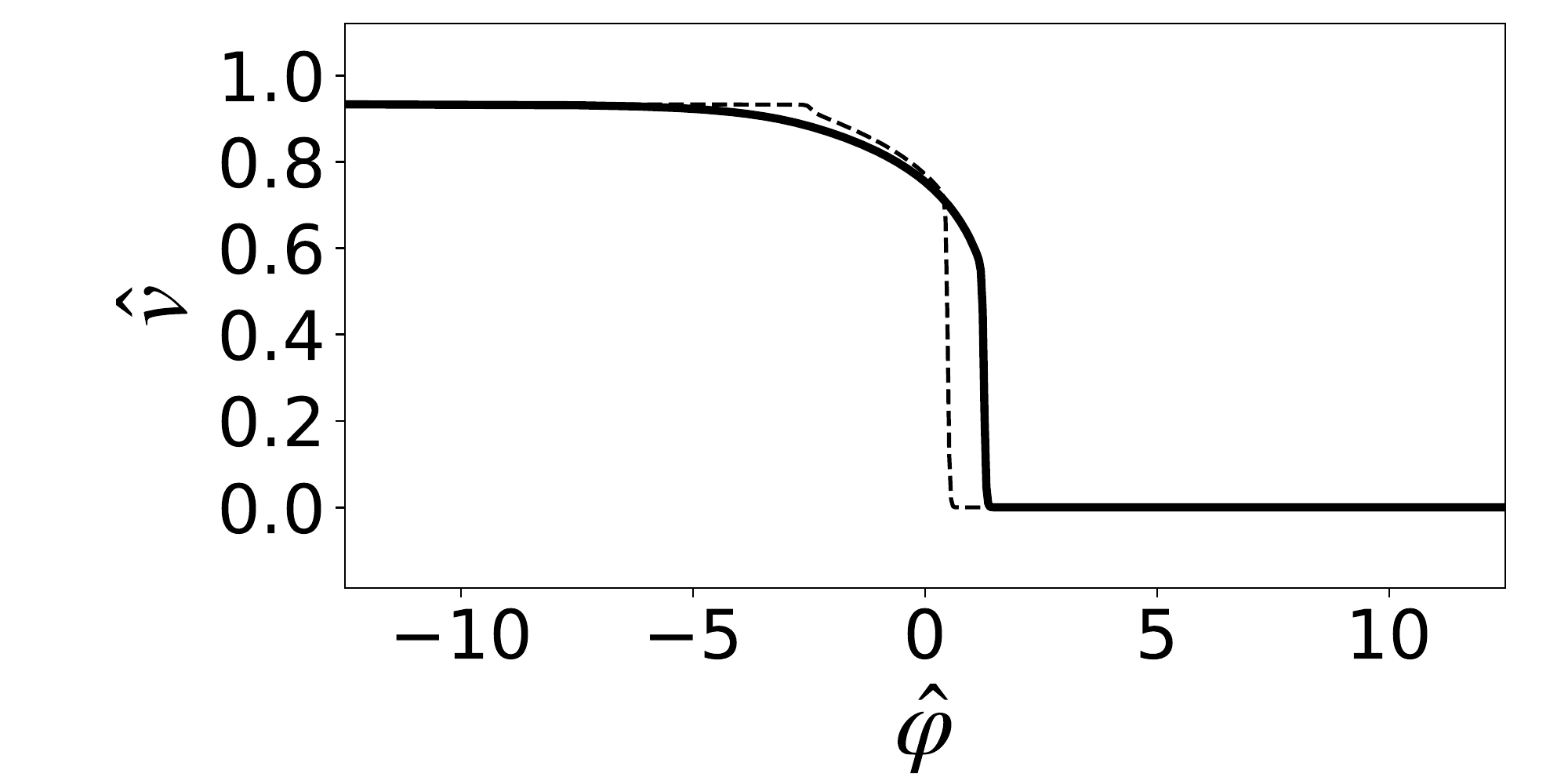}\,%
		\includegraphics[width=0.3\linewidth]{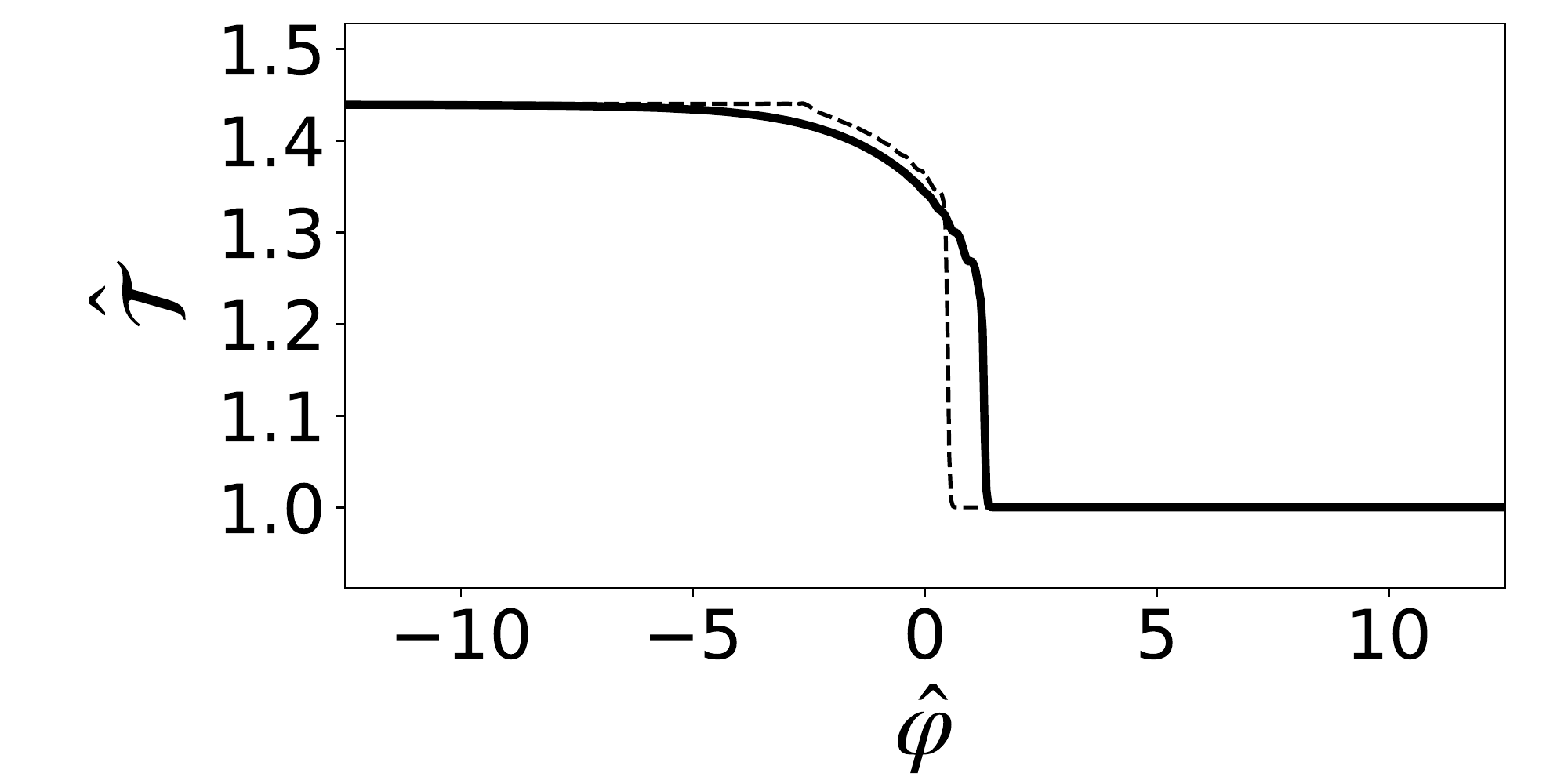}%
	\end{center}
	\caption{Shock structure predicted by the RET$_6$ theory (Solid curves) and the Eulerian theory (dashed curves) with $\gamma_1=7/5$, $\gamma_2=9/7$, $\mu = 0.4$, $c_0 = 0.75$, and $M_0 = 1.5$ indicated by circle No. X in Region $17$ in Figure \ref{fig:subshock_regions_mu04_ET6andEulerian2} (top row), and for $M_0 = 1.7$ indicated by circle No. XI in Region $18$ (bottom row).  
    }
	\label{fig:c075_M0-1_5}
\end{figure}

The last example is the shock structure with $\gamma_1 = 7/5$, $\gamma_2 = 9/7$, $\mu = 0.4$, and $c_0 = 0.7$ for $M_0 = 1.5$ and for $M_0 = 1.7$ shown in Figure \ref{fig:c075_M0-1_5}. 
These parameters are indicated by circles No. X and XI in Figure \ref{fig:subshock_regions_mu04_ET6andEulerian2}. 
According to the theorem, a sub-shock for the shock structure of the constituent $1$ arises in both theories. 
No sub-shock for the shock structure of the constituent $2$ is observed in the prediction by both theories for $M_0=1.5$, and multiple sub-shocks are predicted only in the Eulerian theory for $M_0=1.7$. 

\section{Conclusions}
\label{sec:Summary}
In this paper, we have analyzed the shock structure in a binary mixture of rarefied polyatomic gases based on rational extended thermodynamics. 
We have used  the system of the field equations derived in the context of the multi-temperature model in which the dynamic pressure's contribution is relevant. 
We have classified the possible parameters by considering the necessary conditions to observe the sub-shock formation and have tested the sub-shock formation numerically. As we expect, the shock profile is more regularized with respect to the mixture of Eulerian gas, and in the case that exists sub-shocks, the amplitude of the sub-shock is reduced compared to those of non-dissipative gases. 
We have also observed from the numerical simulation that for very   small Mach numbers in the regions in which the existence is only necessary, the singularity is only apparent, while increasing the Mach number always we have sub-shocks in this kind of region. 

The present results are in agreement with the fact that the present theory satisfies the so-called K-condition \cite{ShockBinaryEulerian_lincei} that, together with the convexity of entropy, take the system in the assumptions  of general theorems for which if the initial data are sufficiently small, there exist smooth global solutions \cite{RuggeriSerre,Yong,ShockBinaryET6_RdM,BookNew,Bianchini}.

It remains open to compare with experimental data for mixtures in which constituents have large bulk viscosity.  
However, at the moment, the authors were not able to find experimental data on the shock-wave profiles for this kind of mixture in the literature. Instead, there are several single gases with this characteristic in which, as we say in precedence, the RET$_6$ theory is very successful.

\section*{Acknowledgments}
This paper is dedicated to Gerald Warnecke on the occasion of his 65th birthday.
The work has been partially supported by JSPS KAKENHI Grant Numbers  JP19K04204 (S.T.). 
The paper has also been carried out in the framework of the activities of the Italian National Group of Mathematical Physics of the Italian National Institute of High Mathematics GNFM/INdAM (T.R.).

\section*{Compliance with Ethical Standards} 

\subsection*{Funding} Partially supported by JSPS KAKENHI Grant Numbers  JP19K04204 (S.T.)

\subsection*{Conflict of Interest} On behalf of all authors, the corresponding author states that there is no conflict of interest 

\subsection*{Ethical Approval in the manuscript} Not Applicable

\bibliographystyle{ytphys}
\bibliography{bibliography} 

\providecommand{\href}[2]{#2}\begingroup\raggedright\begin{thebibliography}{10}

\bibitem{VincentiKruger}
W.~G. Vincenti and C.~H. Kruger, Jr, {\em Introduction to Physical Gas
  Dynamics}.
\newblock John Wiley and Sons, New York, London, Sydney, 1965.

\bibitem{Zeldovich}
Y.~B. Zel'dovich and Y.~P. Raizer, {\em Physics of Shock Waves and
  High-Temperature Hydrodynamic Phenomena}.
\newblock Dover Publications, Mineola, New York, 2002.

\bibitem{Bird}
G.~A. Bird, {\em Molecular Gas Dynamics and the Direct Simulation of Gas
  Flows}.
\newblock Oxford Univ Press, Oxford, 1994.

\bibitem{PavicTorr}
V.~Djordji\'c, M.~Pavi\'c-\u{C}oli\'c, and M.~Torrilhon, {\slshape Consistent,
  explicit, and accessible {B}oltzmann collision operator for polyatomic
  gases,} {\em Physical Review E} {\bfseries 104} (2021) 025309.

\bibitem{PavicGamba}
I.~M. Gamba and M.~Pavi\'c-\u{C}oli\'c, {\slshape On the {C}auchy problem for
  {B}oltzmann equation modeling a polyatomic gas,} {\em J. Math. Phys.}
  {\bfseries 64} (2023) 013303.

\bibitem{Pirner}
M.~Pirner, {\slshape A review on {BGK} models for gas mixtures of mono and
  polyatomic molecules,} {\em Fluids} {\bfseries 6} (2021) 393.

\bibitem{RET}
I.~M\"uller and T.~Ruggeri, {\em Rational Extended Thermodynamics}.
\newblock Springer, New York, 1998.

\bibitem{RuggeriSugiyama}
T.~Ruggeri and M.~Sugiyama, {\em Rational Extended Thermodynamics beyond the
  Monatomic Gas}.
\newblock Springer, Cham, 2015.

\bibitem{BookNew}
T.~Ruggeri and M.~Sugiyama, {\em Classical and Relativistic Rational Extended
  Thermodynamics of Gases}.
\newblock Springer, Cham, 2021.

\bibitem{ET14shock}
S.~Taniguchi, T.~Arima, T.~Ruggeri, and M.~Sugiyama, {\slshape Thermodynamic
  theory of the shock wave structure in a rarefied polyatomic gas: Beyond the
  bethe-teller theory,} {\em Phys. Rev. E} {\bfseries 89} (Jan, 2014) 013025.

\bibitem{ET6shock}
S.~Taniguchi, T.~Arima, T.~Ruggeri, and M.~Sugiyama, {\slshape Effect of the
  dynamic pressure on the shock wave structure in a rarefied polyatomic gas,}
  {\em Physics of Fluids} {\bfseries 26} (2014) 016103.

\bibitem{BetheTeller}
H.~A. Bethe and E.~Teller, {\em Deviations from Thermal Equilibrium in Shock
  Waves}.
\newblock reprinted by Engineering Research Institute. University of Michigan,
  Michigan, 1941.

\bibitem{Gilbarg}
D.~Gilbarg and D.~Paolucci, {\slshape The structure of shock waves in the
  continuum theory of fluids.,} {\em J. Rat. Mech. Anal.} {\bfseries 2} (1953)
  617.

\bibitem{6fields}
T.~Arima, S.~Taniguchi, T.~Ruggeri, and M.~Sugiyama, {\slshape Extended
  thermodynamics of real gases with dynamic pressure: An extension of
  {M}eixner's theory,} {\em Physics Letters A} {\bfseries 376} (2012)
  2799--2803.

\bibitem{6fields2}
T.~Arima, T.~Ruggeri, M.~Sugiyama, and S.~Taniguchi, {\slshape Non-linear
  extended thermodynamics of real gases with 6 fields,} {\em International
  Journal of Non-Linear Mechanics} {\bfseries 72} (2015) 6--15.

\bibitem{NLET6shock}
S.~Taniguchi, T.~Arima, T.~Ruggeri, and M.~Sugiyama, {\slshape Overshoot of the
  non-equilibrium temperature in the shock wave structure of a rarefied
  polyatomic gas subject to the dynamic pressure,} {\em International Journal
  of Non-Linear Mechanics} {\bfseries 79} (2016) 66--75.

\bibitem{Kosuge}
S.~Kosuge, K.~Aoki, and T.~Goto, {\slshape Shock wave structure in polyatomic
  gases: Numerical analysis using a model {B}oltzmann equation,} {\em AIP
  Conference Proceedings} {\bfseries 1786} (2016) 180004.

\bibitem{2018kyoto}
S.~Kosuge and K.~Aoki, {\slshape Shock-wave structure for a polyatomic gas with
  large bulk viscosity,} {\em Phys. Rev. Fluids} {\bfseries 3} (Feb, 2018)
  023401.

\bibitem{GMixture}
T.~Arima, T.~Ruggeri, M.~Sugiyama, and S.~Taniguchi, {\slshape Galilean
  invariance and entropy principle for a system of balance laws of mixture
  type,} {\em Atti Accad. Naz. Lincei Cl. Sci. Fis. Mat. Natur.} {\bfseries 28}
  (2017) 66--75.

\bibitem{Ruggeri1993}
T.~Ruggeri, {\slshape Breakdown of shock-wave-structure solutions,} {\em Phys.
  Rev. E} {\bfseries 47} (Jun, 1993) 4135--4140.

\bibitem{Breakdown}
G.~Boillat and T.~Ruggeri, {\slshape On the shock structure problem for
  hyperbolic system of balance laws and convex entropy,} {\em Continuum
  Mechanics and Thermodynamics} {\bfseries 10} (Oct, 1998) 285--292.

\bibitem{ShRu}
T.~Ruggeri and S.~Taniguchi, {\slshape Shock waves in hyperbolic systems of
  nonequilibrium thermodynamics,} in {\em Applied Wave Mathematics II},
  A.~Berezovski and T.~Soomere, eds., Mathematics of Planet Earth, ch.~8,
  pp.~167--186.
\newblock Springer, Cham, 2019.

\bibitem{Weiss}
W.~Weiss, {\slshape Continuous shock structure in extended thermodynamics,}
  {\em Phys. Rev. E} {\bfseries 52} (Dec, 1995) R5760--R5763.

\bibitem{IJNLM2017}
S.~Taniguchi and T.~Ruggeri, {\slshape On the sub-shock formation in extended
  thermodynamics,} {\em International Journal of Non-Linear Mechanics}
  {\bfseries 99} (2018) 69--78.

\bibitem{FMR}
F.~Conforto, A.~Mentrelli, and T.~Ruggeri, {\slshape Shock structure and
  multiple sub-shocks in binary mixtures of {E}ulerian fluids,} {\em Ricerche
  di Matematica} {\bfseries 66} (Jun, 2017) 221--231.

\bibitem{Bisi1}
M.~Bisi, G.~Martal{\`o}, and G.~Spiga, {\slshape Shock wave structure of
  multi-temperature {E}uler equations from kinetic theory for a binary
  mixture,} {\em Acta Applicandae Mathematicae} {\bfseries 132} (Aug, 2014)
  95--105.

\bibitem{Bisi2}
V.~Artale, F.~Conforto, G.~Martal{\`o}, and A.~Ricciardello, {\slshape Shock
  structure and multiple sub-shocks in grad 10-moment binary mixtures of
  monoatomic gases,} {\em Ricerche di Matematica} {\bfseries 68} (Dec, 2019)
  485--502.

\bibitem{subshock2}
S.~Taniguchi and T.~Ruggeri, {\slshape A 2 $\times$ 2 simple model in which the
  sub-shock exists when the shock velocity is slower than the maximum
  characteristic velocity,} {\em Ricerche di Matematica} {\bfseries 68} (Jun,
  2019) 119--129.

\bibitem{ShockBinaryEulerian_lincei}
T.~Ruggeri and S.~Taniguchi, {\slshape Sub-shock formation in shock structure
  of a binary mixture of polyatomic gases,} {\em Atti Accad. Naz. Lincei Cl.
  Sci. Fis. Mat. Natur.} {\bfseries 32} (2021) 167--179.

\bibitem{ShockBinaryEulerian_PhysFluids}
T.~Ruggeri and S.~Taniguchi, {\slshape A complete classification of sub-shocks
  in the shock structure of a binary mixture of {E}ulerian gases with different
  degrees of freedom,} {\em Physics of Fluids} {\bfseries 34} (2022) 066116.

\bibitem{ShockBinaryET6_RdM}
T.~Ruggeri and S.~Taniguchi, {\slshape Shock structure and sub-shocks formation
  in a mixture of polyatomic gases with large bulk viscosity,} {\em Ricerche di
  Matematica, in press} (2023) .

\bibitem{MRSimic}
D.~Madjarevi\'c, T.~Ruggeri, and S.~Simi\'c, {\slshape Shock structure and
  temperature overshoot in macroscopic multi-temperature model of mixtures,}
  {\em Physics of Fluids} {\bfseries 26} (2014) 106102.

\bibitem{RuggeriSimic2008proceedings}
T.~Ruggeri and S.~Simi\'c, {\slshape Mixture of gases with multi-temperature:
  Identification of a macroscopic average temperature,} in {\em Memorie
  dell'Accademia delle Scienze, Lettere ed Arti di Napoli, Proceedings
  Mathematical Physics Models and Engineering Sciences}, pp.~455--465.
\newblock 2008.
\newblock \url{http://www.societanazionalescienzeletterearti.it/pdf/Memorie}.

\bibitem{GouinRuggeri}
H.~Gouin and T.~Ruggeri, {\slshape Identification of an average temperature and
  a dynamical pressure in a multitemperature mixture of fluids,} {\em Phys.
  Rev. E} {\bfseries 78} (Jul, 2008) 016303.

\bibitem{RuggeriSimic2009}
T.~Ruggeri and S.~Simi\'c, {\slshape Average temperature and {M}axwellian
  iteration in multitemperature mixtures of fluids,} {\em Phys. Rev. E}
  {\bfseries 80} (Aug, 2009) 026317.

\bibitem{SimicMT}
T.~Ruggeri and S.~Simi\'c, {\slshape On the hyperbolic system of a mixture of
  {E}ulerian fluids: a comparison between single- and multi-temperature
  models,} {\em Mathematical Methods in the Applied Sciences} {\bfseries 30}
  (2007) 827--849.

\bibitem{Boillat-1997}
G.~Boillat and T.~Ruggeri, {\slshape Hyperbolic principal subsystems: Entropy
  convexity and subcharacteristic conditions,} {\em Archive for Rational
  Mechanics and Analysis} {\bfseries 137} (Jun, 1997) 305--320.

\bibitem{Brini_Osaka}
F.~Brini and T.~Ruggeri, {\slshape On the {R}iemann problem in extended
  thermodynamics,} in {\em Proceedings of the 10th International Conference on
  Hyperbolic Problems (HYP2004)}, pp.~319--326.
\newblock Yokohama Publisher Inc., Yokohama, 2006.

\bibitem{Brini_Wascom}
F.~Brini and T.~Ruggeri, {\slshape The {R}iemann problem for a binary
  non-reacting mixture of {E}uler fluids,} in {\em Proceedings XII Int.
  Conference on Waves and Stability in Continuous Media}, R.~Monaco,
  S.~Pennisi, S.~Rionero, and T.~Ruggeri, eds., pp.~102--108.
\newblock World Scientific, Singapore, 2004.

\bibitem{MentrelliRuggeri}
A.~Mentrelli and T.~Ruggeri, {\slshape Asymptotic behavior of {R}iemann and
  {R}iemann with structure problems for a 2$\times$2 hyperbolic dissipative
  system,} {\em Suppl. Rend. Circ. Mat. Palermo II} {\bfseries 78} (2006)
  201--225.

\bibitem{Liu_conjecture}
T.-P. Liu, {\em Nonlinear hyperbolic-dissipative partial differential
  equations}, pp.~103--136.
\newblock Springer, Berlin, Heidelberg, 1996.

\bibitem{Liu_struct1}
T.-P. Liu, {\slshape Linear and nonlinear large-time behavior of solutions of
  general systems of hyperbolic conservation laws,} {\em Communications on Pure
  and Applied Mathematics} {\bfseries 30} (1977) 767--796.

\bibitem{Liu_struct2}
T.-P. Liu, {\slshape Large-time behavior of solutions of initial and
  initial-boundary value problems of a general system of hyperbolic
  conservation laws,} {\em Communications in Mathematical Physics} {\bfseries
  55} (Jun, 1977) 163--177.

\bibitem{BriniRuggeri}
F.~Brini and T.~Ruggeri, {\slshape On the {R}iemann problem with structure in
  extended thermodynamics,} {\em Suppl. Rend. Circ. Mat. Palermo II} {\bfseries
  78} (2006) 31--43.

\bibitem{UCS2}
S.~F. Liotta, V.~Romano, and G.~Russo, {\slshape Central schemes for balance
  laws of relaxation type,} \href{http://www.jstor.org/stable/3061925}{{\em
  SIAM Journal on Numerical Analysis} {\bfseries 38} (2001) 1337--1356}.

\bibitem{RuggeriSerre}
T.~Ruggeri and D.~Serre, {\slshape Stability of constant equilibrium state for
  dissipative balance laws system with a convex entropy,} {\em Quart. Appl.
  Math.} {\bfseries 62} (2004) 163–179.

\bibitem{Yong}
W.-A. Yong, {\slshape Entropy and global existence for hyperbolic balance
  laws,} {\em Arch. Rational Mech. Anal.} {\bfseries 172} (2004) 247–266.

\bibitem{Bianchini}
S.~Bianchini, B.~Hanouzet, and R.~Natalini, {\slshape Asymptotic behavior of
  smooth solutions for partially dissipative hyperbolic systems with a convex
  entropy,} {\em Comm. Pure Appl. Math.} {\bfseries 60} (2007) 1559–1622.

\end{thebibliography}\endgroup

\end{document}